\documentclass[screen]{acmart}

\usepackage{threeparttable}
\usepackage[export]{adjustbox}

\usepackage[yyyymmdd]{datetime}

\usepackage{tabularx}
\usepackage{multirow}
\usepackage{dcolumn} 
\newcolumntype{d}[1]{D{.}{.}{#1}}

\usepackage{subcaption}
\usepackage{enumitem}

\usepackage{todonotes}
\let\xtodo\todo
\renewcommand{\todo}[1]{\xtodo[inline,color=green!50]{#1}}

\newcommand{\beforeparagaph}{\vspace{-0.5em}}

\newcommand{\pquote}[1]{\textit{``#1''}}
\newcommand{\sysname}{CoGen3D}
\AtBeginDocument{%
  }

\setcopyright{none}

\begin{document}

\title{\sysname{}: An Agentic Human-AI Co-Design Pipeline for 3D Asset Generation for Virtual Reality}

\author{Weiwei Jiang}
\affiliation{%
  \institution{Nanjing University of Information Science and Technology}
  \city{Nanjing}
  \country{China}
}
\email{weiweijiangcn@gmail.com}

\author{Wanyu He}
\affiliation{%
  \institution{Nanjing University of Information Science and Technology}
  \city{Nanjing}
  \country{China}
}

\author{Zheyu Tan}
\affiliation{%
  \institution{Japan Advanced Institute of Science and Technology}
  \city{Nomi}
  \country{Japan}
}
\email{zheyutan@jaist.ac.jp}

\author{Zheyuan Kuang}
\affiliation{%
  \institution{The University of Sydney}
  \city{Sydney}
  \country{Australia}
}

\author{Difeng Yu}
\affiliation{%
  \institution{University of Copenhagen}
  \city{Copenhagen}
  \country{Denmark}
}

\author{Shinobu Hasegawa}
\affiliation{%
  \institution{Japan Advanced Institute of Science and Technology}
  \city{Ishikawa}
  \country{Japan}
}

\author{Sven Mayer}
\affiliation{%
  \institution{TU Dortmund University}
  \city{Dortmund}
  \country{Germany}
}
\affiliation{%
  \institution{Research Center Trustworthy Data Science and Security}
  \city{Dortmund}
  \country{Germany}
}
\email{info@sven-mayer.com}

\author{Zhanna Sarsenbayeva}
\affiliation{%
  \institution{The University of Sydney}
  \city{Sydney}
  \country{Australia}
}

\renewcommand{\shortauthors}{W. Jiang et al.}

\begin{abstract}
Creating 3D assets for virtual reality requires modeling expertise, which restricts the authorship of immersive experiences. 
Existing generative AI tools rely on unconstrained, command-driven prompting, lacking the conversational scaffolding needed for users to articulate their intent and validate designs prior to rendering. 
To address this, we introduce \sysname{}, an agentic human-AI co-design pipeline that proactively guides users through conversational intent elicitation, a concept image confirmation, and image-to-3D generation that directly deploys to immersive scenes. 
We evaluated this system through a user study (N=120) across six affectively diverse immersive scenes, observing 60 \textit{Design group} participants who co-created 3D assets for the scenes, and 60 \textit{Validation group} participants who experienced the scenes with generated assets.
Our findings show that co-designed assets are associated with higher scene engagement and shifted affective responses, while participants generally preferred concept images over the final 3D assets, with no increased leniency toward degradation in their own creations. Analysis of the human-AI conversations further shows that target environments shape users' conversational patterns.
Our results suggest that our staged, intent-based co-design can democratize virtual reality authoring and shift immersive content creation from technical execution toward collaborative spatial design.
\end{abstract}




\maketitle

\section{Introduction}
\label{sec:intro}
Immersive technologies have the potential to facilitate engaging and transformative experiences across a multitude of domains. Nevertheless, the realization of this potential is fundamentally contingent upon the availability of rich, scene-congruent 3D content. In particular, creating a 3D asset requires manual effort and proficiency in professional modeling software. This barrier effectively confines the authorship and personalization of immersive environments to experts who have proficient experience. In the meantime, communicating between end-users and VR developers remains a  challenge~\cite{ashtari2020creating}. This makes designing and deploying virtual reality (VR) experiences time-consuming, limiting the accessibility of immersive content creation.

Recent advancements in artificial intelligence (AI) have catalyzed a paradigm shift, demonstrating promising efficacy in AI-assisted extended reality (XR) authoring to ease the immersive content creation. For example, recent systems such as VRCopilot~\cite{zhang2024vrcopilot} and ImaginateAR~\cite{lee2025imaginatear} have successfully integrated large language models (LLMs) and generative pipelines to accelerate spatial prototyping and augmented reality scene composition. Other approaches have explored gesture-driven 3D prop specification~\cite{yao2026gestuprop}, the dynamic synthesis of 2D panoramic skyboxes, ambient audio, and pre-authored conversational agents for therapeutic VR sanctuaries~\cite{wen2026prompt}, hybrid natural-language and direct-manipulation radiance-field editing~\cite{vachha2025dreamcrafter}, and AI-assisted co-design for situated safety authoring in Augmented Reality (AR)~\cite{li2026safetybuilder}. 
With the continuous evolution of generative methods, including text-to-image~\cite{rombach2022high-resolution, esser2024scaling} and image-to-3D synthesis~\cite{liu2023zero-1-to-3, zhao2025hunyuan3d}, the technical feasibility of generating 3D assets from natural language or visual prompts has been established, and the research community has begun to explore how these capabilities can be integrated into interactive authoring tools.
Nevertheless, a critical limitation of existing immersive generative systems is their persistent reliance on command-driven interaction models, wherein users interface with an unconstrained prompt box, and the system passively executes the directive. 
This paradigm lacks the scaffolding to enable users to articulate their design intentions, iteratively negotiate design parameters, or validate conceptual formulations before initiating computationally intensive 3D generation. 

We present \sysname{}, an agentic human-AI co-design pipeline developed to situate natural language conversation as the central locus of the 3D authoring process to further simplify immersive content creation. \sysname{} scaffolds non-expert users through three deliberate stages to address the limitations of single-shot prompting. Initially, an LLM-driven conversational agent actively elicits user preferences pertaining to object semantics, stylistic attributes, and virtual environmental consistency to establish a shared conceptualization of the target asset.
Subsequently, the system synthesizes a 2D concept image, facilitating user review, iterative refinement, and definitive confirmation. 
Finally, upon user confirmation, the system executes the computationally intensive image-to-3D conversion and deploys a textured mesh directly into a Unity-based VR runtime.
Our staged architecture strategically fronts human judgment during the rapid 2D ideation phase, thereby deferring the high-latency 3D modeling process until the creative trajectory is unequivocally established.

To evaluate our system, we design a comprehensive user study with specific design goals for the participants. In particular, we focus on designing assets for affective scenes using an open-sourced dataset that has been evaluated by multiple studies~\cite{jiang2024immersive,kuang2026understanding}. For the user study ($N=120$), we used six VR environments drawn from a validated affective stimulus set~\cite{jiang2024immersive,kuang2026understanding} to validate our system. Here, 60 participants designed assets for the scenes, and an additional 60 participants validated them in VR.

Experimental results show that our system successfully enables non-expert users to co-create and deploy 3D assets in VR, with co-designed assets significantly increasing scene engagement time and shifting emotional responses. In summary, our work makes the following contributions:
\begin{itemize}[itemsep=0em,leftmargin=*,topsep=0em]
    \item \textit{System:} We contribute \sysname{}, an open-source agentic co-design pipeline connecting staged conversational agents and generative endpoints, and an instrumented VR runtime for \textit{in-situ} applications and evaluations.
    \item \textit{Empirical evidence:} We contribute quantitative and qualitative findings detailing the impact of AI-generated assets on generation satisfaction, emotional shifts, human-AI interaction patterns, and in-scene engagement behaviors with AI-generated 3D assets.
    \item \textit{Design implications:} We articulate actionable insights for future VR authoring tools, highlighting the necessity of 2D confirmation gates and the cognitive influence of environmental emotion on the design process.
\end{itemize}

\section{Related Work}
\label{sec:related}

Prior studies in generative 3D modeling, AI-assisted XR authoring, and human-AI co-design show great potential of generative AI for 3D content creation but also reveal persistent challenges around various aspects. In this section, we summarize recent progress in these areas and position our work at the intersection of these strands with a focus on the transition points between conversational preference elicitation, 2D concept confirmation, and final 3D asset deployment in VR.

\subsection{Generative Models for 3D Asset Creation}
\label{sec:related:gen3d}

The landscape of 3D asset generation has been fundamentally reshaped by advancements in neural implicit representations and volumetric rendering. For example, early works like Vox-Fusion established viable pathways for integrating dense tracking and mapping into extended reality applications~\cite{yang2022vox-fusion}, while CoNeRF introduced critical user-controllability over latent radiance fields~\cite{kania2022conerf}. Following these foundational shifts, a cluster of subsequent research has rapidly refined the scalability and visual fidelity of these methods. These follow-ups include adaptive mesh recovery techniques~\cite{tang2023delicate} and Gaussian splatting optimizations that push toward dynamic tracking and physically based rendering~\cite{bao20253d, matsuki20254dtam, xiong2025texgaussian}. 

These works provide the robust rendering substrate necessary for interactive spatial systems. Thereafter, the field has moved toward semantic generation driven by natural language and visual prompts. Existing approaches have targeted the direct synthesis of high-resolution models and subject-driven personalization. For example, by utilizing coarse-to-fine diffusion strategies seen in Magic3D~\cite{lin2023magic3d} or the unified generation architectures of SSDNeRF and DreamBooth3D~\cite{chen2023single-stage, raj2023dreambooth3d}. To make these generations practical for real-time engines, subsequent works have focused on rapid mesh extraction and structural decomposition. By separating global semantics from local geometry or leveraging single-image priors, systems such as TextMesh~\cite{tsalicoglou2024textmesh}, One-2-3-45~\cite{liu2023one-2-3-45}, and SeparateGen~\cite{li2026separategen} have drastically reduced the computational overhead required to produce discrete, textured 3D assets~\cite{xiang2025structured}. Parallel deployment-focused pipelines have scaled these priors to room-level environments~\cite{hllein2023text2room} or utilized 2D supervision to bypass expensive 3D training data entirely~\cite{henderson2020leveraging, wen2022pixel2mesh++}.

Despite these algorithmic leaps, converting rapid 2D ideations into 3D geometry often degrades texture and structural fidelity. As prior research focuses on offline benchmarks rather than interactive workflows, optimal user intervention strategies remain unclear. To address this, our staged pipeline replaces one-shot generation with a mandatory 2D confirmation phase, allowing us to empirically measure how this human-in-the-loop checkpoint impacts final VR user satisfaction.

\subsection{AI-Assisted XR Authoring and Interaction}
\label{sec:related:xr}

Recent literature highlights a growing interest in embedding generative capabilities directly into extended reality (XR) authoring workflows. A notable milestone in this space is VRCopilot, which demonstrated that providing users with scaffolded intermediate representations yields significantly greater creative agency than fully automated scene generation~\cite{zhang2024vrcopilot}. Building on this paradigm of hybrid control, subsequent systems have explored diverse modalities to support in-situ creation. For instance, ImaginateAR facilitates rapid language-driven spatial prototyping in mobile augmented reality environments~\cite{lee2025imaginatear}, while systems like Dreamcrafter and RealityCrafter blend natural language prompting with direct manipulation proxies to enable the editing of high-latency radiance fields and mixed-reality reconstructions~\cite{vachha2025dreamcrafter, kim2025realitycrafter}. These efforts underscore a consensus that balancing AI automation with granular, user-directed interventions is critical for effective spatial editing~\cite{chen2026ai4xr, caetano2025agentic}.

Beyond high-level system architecture, effective immersive authoring requires translating user intent into spatially logical interactions. Systems such as GestuProp~\cite{yao2026gestuprop} and MIMIC~\cite{wanniarachchi2025mimic} emphasize that purely verbal commands are often insufficient for 3D contexts, demonstrating that coupling speech with gestures or multimodal relative instructions vastly improves how users specify object placement and properties. Complementary research has addressed the visual and physical coherence of these generated inserts, ranging from the extraction of robust generative lighting to ensure believable environmental blending~\cite{zhao2025clear, safaribazargani20262d}, to leveraging fine-grained microgestures and shared physical surfaces to anchor virtual artifacts~\cite{cai2025hpipainting, huang2024surfshare}. 

A parallel strand of research leverages these interactive authoring capabilities to craft deeply contextual and emotionally resonant experiences, such as co-creating personalized therapeutic sanctuaries~\cite{wen2026prompt}, reminiscence environments for older adults~\cite{li2025remverse}, or situated hazard simulations~\cite{li2026safetybuilder}. However, despite this robust ecosystem of interaction techniques and application domains~\cite{rahimi2025generative}, a research gap remains. Existing tools typically optimize scene layout logistics, specialized input modalities, or single-stage generation quality, yet rarely encompass the full lifecycle of conversational planning, iterative visual confirmation, and immersive deployment. By formalizing and explicitly evaluating this multi-stage pipeline against user-centered behavioral outcomes, \sysname{} serves as a crucial bridge between interaction-centric XR toolkits and model-centric generative pipelines.

\subsection{Human-AI Co-Design and Creativity Support}
\label{sec:related:codesign}

Human-AI co-design research consistently emphasizes that effective creativity support emerges from iterative dialogue rather than one-shot automation. Several foundational frameworks formalize collaborative design as a continuous cycle of exploration, interpretation, and evaluation~\cite{shneiderman2007creativity,davis2013human-computer}. Empirical studies of AI-assisted authoring reinforce this paradigm, demonstrating that creators struggle less with generating raw outputs than with articulating nuanced intent and preserving control over the model's behavior~\cite{gmeiner2023exploring, yang2020re-examining, zhou2024understanding}. This necessity for collaborative iteration is not unique to spatial design. In particular, it has proven essential for aligning AI capabilities with human constraints across a wide array of disciplines, including educational curriculum design, music co-creation, clinical decision-making, and social work~\cite{lin2021engaging, krol2025exploring, fouad2026human-centered, christensen2026discretionary, madaio2020co-designing, carlon2025bayesian}.

Within the specific context of 3D modeling, Req2CAD offers a highly relevant precedent. It demonstrated that progressive decomposition—translating functional requirements into editable CAD components—effectively helps users bridge the gap between ambiguous goals and concrete geometry~\cite{jing2026req2cad}. This logic directly inspires \sysname{}'s architectural choice to scaffold 3D generation through an initial conversational elicitation phase and a 2D confirmation gate. However, while prior literature establishes conversation as a methodological requirement for co-creation, it leaves open the question of how conversational structures interact with multimodal generation pipelines. Specifically, it underexplores how users navigate the inevitable quality loss between intermediate 2D concepts and final 3D outputs. We position our work at this intersection, explicitly evaluating how a multi-stage conversational pipeline shapes user satisfaction, affective outcomes, and behavioral engagement in VR.

\subsection{Affective Evaluation in Immersive Environments}
\label{sec:related:affect}

Emotion is a fundamental dimension of immersive experiences, where virtual presence and agency can significantly intensify or reshape affective responses~\cite{riva2007affective, estupinan2014can, somarathna2022virtual, jicol2021effects, chirico2018designing, li2017public}. Because this emotional resonance is highly sensitive to environmental context, prior frameworks emphasize multimodal design factors, such as visual styling, ambient audio, embodiment, and haptics, as active regulators of user affect rather than mere aesthetic choices~\cite{lottridge2011affective, bartram2017affective, jun2018full, kern2020influence, elor2021understanding, wagener2024moodshaper}. To study these dynamics, the research community has increasingly adopted standardized VR stimulus libraries, which provide replicable, pre-validated baselines for emotional elicitation across various experimental setups~\cite{schone2023library, jiang2024immersive, kuang2026understanding}.

To evaluate these experiences \textit{in situ} without disrupting ecological validity, the Self-Assessment Manikin (SAM) remains the practical and established standard for capturing the core dimensions of valence, arousal, and dominance~\cite{bradley1994measuring, russell1977evidence, xie2020applying, csikszentmihalyi1987validity}. These methodological foundations directly inform our evaluation strategy. However, while existing affective VR studies thoroughly examine pre-authored, static environments, they rarely feature scenes populated with user-generated or AI-assisted content. This leaves a critical gap in understanding how psychological co-authorship and the visual fidelity of generated 3D objects jointly shape emotional engagement. By deploying the \sysname{} pipeline within validated affective scenes, our work bridges this gap, providing controlled empirical evidence on how multi-stage AI co-design impacts spatial perception, user interaction behavior, and emotional outcomes in virtual reality.

\section{System Design}
\label{sec:system}
We developed \sysname{}, a system that connects a conversational LLM agent with text-to-image and image-to-3D generative models, and integrates the generated assets into a Unity-based VR runtime. The system is designed to support a staged co-design process where users first engage in a dialogue to elicit their preferences, then review and confirm a 2D concept image, and finally receive a generated 3D asset that can be manipulated in VR.

\begin{figure}
  \centering
  \includegraphics[width=\linewidth]{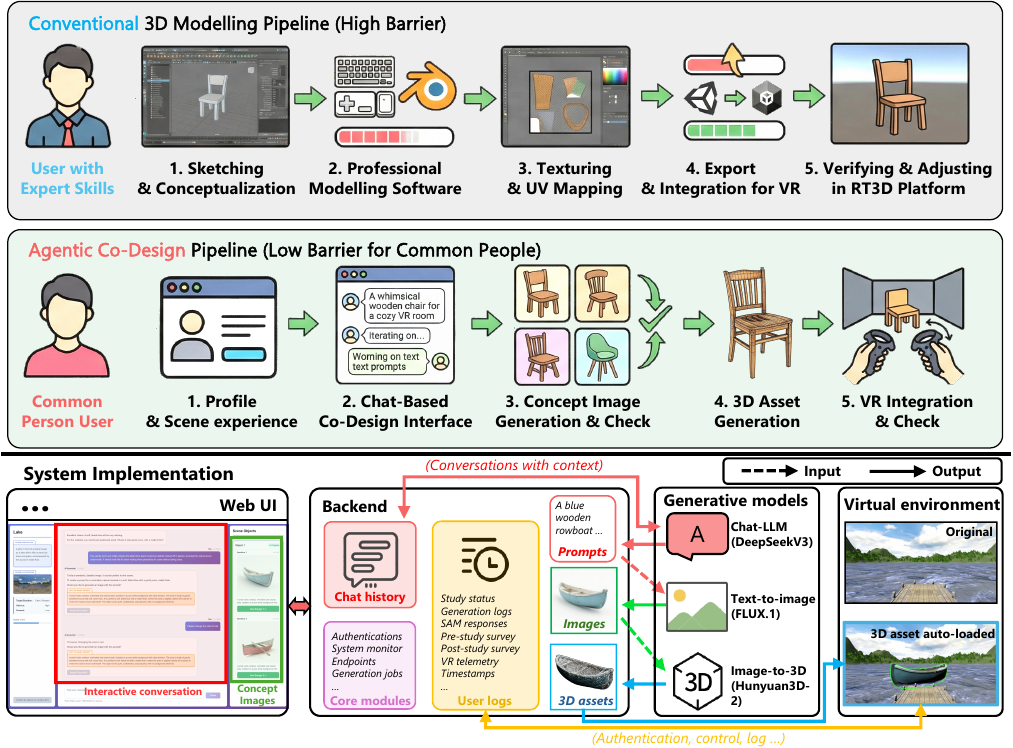}
  \caption{System design of \sysname{}. Top: the design pipeline of our system, compared with a conventional manual design process. Bottom: the implementation architecture of our system, which consists of a web-based co-design interface for users to interact with LLM agents, a custom backend connecting to a chat-based LLM (DeepSeek V3), a text-to-image model (FLUX.1), and an image-to-3D model (Hunyuan3D-2), and a Unity-based VR client that integrates the generated 3D assets for immersive experience.}
  \label{fig:system_design}
  \Description{Two-part architecture diagram. Top path shows manual 3D modeling versus a three-stage co-design pipeline. Bottom shows web client, Django backend, three model endpoints, and Unity VR runtime loading GLB assets.}
\end{figure}

In particular, \sysname{} comprises three coupled components: 1)~a browser-based co-design client where participants chat with an LLM agent, confirm prompts, and trigger generations; 2)~a Django backend that orchestrates DeepSeek V3, a self-hosted FLUX.1 endpoint, and a self-hosted Hunyuan3D-2 worker; and 3)~a Unity VR client that loads generated GLB assets at runtime.
\autoref{fig:system_design} summarizes the conventional manual workflow versus our staged pipeline and the deployment topology used in the user study.

\subsection{Pipeline Implementation}
\label{sec:system:implementation}

\subsubsection{Design Pipeline} 
As illustrated in \autoref{fig:system_design}, compared to the conventional manual design process, which requires users to have expertise in 3D modeling software and to manually create assets from scratch, our system implements a structured co-design pipeline that can understand users' intents from conversational dialogues and generate 2D concept images prior to 3D asset generation. In particular, our system guides users through three stages: 1) intent elicitation through agent-guided conversation, where the system asks targeted questions to understand the user's intent and preferences for the desired asset; 2) constrained image generation, where the system synthesizes a text prompt based on the elicited preferences and generates a 2D concept image, which is then confirmed by the user; and 3) image-to-3D conversion, where the confirmed 2D image is used as input to generate a 3D asset that can be integrated into VR scenes. This pipeline allows users to create 3D assets without requiring expertise in 3D modeling or prompt engineering, and it significantly reduces the time and effort required compared to manual design.

\subsubsection{System Implementation} 
Our system is a multi-agent architecture with a custom web-based interface for user interaction and a custom backend for instrumenting the co-design pipeline. The web-based interface is implemented using React.js, and the backend is implemented using Python with Django framework. The backend connects to three generative models through their respective APIs. For chat-based LLM, we use DeepSeek V3.2 through its official API endpoint, as it is resource-demanding for deploying a full-size model in a local machine (expected $\sim$$1543~GB$ GPU memory), while a distilled model may suffer from long-context understanding with limited capabilities~\cite{deepseek-ai2025deepseek-v3}. For both text-to-image generation and image-to-3D generation, we deployed the open-source models (FLUX.1~\cite{labs2025flux1kontextflowmatching} and Hunyuan3D-2~\cite{zhao2025hunyuan3d}) on our local servers with GPU acceleration (Nvidia A6000 48GB) with custom API endpoints. This allowed us to have more control over the generation process and to avoid unexpected API changes or rate limits that could arise with third-party services. 

\subsubsection{Agentic Elicitation and Conversation Design}
The DeepSeek V3.2 agent follows structured pre-design survey content (scene descriptions and design preferences such as objects, colors, and shapes~\cite{dresp2016affine,schloss2018modeling}). The agent is instructed to exhibit agentic behavior by acting as a collaborative designer through a structured system prompt. Rather than passively awaiting instructions, it proactively asks short, incremental questions about object type, materials, silhouette, palette, and narrative fit, and only proposes a draft text-to-image prompt after the user confirms that the major axes are captured.
The backend stores full chat transcripts and timestamps for each assistant and user turn for later latency analysis.
When the user accepts a prompt, the backend forwards it to FLUX.1. Alternatively, the user can choose to revise the design conversationally before the 3D asset generation job is enqueued.

\subsubsection{VR Client Integration}
The Unity client receives a GLB URL from the backend server, downloads the mesh at runtime, instantiates it with a default scale and orientation heuristic, and attaches VR manipulation components (grab, rotation, and uniform scale).
Users can reposition assets inside the tracked virtual play space. Telemetry streams transforms back to the study database for the engagement analyses reported later for every 100 milliseconds.
We did not run any offline mesh repair. Consequently, the loaded assets in VR scenes show the exact exported Hunyuan3D-2 meshes with textures.

\begin{figure}[t]
  \centering
  \begin{minipage}[t]{0.49\linewidth}
    \vspace{0pt}
    \centering
    \includegraphics[width=\linewidth]{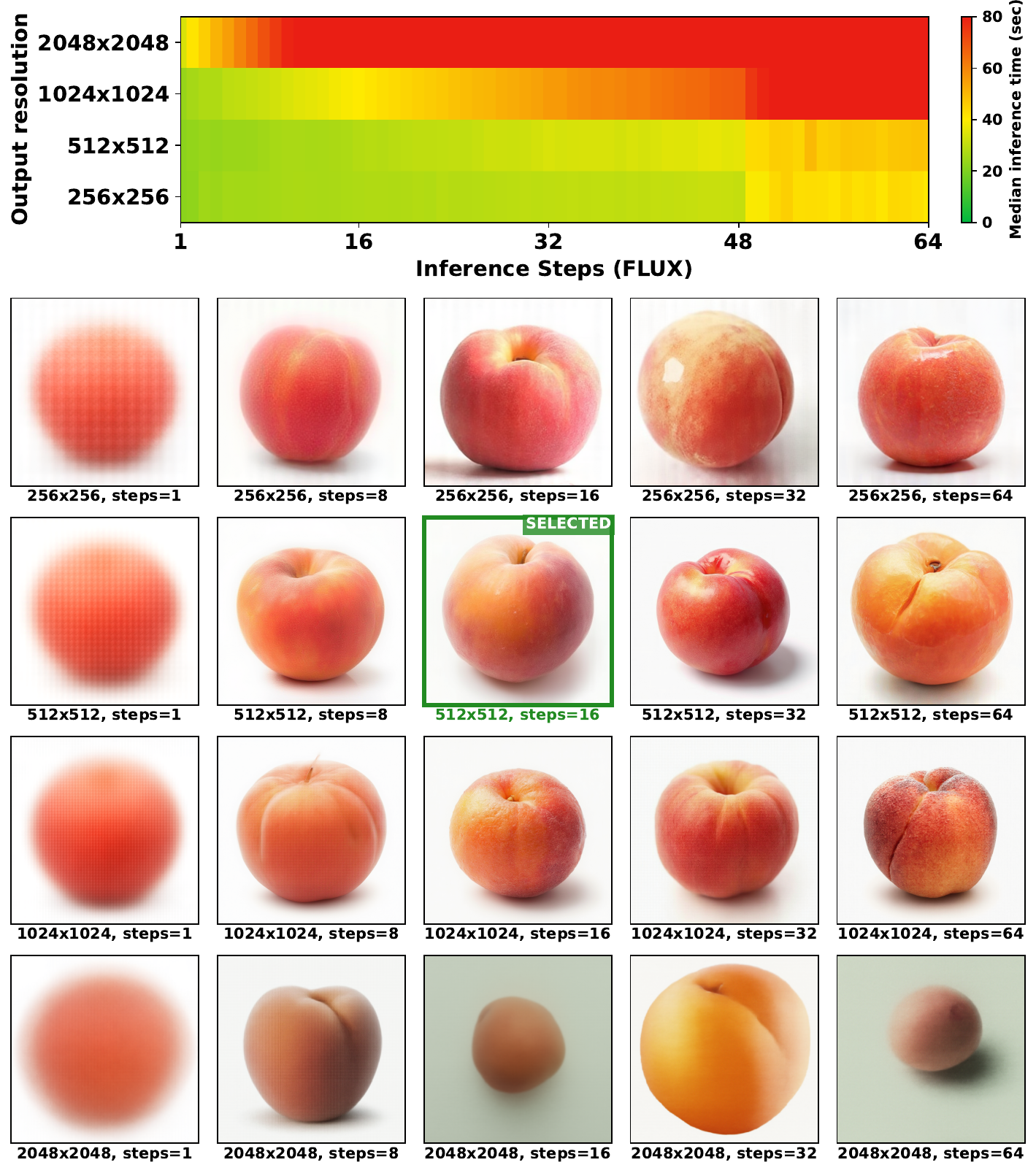}
  \end{minipage}\hfill
  \begin{minipage}[t]{0.49\linewidth}
    \vspace{0pt}
    \centering
    \includegraphics[width=\linewidth]{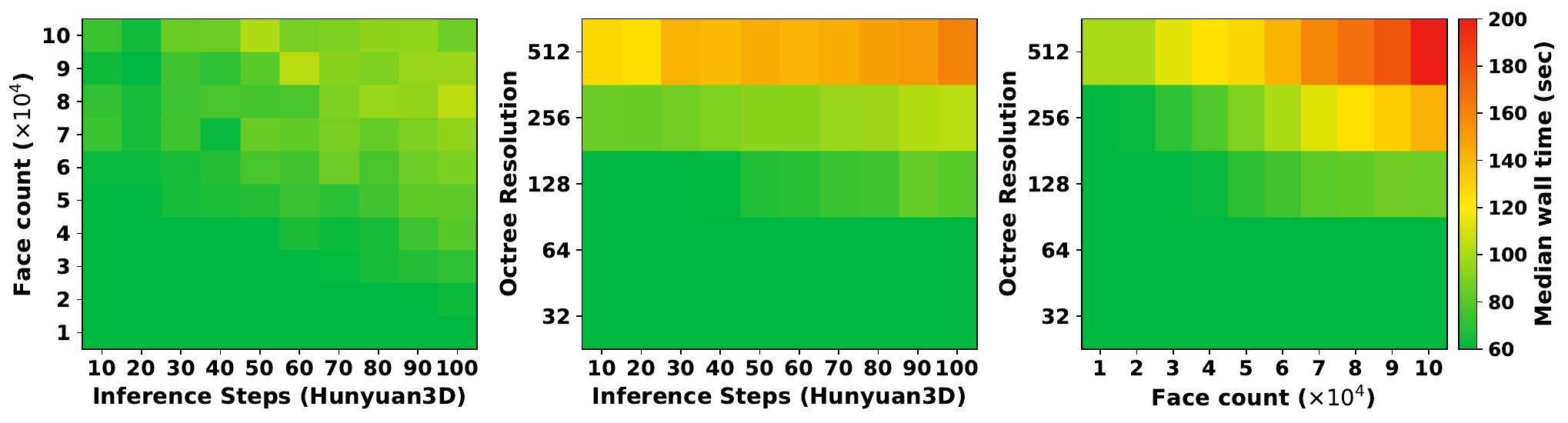}
    \\\smallskip
    \vspace{-1pt}
    \includegraphics[width=0.965\linewidth]{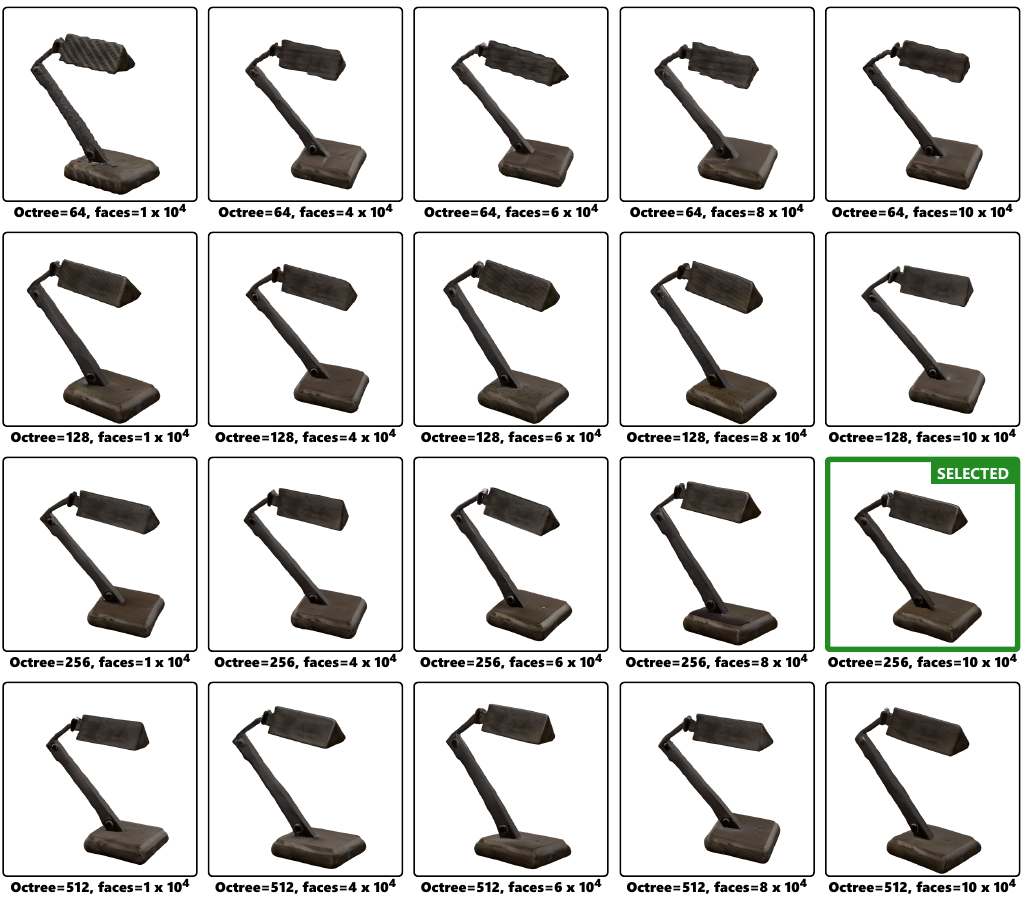}
  \end{minipage}
  \caption{Generation endpoint benchmarks. \textbf{Left:}~FLUX.1 wall-clock latency versus diffusion steps, faceted by output resolution. \textbf{Right}~Hunyuan3D-2 wall-clock latency with parameters between pairs of inference steps, octree resolution, and number of faces. The bottom parts of both panels show example outputs at the selected parameter settings.}
  \Description{Benchmarks generation latency against quality for FLUX.1 (left) and Hunyuan3D-2 (right). Heatmaps reveal that inference time scales with resolution, diffusion steps, and geometric complexity, while the corresponding image grids illustrate the resulting gains in visual fidelity. These metrics justify the selected endpoints, notably 512px at 16 steps for FLUX.1 and 256 octree resolution for Hunyuan3D-2, as the optimal balance between computational cost and output detail.}
  \label{fig:benchmark_latency}
\end{figure}

\subsection{Performance Test and Optimization}
\label{sec:system:performance}
We conducted server-side performance tests on the same hardware (an A6000 48~GB GPU) used in the study with a grid search to validate our system and select the optimal configuration for the user study. 
For FLUX.1, we varied the number of diffusion steps and the output resolution. For Hunyuan3D-2, we varied the number of diffusion steps, octree resolution, and face count.
\autoref{fig:benchmark_latency} reports the resulting wall-clock latency trends and samples of generated outputs at the selected parameter settings.
Benchmark sweeps indicated that moderate FLUX.1 settings preserve interactive responsiveness, whereas Hunyuan3D-2 requires higher compute budgets to stabilize geometry and texture quality.
Accordingly, we fixed the deployed study configuration as follows to balance quality, latency, and reproducibility:
\begin{itemize}
  \item \textbf{DeepSeek V3.2 (conversation):} temperature $=0.7$, max tokens $=2000$ for design dialogue (and $1000$ for preference summarization), to preserve creative diversity while maintaining responsive, parseable prompt outputs.
  \item \textbf{FLUX.1 (text-to-image):} resolution $512\times512$, guidance scale $=1.2$, inference steps $=16$, default seed $=42$, selected to keep concept generation responsive for interactive co-design.
  \item \textbf{Hunyuan3D-2 (image-to-3D):} GLB output with textures enabled, inference steps $=100$, guidance scale $=5.5$, octree resolution $=256$, target face count $=100{,}000$, seed $=42$, selected as the best quality-latency trade-off for VR-ready mesh export.
\end{itemize}

End-to-end logs also showed that median image generation remained substantially faster than median image-to-3D generation, consistent with users perceiving 2D generation as interactive and 3D generation as batch processing. In addition, we observed that Hunyuan3D-2 latency was more sensitive to input content due to the complexity of 3D geometry and texture synthesis. Under particular cases, image-to-3D conversion could take up to tens of minutes, causing long waits and potential workflow disruption. To mitigate this issue and ensure a smoother user experience, we used asynchronous job queues, stale-job detection, bounded retries, and periodic service health checks. A batched design process can further reduce perceived waiting by overlapping co-design and 3D processing, \textit{i.e.}, once a user confirms concept image~1, asset~1 can generate in the background while the user starts co-designing image~2. Subsequent confirmed images can then be queued in the same manner. This staggered workflow keeps users engaged while earlier 3D jobs run asynchronously.

\section{User Study}
\label{sec:method}


We conducted a between-subject user study to evaluate \sysname{} and understand how AI-assisted asset generation influences collaborative 3D design, content quality perception, and affective experience in immersive environments. In particular, we aim to evaluate the effectiveness of \sysname{} in supporting non-expert, scene-specific co-design from conversation to deployable VR assets, and to understand how the resulting generated content influences users' affective experience and engagement when integrated into immersive environments. 

To ground our evaluation, we designed our study around adding AI-generated assets to an open-source, previously validated framework of immersive VR environments. This allows us to leverage established emotional baselines and enable rigorous comparison of how generated content modulates affective response in controlled, replicable contexts. However, we emphasize that \sysname{} is not limited to these scenes. The co-design pipeline is generalizable to other VR environments. 

\textbf{Research Questions.}
Based on our objectives above, the study was designed to answer the following research questions:

\begin{enumerate}[label=\textbf{RQ\arabic*},topsep=1em]
\item How effectively can \sysname{} pipeline support non-expert, scene-specific co-design from conversation to deployable VR assets?

\item How do users perceive differences in output quality between intermediate 2D concept images and final 3D assets, and to what extent is this gap moderated by authorship context?

\item How does integrating generated assets into immersive scenes affect user's perceptions and behaviors, including affective responses (SAM~\cite{bradley1994measuring}), scene engagement time, and in-scene asset interaction patterns?
\end{enumerate}

\subsection{Study Protocol}
\label{sec:method:protocol}
\begin{figure}[t]
  \centering
  \includegraphics[width=\linewidth]{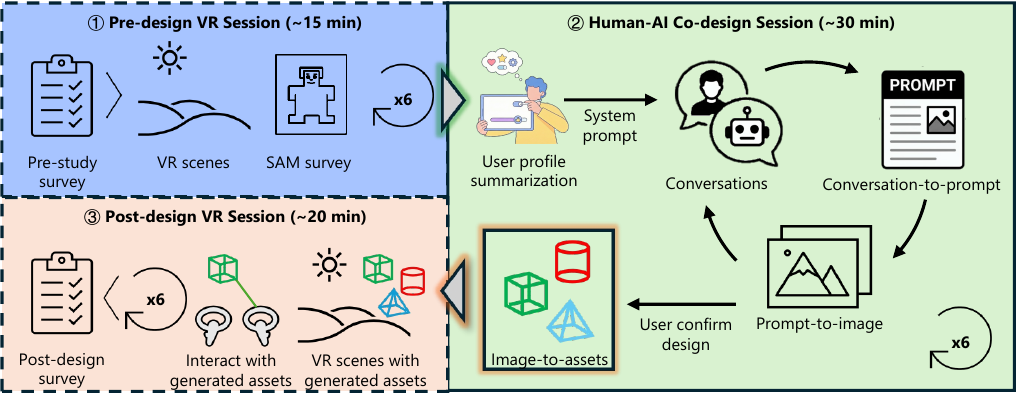}
  \caption{Study protocol for the experiments. For participants in the \textit{Design group}, the protocol consists of: 1) a pre-design VR session with a pre-study survey and a pre-design VR experience without generated assets; 2) a co-design session for generating an asset for each scene respectively; 3) a post-design VR session by interacting with generated assets in the corresponding scenes respectively, with a post-study survey. The participants in the \textit{Validation group} only experience the VR session with generated assets, with a pre-study survey and a post-study survey, respectively.}
  \Description{A horizontal flowchart with two parallel timelines. The top timeline (\textit{Design group}, N=60) runs left to right through four stages: a pre-study survey, a pre-design VR session in which participants experience the six scenes without any generated assets and submit SAM ratings, a co-design session on a web desktop with the LLM agent producing a text prompt, a 2D concept image, and a 3D asset for each scene, and a post-design VR session revisiting the six scenes populated with the participant's own assets, followed by a post-study survey. The bottom timeline (\textit{Validation group}, N=60) is shorter and consists of only a pre-study survey, a single VR session in which participants experience the six scenes populated with assets authored by two unrelated participants from the \textit{Design group} per scene and submit per-scene Self-Assessment Manikin (SAM)~\cite{bradley1994measuring} ratings, and a post-study survey with image and asset satisfaction items. Arrows connect the stages within each group, and both timelines share a common survey baseline at the far left.}
  \label{fig:study_protocol}
\end{figure}

We designed our study protocol to answer these questions as illustrated in \autoref{fig:study_protocol}. In particular, we assigned our participants into two groups: a \textit{Design group}, we refer to them as \textit{designers} throughout the paper; and a \textit{Validation group} (\textit{validators}), as detailed below. This design allows us to evaluate how authorship context relates to affective experience and content perception~\cite{norton2012ikea}. The study ethics was approved by our institutional review board, and all participants provided informed consent.

\subsection{Task}
\subsubsection{Design Group Task}
\label{sec:method:protocol:designer}
We asked participants in the \textit{Design group} to complete three subsequent sessions: a \textit{pre-design session}, a \textit{co-design session}, and a \textit{post-design session}. Each session was designed to capture different aspects of the co-design process and its impact on affective experience.

\textbf{Pre-design session ($\sim$15~min).} 
  After consent, participants were asked to complete a pre-study survey in the Web UI. The survey includes demographics, VR experience questions, and design preferences. 
  Then, participants were asked to experience the \emph{original} six VR scenes, including a short tutorial to learn basic controls and practice in-VR navigation. SAM~\cite{bradley1994measuring} (valence, arousal, dominance) ratings were collected after each scene. This part of the protocol is identical to previous studies~\cite{jiang2024immersive,kuang2026understanding}.

\textbf{Co-design session ($\sim$30~min).} 
  After the VR experience, participants returned to the Web UI. 
  The Web UI presented a scene description and keyword anchors for each of the six scenes. 
  They were instructed to design one asset per scene, with the goal of enhancing the scene's emotional impact based on their pre-design VR experience. 
  For each scene, they engaged in a conversation with the co-design agents, which asked targeted questions to elicit their design intents and synthesized a text prompt for text-to-image generation. Participants reviewed the generated 2D concept image and could request revisions before confirming it for 3D generation. This iterative process allowed participants to refine their vision for the asset while maintaining control over the creative direction. Upon commitment, the system queued the image-to-3D generation job and moved on to the next scene, allowing for a seamless workflow across all six scenes.

\textbf{Post-design session ($\sim$20~min).} 
  Participants re-entered VR to experience the same six scenes, integrated with the assets they designed. They were allowed to interact with the assets using standard VR manipulation (grabbing, scaling, rotating). 
  After the VR experience \emph{with} generated assets, they were asked to complete a post-study survey using the Web UI. The post-study survey includes per-scene satisfaction items for both the 2D concept images and the final 3D assets, Likert-scale items about their co-design experience (ease of use, helpfulness of the agent, creativity support, engagement), and free-text feedback on the overall workflow and any challenges they encountered.

\subsubsection{Validation Group Task}
\label{sec:method:protocol:validation}
Participants in the \textit{Validation group} completed a simplified protocol in which they experienced only VR scenes with generated assets. In particular, they first filled in the \textit{pre-survey}, which was the same as the \textit{Design group}. Then, participants experienced all six scenes, each containing two assets generated by two design participants, and provided in-VR SAM ratings after each scene. Finally, the participants completed a post-study survey that included ratings of 2D image and 3D asset satisfaction for each scene (ease of use, helpfulness of the agent, creativity support, engagement).

\begin{table*}[t]
  \centering
  \renewcommand{\arraystretch}{1.2} 
  \caption{Summary of Six Immersive VR Environments Spanning the Valence-Arousal Circumplex. Four quadrants are represented: High Arousal-High Valence (HAHV), High Arousal-Low Valence (HALV), Low Arousal-High Valence (LAHV), and Low Arousal-Low Valence (LALV). Each scene is designed to elicit distinct emotional responses, with brief descriptions and representative screenshots.}
  \label{tab:scenes_summary}
  \small
  \begin{tabularx}{\textwidth}{l l X c}
    \toprule
    \textbf{Scene Name} & \textbf{Quadrant} & \textbf{Description} & \textbf{Screenshot} \\
    \midrule
    \textit{Shouting Man with Gun} & \begin{tabular}[t]{@{}l@{}}High Arousal\\Low Valence\end{tabular} & 
    A furnished and well-illuminated attic where, after certain time, a man breaches the front door screaming loudly and aims a pistol at the player. & 
    \includegraphics[width=3.0cm, valign=t]{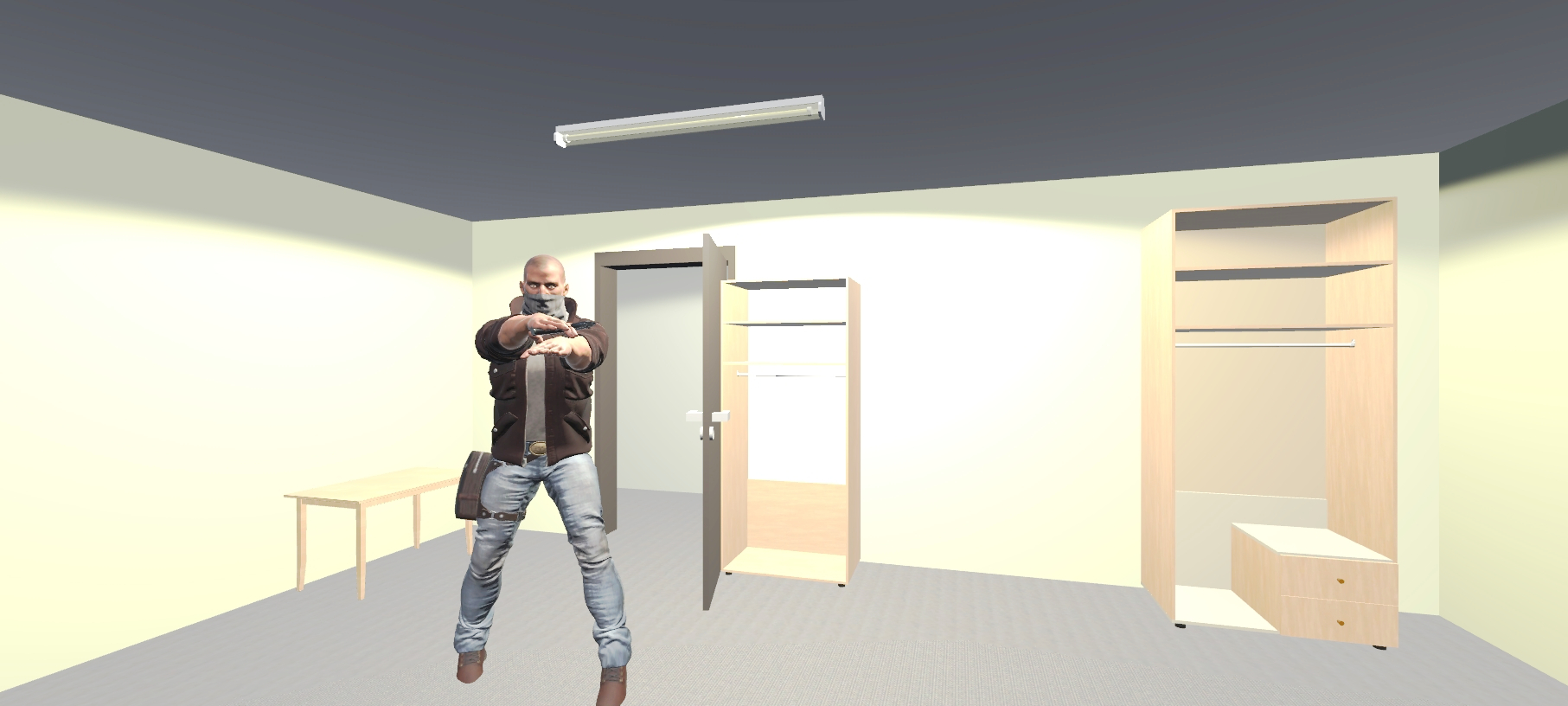} \\
    \midrule
    \textit{Surrounded by Elephants} & \begin{tabular}[t]{@{}l@{}}High Arousal\\High Valence\end{tabular} & 
    A open grassland with distant hills and a cloudy sky, where a herd of elephants gradually approaches, accompanied by wind and elephant trumpeting. & 
    \includegraphics[width=3.0cm, valign=t]{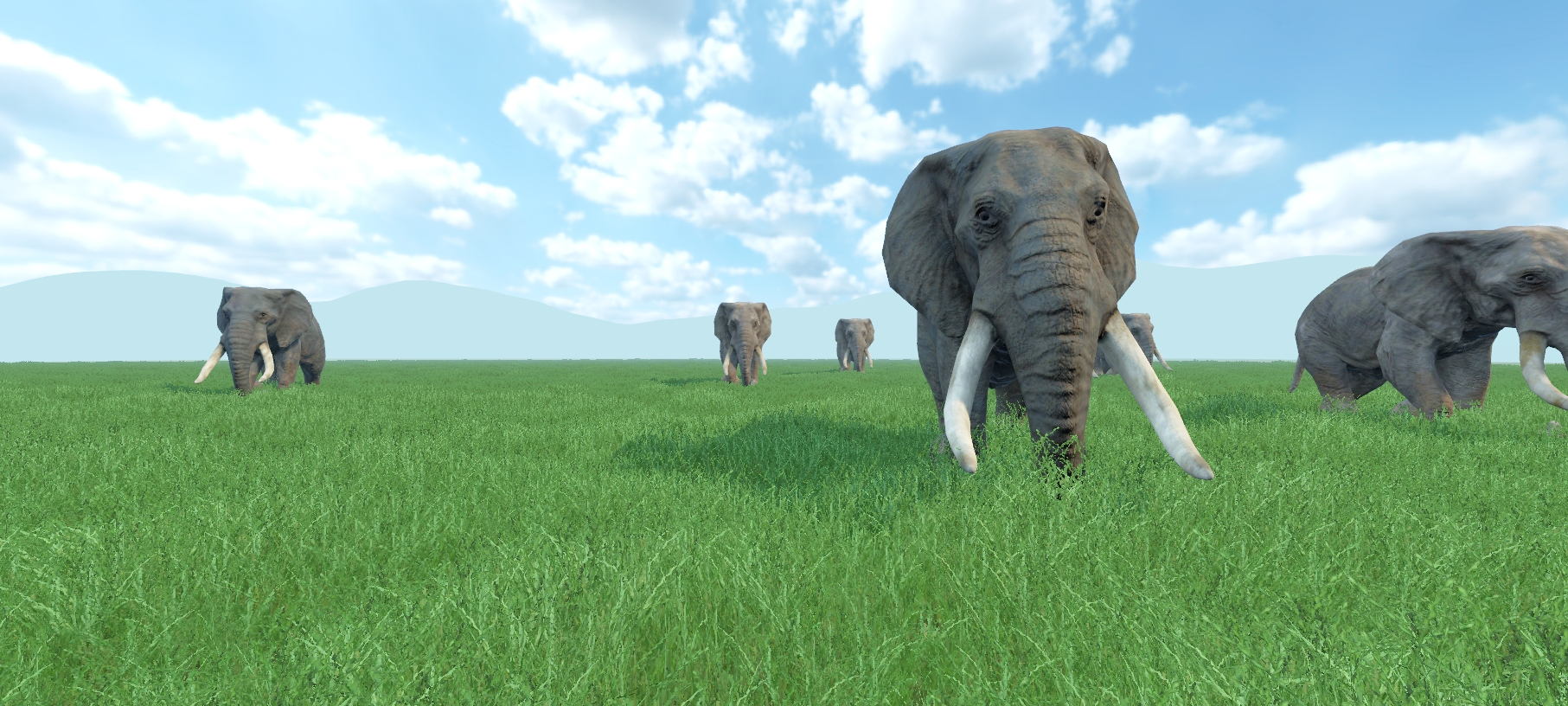} \\
    \midrule
    \textit{Jetty at Lake} & \begin{tabular}[t]{@{}l@{}}Low Arousal\\High Valence\end{tabular} & 
    A jetty in front of a stone house by a lake within hills covered by green trees and grass, accompanied by the sound of water flow. & 
    \includegraphics[width=3.0cm, valign=t]{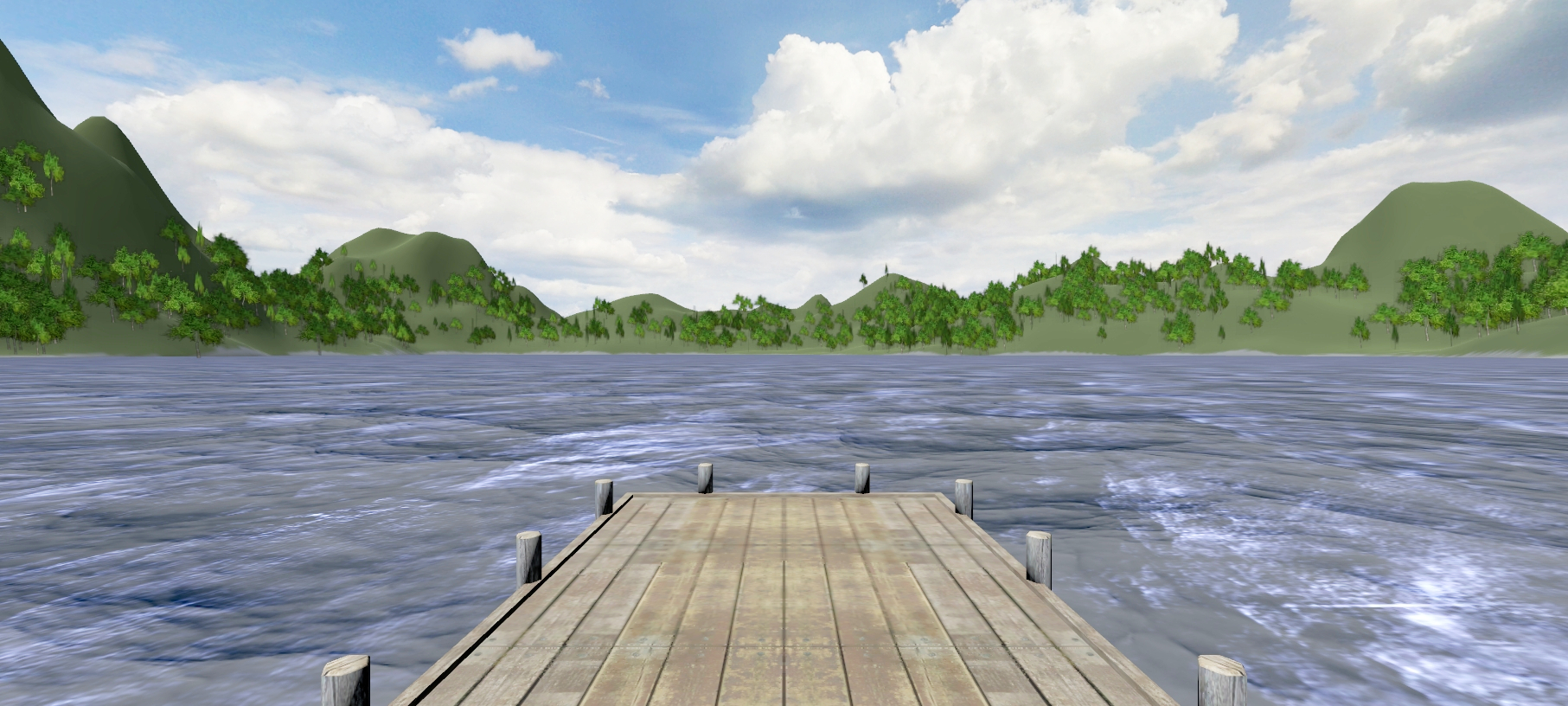} \\
    \midrule
    \textit{Puppies} & \begin{tabular}[t]{@{}l@{}}Low Arousal\\High Valence\end{tabular} & 
    A spacious and furnished room inside a house where several puppies walk playfully around. The scene is quiet. & 
    \includegraphics[width=3.0cm, valign=t]{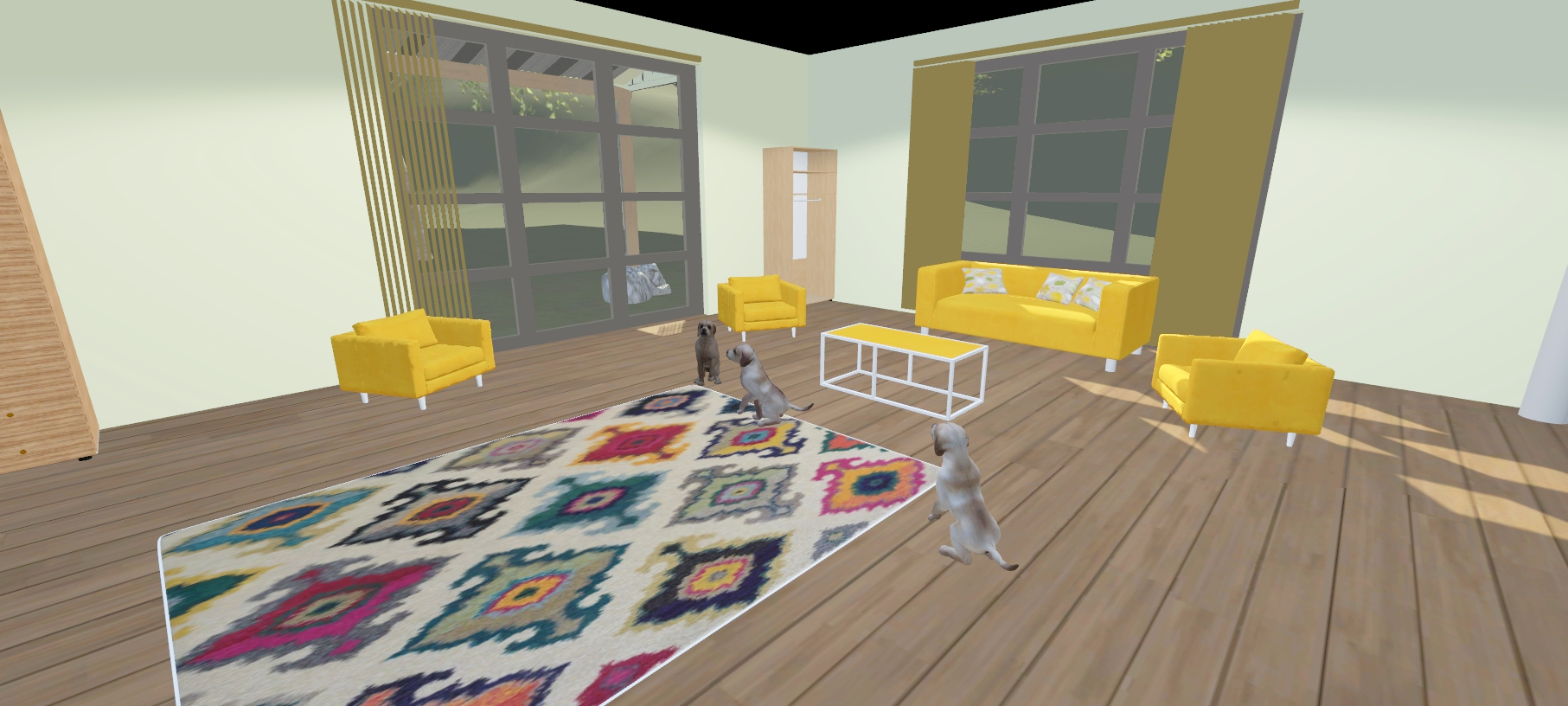} \\
    \midrule
    \textit{Solitary Confinement} & \begin{tabular}[t]{@{}l@{}}Low Arousal\\Low Valence\end{tabular} & 
    A gloomy cell containing a flickering light, a toilet set, and a single bed. The player hears water dropping and electricity flickering. & 
    \includegraphics[width=3.0cm, valign=t]{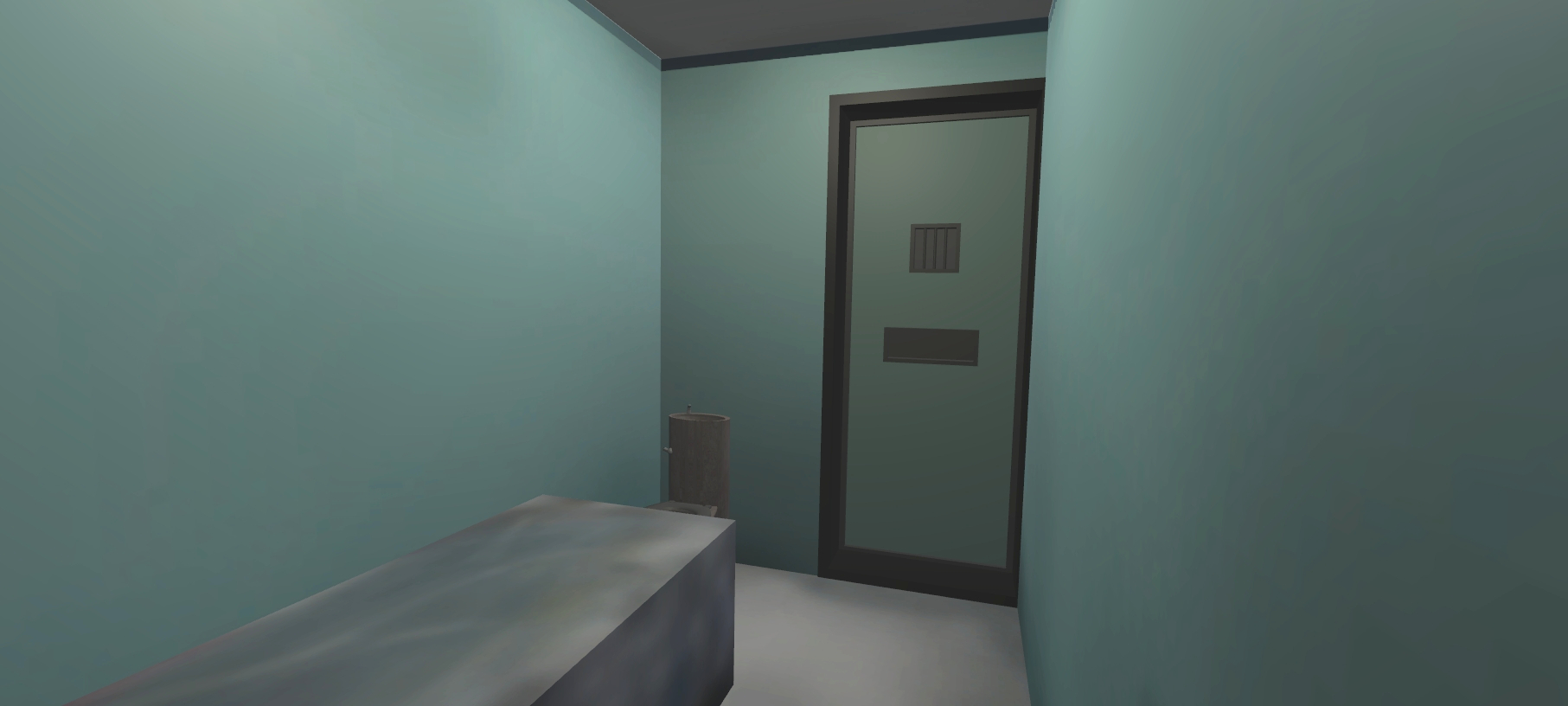} \\
    \midrule
    \textit{Tunnel} & \begin{tabular}[t]{@{}l@{}}Low Arousal\\Low Valence\end{tabular} & 
    A long tunnel illuminated by yellowish lights with a few pedestrians passing by, accompanied by the sound of footsteps. & 
    \includegraphics[width=3.0cm, valign=t]{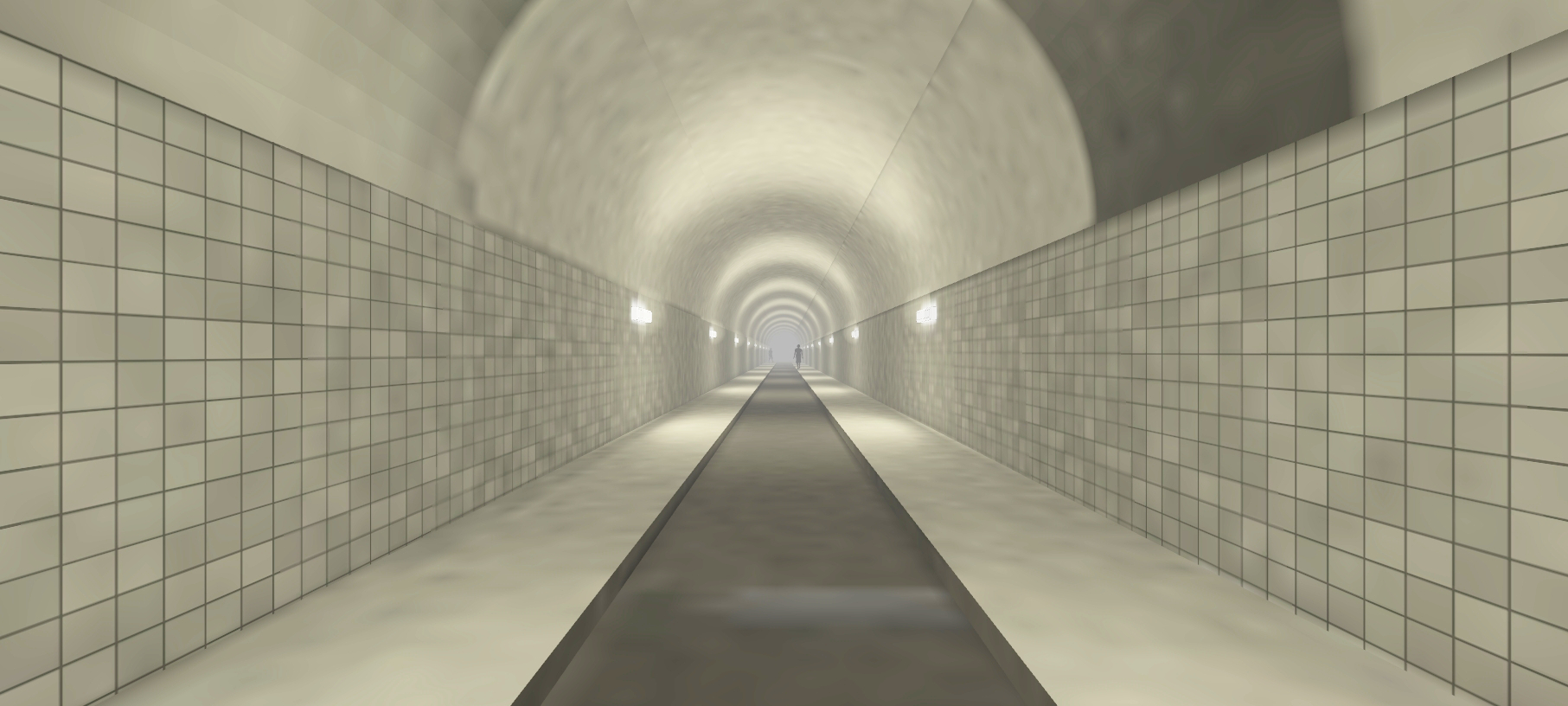} \\
    \bottomrule
  \end{tabularx}
\end{table*}

\subsection{Apparatus} 
\label{sec:method:scenes}

For the user study, we adopted six immersive environments previously validated for emotion elicitation~\cite{jiang2024immersive,kuang2026understanding}. Overall, these scenes span all four quadrants of Russell's valence--arousal circumplex~\cite{russell1980circumplex}: Low-Arousal High-Valence (LAHV), High-Arousal High-Valence (HAHV), Low-Arousal Low-Valence (LALV), and High-Arousal Low-Valence (HALV).

\autoref{tab:scenes_summary} provides a comprehensive summary of the six immersive environments. All scenes were built in Unity 6000.3.0f1 LTS against the Meta Quest runtime with OpenXR Plugin 1.16.1. The generated 3D assets were loaded with glTFast 6.15.1 and the Unlit Render Pipeline.

To probe potential effects of \emph{designer-authored} generated content (\textit{i.e.}, IKEA effect~\cite{norton2012ikea}), we presented each scene in an \emph{asset-free} form during the designers' pre-design phase, then re-used the same environments with inserted co-designed assets during the designers' post-design phase and the validation sessions.
To avoid confounding interaction telemetry with scene-native affordances, we did not reinstate interactive objects from the original benchmark scenes (\textit{e.g.}, flashlights, balls, riot shield in \cite{kuang2026understanding}), so recorded manipulation behavior can be attributed to \sysname{} assets.

\subsection{Measuremetns} 
\label{sec:data_collection}

\paragraph{Generated Content Ratings} 
Two items measured output quality perception: \emph{Image Satisfaction} (``How satisfied are you with the generated 2D image?'') and \emph{Asset Satisfaction} (``How satisfied are you with the generated 3D model?''), both on 5-point Likert scales (1 = very dissatisfied, 5 = very satisfied). 
\beforeparagaph
\paragraph{Behavioral Telemetry and Logs}
Unity clients streamed timestamped headset and object transforms, enabling scene dwell time, translation paths, rotation paths, and cumulative scale edits for users.
Generation workers logged FLUX.1 and Hunyuan3D-2 queue times for the analyses in \autoref{sec:results}.

\beforeparagaph
\paragraph{Chat History}
For each co-design session in the \textit{Design group}, we archived the complete human-AI conversation history, including message role (participant or assistant), message text, scene identifier, turn index, and server-side timestamps. We additionally logged prompt revisions and prompt-confirmation events that trigger generation requests, enabling reconstruction of the iterative preference-elicitation process and scene-level analyses of conversational pacing and design intent.

\beforeparagaph
\paragraph{Prompts and Qualitative Feedback}
The final confirmed text-to-image prompts were archived for semantic coding ($n{=}360$ lines across designers and scenes).
Participants in the \textit{Design group} also answered four Likert items about the co-design agent (ease of use, helpfulness, creativity support, engagement) and provided free-text reflections on the overall workflow.

\beforeparagaph
\paragraph{Emotional Responses} 
We used the 9-point pictographic SAM~\cite{bradley1994measuring} in VR after each scene for both groups.
For the \textit{primary} affective models reported below, we compare designers' \emph{pre-design} SAM (scenes without user-generated props) to validators' SAM collected after experiencing the same scenes \emph{with} assets authored by unrelated designers.

\subsection{Participants} 
In total, we recruited 120 participants from the local university community and assigned them to two groups. We first recruited a \textit{Design group} ($N=60$) that experienced them in VR and co-designed assets using \sysname{} (30 females, aged between 18 and 51, $M=23.87 \pm 7.13$), and then a \textit{Validation group} ($N=60$) that experienced the same VR scenes with generated assets they did not co-design (33 females, aged between 19 and 32, $M=22.27 \pm 3.44$). The experiments were conducted in a quiet office using a Meta Quest 3 headset in tethering mode. Participants received approximately 5 USD in compensation, consistent with the local hourly wage. 

\begin{figure}[t]
  \centering
  \includegraphics[width=\linewidth]{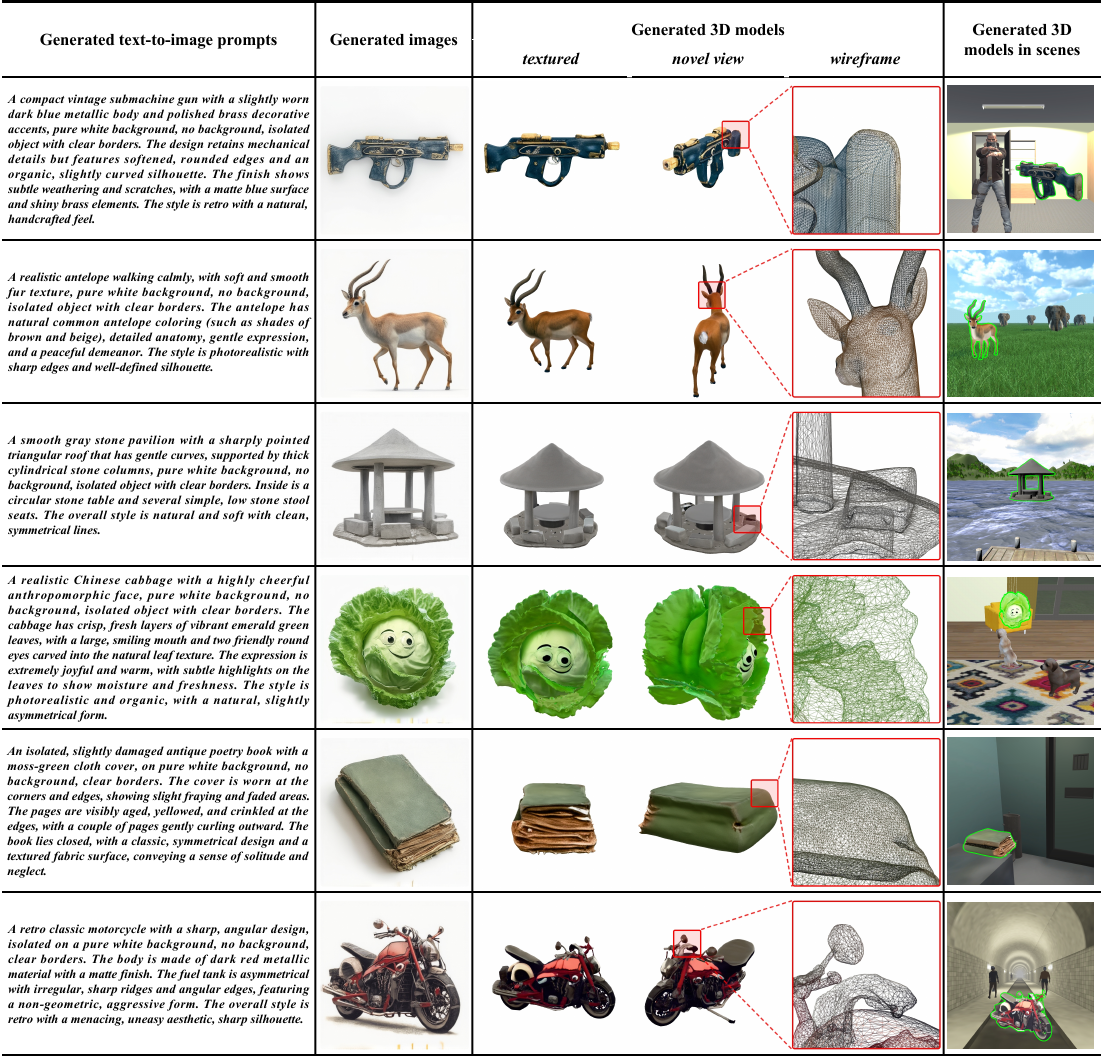}
  \caption{Examples of generated text-to-image prompts, images, and 3D assets for each of the six scenes. The left column shows the final prompts synthesized by the LLM agent confirmed by the participants; the second column shows the confirmed 2D concept images generated based on the prompts; the middle columns show the resulting 3D assets generated based on the 2D concept images, with a textured view, a novel view, and a part of a detailed wireframe view, respectively. The right column shows the final VR scenes with the generated 3D assets integrated with a green outline. The examples demonstrate that our system can generate semantically appropriate 2D images and 3D assets based on participants' intent.}
  \Description{A six-row grid in which each row corresponds to one of the six VR scenes. Row one (Tunnel) shows a text prompt describing a retro motorcycle with an asymmetrical fuel tank, a photorealistic 2D concept image of that motorcycle, three rendered views of the generated 3D motorcycle (textured, novel viewpoint, wireframe detail), and a screenshot of the Tunnel scene with the motorcycle placed inside and outlined in green. Subsequent rows follow the same five-column structure for Puppies (a stylized tennis ball), Jetty at Lake (a stone pavilion with a pointed roof), Solitary Confinement (a damaged antique poetry book with yellowed pages), Shouting Man with Gun (a compact vintage submachine gun with brass accents), and Surrounded by Elephants (an anthropomorphic Chinese cabbage figure). Across rows, the 2D concept images appear high-fidelity while the corresponding 3D assets show visibly reduced texture detail and smoother geometry, illustrating the image-to-3D quality bottleneck discussed in the text.}
  \label{fig:generation_samples}
\end{figure}

\section{Results}
\label{sec:results}
We analyzed satisfaction ratings with an ordinal cumulative-link mixed model. We analyzed SAM ratings with aligned rank transform (ART) ANOVA.
We analyzed time- and interaction-based outcomes with mixed-effects models, primarily Gamma GLMMs with participant random intercepts. We report likelihood-ratio tests for omnibus effects and use FDR-adjusted pairwise comparisons for post-hoc contrasts. Analyses were conducted in R (4.5.3) using \texttt{ordinal}, \texttt{ARTool}, \texttt{lme4}, \texttt{glmmTMB}, and \texttt{emmeans}.

\subsection{Generated Contents}
\label{sec:results_contents}

\autoref{fig:generation_samples} provides representative examples of the generated content across the six scenes, illustrating the progression from textual intent to deployable 3D asset. The examples demonstrate that the system successfully guides users to generate semantically appropriate objects with diverse aesthetic requirements. The detailed textual prompts reflect a highly successful preference elicitation process that captures specific user intents. For instance, participants successfully co-designed highly varied objects, ranging from an aggressive, retro-classic motorcycle with an asymmetrical fuel tank to playful scene-congruent objects such as a stylized tennis ball. The system also accommodated highly specific stylistic and structural requests, such as a smooth gray stone pavilion with a sharply pointed triangular roof, a damaged antique poetry book with yellowed pages, a compact vintage submachine gun with polished brass accents, and even an organic, cheerful anthropomorphic Chinese cabbage. 

While the generated 2D concept images are generally high-quality and accurately reflect the semantic themes of the prompts, \autoref{fig:generation_samples} also visually exposes a key bottleneck in the current generation pipeline. As shown in the transition from the 2D images to the 3D asset columns, there is a drop in output quality during the image-to-3D conversion stage. Specifically, the 3D assets often exhibit a loss of detailed texture, and the generated geometry sometimes fails to capture the finer structural features present in the original 2D concept images. This pattern is visually consistent across the diverse range of generated objects.

Furthermore, the final column of \autoref{fig:generation_samples} shows the generated 3D assets integrated into the VR scenes, demonstrating that they are functional and appropriately spawned within the immersive environments. Participants interacted with these assets in VR, and the system successfully supported the end-to-end co-design process from conversational intent elicitation to deployable content. 

\subsection{Generation Quality Assessment}
\label{sec:results_satisfaction}

\begin{figure}[t]
  \centering
  \includegraphics[width=\linewidth]{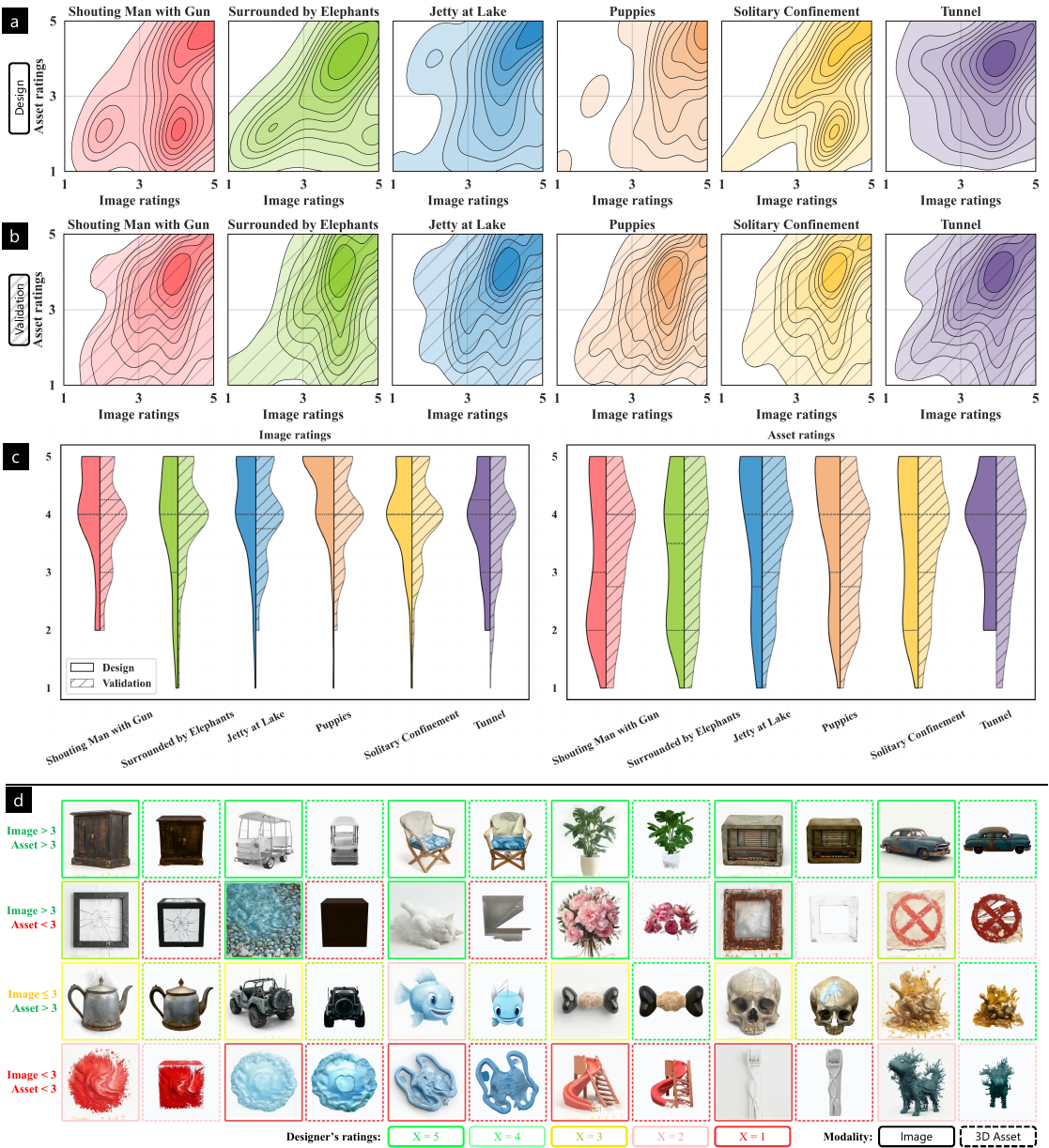}
  \caption{Generation quality assessments. a)--b) Kernel density plots of the joint distribution of image vs. asset satisfaction ratings for both \textit{Design} and \textit{Validation} groups across six VR scenes, respectively. c) Satisfaction rating comparison between groups. The lower 3D asset scores confirm a quality drop during the image-to-3D conversion. d) Representative pairs of 2D concept images and corresponding 3D assets, categorized by satisfaction thresholds ($X>$3 or $X<$3). Border colors indicate the designer's original rating.}
  \Description{A four-panel composite figure. Panels a and b are 5-by-5 grids of kernel-density contour plots with 2D image satisfaction (1--5) on the horizontal axis and 3D asset satisfaction (1--5) on the vertical axis, one panel for the \textit{Design group} and one for the \textit{Validation group}, with per-scene density blobs concentrated in the upper-right quadrant (both ratings high) but with heavier mass spreading toward the right edge than the top edge, indicating image satisfaction tends to exceed asset satisfaction. Panel c is a set of paired violin or box plots contrasting image-satisfaction and asset-satisfaction distributions by scene and by group, where asset-satisfaction distributions sit systematically lower than image-satisfaction distributions. Panel d is a matrix of paired thumbnails: each cell shows a 2D concept image next to its rendered 3D counterpart, grouped into quadrants defined by whether the image and asset satisfaction ratings were above or below 3, with each pair outlined by a color bar running from green (rating 5) through yellow to red (rating 1) to indicate the designer's score.}
  \label{fig:generation_satisfactions}
\end{figure}

Both the \textit{Design group} ($N = 60$) and the \textit{Validation group} ($N = 60$) evaluated the quality of the generated outputs on 5-point Likert scales for two modalities: the 2D concept image and the resulting 3D asset. In rating-level terms, this yielded 720 observations from the \textit{Design group} ($60 \times 6 \times 2$ modalities) and 1{,}440 from the \textit{Validation group} ($60 \times 6 \times 2$ assets $\times 2$ modalities). To account for the ordinal response scale and the crossed structure of these ratings, we fitted a cumulative-link mixed model with fixed effects of Group, Scene, Modality, and their interactions, adjusting for age, gender, and prior VR experience, with crossed random intercepts for participant and asset. \autoref{fig:generation_satisfactions} shows the scene-level satisfaction distributions across both groups and modalities.

\begin{table*}[t]
    \centering
    \begin{minipage}[t]{0.40\linewidth}
        \centering
        \captionof{table}{Type III omnibus tests for a cumulative-link mixed model (logit link) of satisfaction ratings. The model was Satisfaction $\sim$ Group $\times$ Scene $\times$ Modality + Age + Gender + VR Experience + (1\,|\,Participant) + (1\,|\,Asset).}
        \label{tab:satisfaction_glmm_type3_results}
        \scriptsize
        \resizebox{\linewidth}{!}{%
        \begin{tabular}{lrrr}
            \toprule
            \textbf{Estimator} & \textbf{$\chi^2$} & \textbf{df} & \textbf{$p$} \\
            \midrule
            Group & 2.20 & 1 & .138 \\
            \textbf{Scene} & \textbf{19.82} & \textbf{5} & \textbf{.0013} \\
            \textbf{Modality} & \textbf{17.46} & \textbf{1} & \textbf{< .001} \\
            Age & 0.00 & 1 & .973 \\
            Gender & 0.24 & 1 & .626 \\
            VR Experience & 6.81 & 3 & .078 \\
            \textbf{Group $\times$ Scene} & \textbf{14.01} & \textbf{5} & \textbf{.016} \\
            Group $\times$ Modality & 2.46 & 1 & .117 \\
            Scene $\times$ Modality & 8.07 & 5 & .152 \\
            Group $\times$ Scene $\times$ Modality & 3.26 & 5 & .660 \\
            \bottomrule
        \end{tabular}%
        }
    \end{minipage}
    \hfill
    \begin{minipage}[t]{0.58\linewidth}
        \centering
        \captionof{table}{Fixed-effects estimates from the CLMM for generation satisfaction on the cumulative logit scale. Baseline levels are Design cohort, Shouting Man with Gun scene, Image modality, female gender, and highest-frequency VR use category. 
        }
        \label{tab:satisfaction_glmm_fixed_effects}
        \scriptsize
        \resizebox{\linewidth}{!}{%
        \begin{tabular}{lccc}
            \toprule
            \textbf{Estimator} & \textbf{Est.} & \textbf{SE} & \textbf{95\% CI} \\
            \midrule
            \textbf{Modality: Asset} & \textbf{$-$1.520{***}} & \textbf{0.364} & \textbf{[$-$2.23, $-$0.81]} \\
            Scene: Surrounded by Elephants & $-$0.411 & 0.378 & [$-$1.15, 0.33] \\
            Scene: Jetty at Lake & +0.132 & 0.380 & [$-$0.61, 0.88] \\
            \textbf{Scene: Puppies} & \textbf{+1.018{*}} & \textbf{0.396} & \textbf{[0.24, 1.79]} \\
            Scene: Solitary Confinement & $-$0.194 & 0.375 & [$-$0.93, 0.54] \\
            Scene: Tunnel & $-$0.545 & 0.374 & [$-$1.28, 0.19] \\
            Validation $\times$ Surrounded by Elephants & +0.653 & 0.428 & [$-$0.19, 1.49] \\
            Validation $\times$ Jetty at Lake & +0.107 & 0.429 & [$-$0.73, 0.95] \\
            Validation $\times$ Puppies & $-$0.730 & 0.444 & [$-$1.60, 0.14] \\
            Validation $\times$ Solitary Confinement & +0.496 & 0.425 & [$-$0.34, 1.33] \\
            Validation $\times$ Tunnel & +0.619 & 0.424 & [$-$0.21, 1.45] \\
            \bottomrule
        \end{tabular}%
        }
        \vspace{0.5ex}
        \raggedright\footnotesize
        \textit{Note.} * $p < .05$, ** $p < .01$, *** $p < .001$. Threshold parameters (cut-points) are omitted. Significant estimates are bolded.
    \end{minipage}
\end{table*}

The ordinal GLMM retained the central pattern observed in the descriptive distributions. Modality was a strong predictor of satisfaction ($\chi^2(1)=17.46$, $p<.001$): across scenes and groups, 2D concept images were associated with higher satisfaction than the final 3D assets. Image satisfaction means ranged from 3.87 to 4.45, whereas asset satisfaction means ranged from 3.33 to 3.85, reflecting a systematic quality drop introduced by the image-to-3D conversion stage. Scene also had a significant effect ($\chi^2(5)=19.82$, $p=.001$), indicating that output evaluations varied with environmental context. In contrast, the main effect of Group was not significant ($\chi^2(1)=2.20$, $p=.138$), suggesting no overall difference in how designers and validators used the satisfaction scale after accounting for scene, modality, covariates, and crossed random effects.

The interaction terms further support this interpretation. Group $\times$ Modality was not significant ($\chi^2(1)=2.46$, $p=.117$), nor was the three-way Group $\times$ Scene $\times$ Modality interaction ($\chi^2(5)=3.26$, $p=.660$), indicating that the image-to-3D quality gap was not reliably moderated by authorship. Scene $\times$ Modality was also not significant in the ordinal model ($\chi^2(5)=8.07$, $p=.152$), while Group $\times$ Scene was significant ($\chi^2(5)=14.01$, $p=.016$). Type III CLMM results and main fixed-effects estimates are summarized in \autoref{tab:satisfaction_glmm_type3_results} and \autoref{tab:satisfaction_glmm_fixed_effects}.

FDR-adjusted post-hoc contrasts clarified where the modality gap was strongest. Within the \textit{Validation group}, images were rated significantly higher than assets in every scene, including \textit{Shouting Man with Gun} ($p<.001$), \textit{Surrounded by Elephants} ($p<.001$), \textit{Jetty at Lake} ($p=.026$), \textit{Puppies} ($p<.001$), \textit{Solitary Confinement} ($p=.010$), and \textit{Tunnel} ($p=.004$). Within the \textit{Design group}, the same image-over-asset pattern was significant in five of six scenes: \textit{Shouting Man with Gun} ($p<.001$), \textit{Surrounded by Elephants} ($p=.004$), \textit{Jetty at Lake} ($p=.041$), \textit{Puppies} ($p<.001$), and \textit{Solitary Confinement} ($p=.032$), but not in \textit{Tunnel} ($p=.370$). Direct group contrasts were significant only in \textit{Puppies}, where participants in the \textit{Design group} gave higher ratings than in the \textit{Validation group} for both images ($p<.001$) and assets ($p=.004$); all other scene-by-modality group contrasts were non-significant.

These results highlight a clear bottleneck in the current generation pipeline. In particular, while the text-to-image stage produces outputs that both groups find moderately satisfying, the subsequent image-to-3D conversion introduces a consistent and perceptible quality loss. 

In addition, the non-significant Group main effect and Group $\times$ Modality interaction suggest that the observed quality gap is \textit{not} substantially influenced by authorship bias (often termed the ``IKEA effect''~\cite{norton2012ikea}). Participants who co-designed the assets were generally just as critical of the image-to-3D quality drop as independent validators encountering the assets for the first time. This suggests that the quality gap is perceived as an inherent property of the pipeline, not primarily as a consequence of psychological ownership in co-creation.

\subsection{Emotional Responses}
\label{sec:results_sam}
\begin{figure}[t]
  \centering
  \begin{minipage}[t]{0.13\linewidth}
    \vspace{0pt}
    \includegraphics[height=11.2em,keepaspectratio]{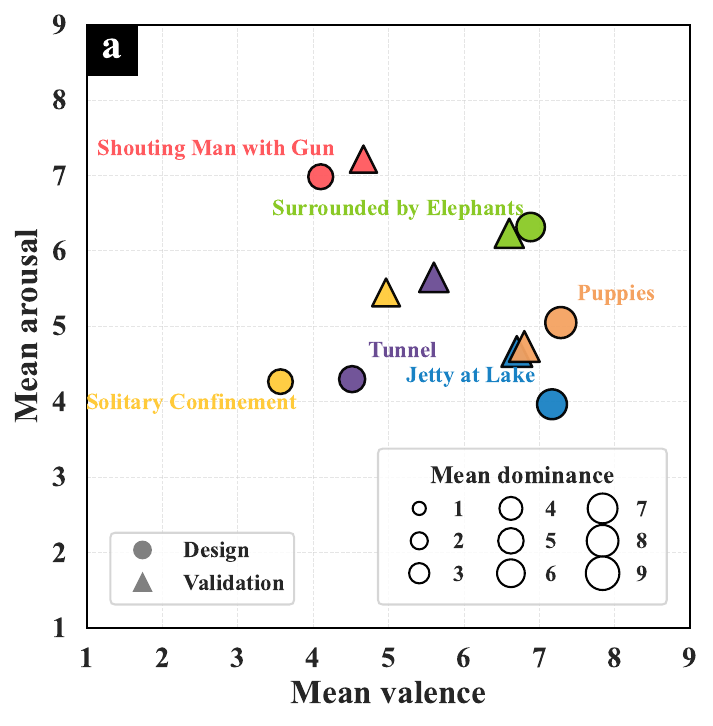}
  \end{minipage}\hfill
  \begin{minipage}[t]{0.80\linewidth}
    \vspace{1pt}
    \includegraphics[height=14.3em,keepaspectratio]{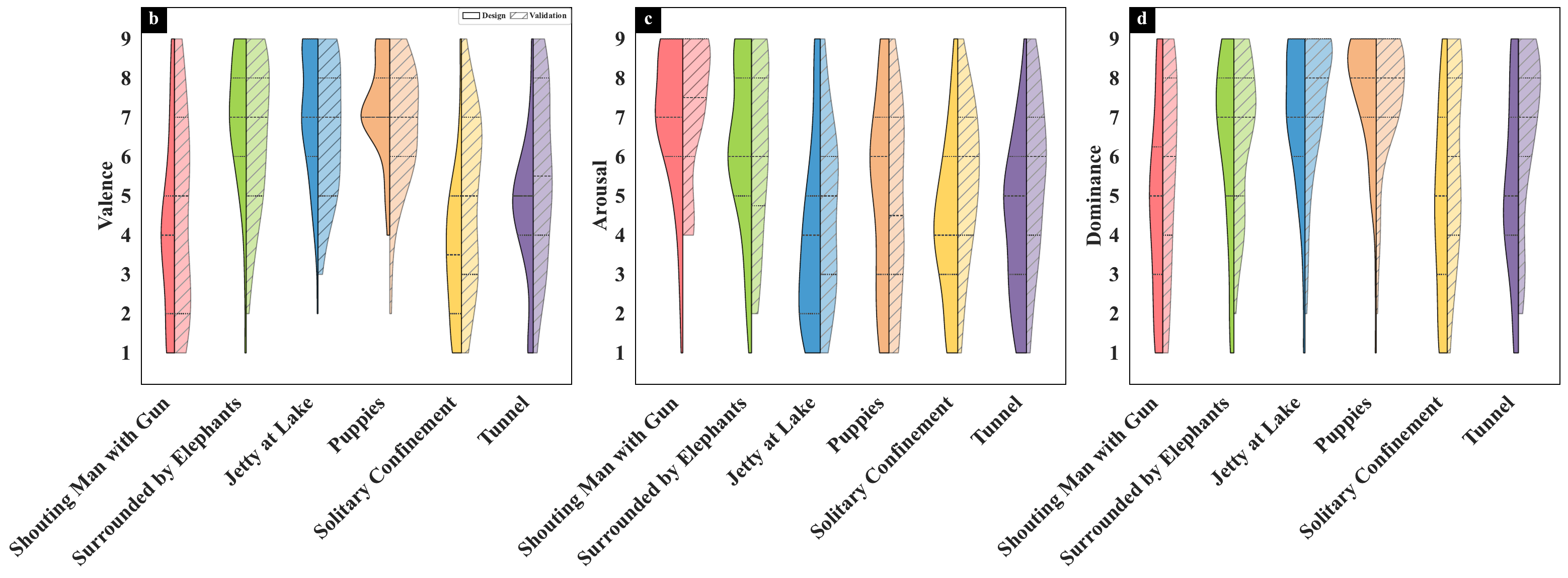}
  \end{minipage}
  \caption{SAM results for \textit{Design group} and \textit{Validation group}. a) Scatter plot of mean SAM ratings for each scene and group. b)-d) Violin plots showing the distribution of SAM ratings for each scene and group across the three dimensions (valence, arousal, dominance). The plots illustrate how emotional responses varied across scenes and between groups.}
  \Description{A four-panel figure summarizing SAM ratings across six scenes. Panel a is a 2D scatter of per-scene mean valence (horizontal axis, 1--9) against mean arousal (vertical axis, 1--9), with one Design group marker and one Validation group marker per scene. Puppies, Jetty at Lake, and Surrounded by Elephants cluster in the high-valence region, Shouting Man with Gun sits in the high-arousal low-valence corner, Tunnel lies in the low-valence low-arousal region, and Solitary Confinement shifts from low valence in the \textit{Design group} to moderately higher valence in the \textit{Validation group}. Panels b, c, and d show paired violin plots of valence, arousal, and dominance, respectively, with each scene occupying a vertical column on the x-axis and two violins per scene (Design pre-design versus Validation). Across panels, the Validation violins are systematically shifted upward in Tunnel and Solitary Confinement, indicating higher reported valence, arousal, and dominance when generated assets are present; the high-valence scenes (Puppies, Jetty at Lake, Surrounded by Elephants) show near-identical distributions across groups with heavy mass at the top of the scale.}
  \label{fig:sam_results}
\end{figure}

In our user study, we assessed participants' emotional responses using the SAM across three dimensions -- valence, arousal, and dominance. In particular, we focused on the \textit{primary contrast} between tha \textit{Designer group}'s pre-design SAM ratings (collected in asset-free scenes) and \textit{Validation group} participants' SAM ratings (collected after experiencing the same scenes with assets generated by unrelated designers, $N=60~participants \times 6~scenes=360$ for both groups). This contrast provides an informative between-group comparison of affective responses across asset-free versus populated scene conditions, while controlling for baseline emotional differences among scene layouts.

\autoref{fig:sam_results} visualizes the results of these two groups.
To test whether populated scenes shifted affect relative to the asset-free baseline, we employed an Aligned Rank Transform (ART) ANOVA with Scene (six virtual environments) as the within-subject factor and Group (Design group's pre-design vs.\ Validation) as the between-subject factor.

The ART ANOVA revealed significant Group $\times$ Scene interaction effects for all three SAM dimensions: valence ($F(5, 590) = 9.36$, $p < .001$), arousal ($F(5, 590) = 4.18$, $p < .001$), and dominance ($F(5, 590) = 7.16$, $p < .001$). The main effect of Scene was significant across all dimensions (all $F(5, 590) > 38.79$, all $p < .001$), confirming that the emotional character of each environment was the primary driver of affective variation. These interactions indicate that between-condition differences were not uniform but varied systematically with scene content.

\textit{Valence.} No overall Group main effect was observed for valence ($F(1, 118) = 1.59$, $p = .210$), indicating that the generated assets did not shift global valence ratings. The interaction, however, localized significant differences to specific scenes. Post-hoc Mann--Whitney contrasts (FDR-adjusted within each SAM dimension) revealed that in \textit{Solitary Confinement} the \textit{Validation group} reported substantially higher valence ($M = 4.97$, $SD = 2.19$) than the \textit{Design group} ($M = 3.57$, $SD = 1.89$; $p = .003$), suggesting that the generated assets mitigated the distressing atmosphere of this environment. A parallel uplift was observed in the \textit{Tunnel} scene (Validation $M = 5.60$, $SD = 2.22$ vs.\ Design $M = 4.52$, $SD = 1.87$; $p = .025$). In contrast, scenes with inherently high valence--\textit{Puppies}, \textit{Jetty at Lake}, and \textit{Surrounded by Elephants}--showed consistently pleasant ratings across both groups with no significant differences (all FDR $p > .17$), indicating a ceiling-like pattern for positively valenced environments.

\textit{Arousal.} A significant Group main effect indicated that the \textit{Validation group} reported generally higher arousal than the \textit{Design group} ($F(1, 118) = 6.03$, $p = .015$). Post-hoc contrasts localized this effect primarily to the \textit{Tunnel} scene, where the \textit{Validation group}'s arousal ($M = 5.65$, $SD = 2.12$) significantly exceeded that of the \textit{Design group} ($M = 4.30$, $SD = 2.02$; FDR $p = .004$), and to \textit{Solitary Confinement}, where \textit{Validation group}'s  arousal ($M = 5.45$, $SD = 1.95$) also exceeded the \textit{Design group}'s baseline ($M = 4.27$, $SD = 2.12$; FDR $p = .005$), suggesting that the generated assets enhanced the stimulatory quality of the low-arousal environments. Notably, the \textit{Shouting Man with Gun} scene produced the highest arousal levels in both groups (\textit{Design group} $M = 6.98$; \textit{Validation group} $M = 7.22$; FDR $p = .66$), pointing to the primacy of scene layout over asset content in driving high-intensity arousal in that context.

\textit{Dominance.} The \textit{Validation group} reported significantly higher dominance overall ($F(1, 118) = 7.69$, $p = .006$), indicating a stronger sense of environmental control when scenes contained generated assets. Post-hoc contrasts identified significant effects in the \textit{Tunnel} scene (\textit{Validation group} $M = 6.80$, $SD = 2.07$ vs.\ \textit{Design group} $M = 5.27$, $SD = 2.39$; FDR $p = .003$) and in \textit{Solitary Confinement} (\textit{Validation group} $M = 6.02$, $SD = 2.35$ vs.\ \textit{Design group} $M = 4.62$, $SD = 2.48$; FDR $p = .009$), with a marginal trend in \textit{Shouting Man with Gun} (\textit{Validation group} $M = 5.73$ vs.\ \textit{Design group} $M = 4.75$; FDR $p = .066$). Scenes with already-high \textit{Design group}'s dominance (\textit{Puppies}, \textit{Jetty at Lake}, \textit{Surrounded by Elephants}) showed no significant between-group differences, consistent with the ceiling effects noted above.

These results indicate that the affective impact of AI-generated 3D assets was concentrated in scenes with neutral or negative emotional tone. In \textit{Tunnel} and \textit{Solitary Confinement}, the presence of generated assets was associated with significantly higher pleasantness, arousal, and perceived control relative to the asset-free condition. Inherently pleasant scenes (\textit{Puppies}, \textit{Jetty at Lake}, \textit{Surrounded by Elephants}) were rated near the top of the scale in both groups, leaving limited room for asset-driven improvement. The significant Scene main effects across all dimensions further underscore that the emotional character of each environment remained the dominant factor in shaping affective responses, with the generated assets modulating those responses in a scene-dependent manner.

\subsection{Scene Engagement and Asset Interaction}

\label{sec:results_engagement}
We then investigated how the presence of generated assets influenced participants' engagement with the virtual environments, as measured by scene dwell time and in-scene asset manipulation patterns. These behavioral metrics provide complementary evidence to the self-reported SAM ratings, offering insight into how generated content affects not only subjective experience but also observable interaction dynamics within the VR contexts.

\vspace{1em}
\subsubsection{Scene Engagement Time}
~\\
To investigate whether the presence of generated 3D assets influenced how long participants engaged with the virtual environments, we analyzed the scene engagement time using a Gamma generalized linear mixed-effects model (GLMM) with a log link. We evaluated the fixed effects of Group (Designer Pre-design vs. Validation), Scene, and their interaction, while adjusting for covariates including age, gender, and prior VR experience.  

The analysis revealed a highly significant main effect of Group ($\chi^2(1) = 82.19$, $p < .001$), indicating that participants in the \textit{Validation group}—who experienced the scenes populated with the AI-generated 3D assets -- spent systematically more time in the environments compared to the \textit{Design group}'s baseline empty scenes. We also observed a significant main effect of Scene ($\chi^2(5) = 46.64$, $p < .001$), demonstrating that engagement times varied naturally depending on the environmental context. 

Furthermore, the model identified a significant Group $\times$ Scene interaction ($\chi^2(5) = 11.83$, $p = .037$), indicating that the magnitude of the engagement increase depended on the specific scene being explored. Interpreting the model on the response scale clarifies this pattern. Within the \textit{Validation group}, participants spent the most time in the \textit{Puppies} scene ($M = 104.08 \pm 9.70$~s) and the least time in \textit{Solitary Confinement} ($M = 68.95 \pm 6.40$~s). FDR-adjusted post-hoc comparisons within the \textit{Validation group} showed that \textit{Puppies} elicited significantly longer engagement than \textit{Solitary Confinement} ($p < .001$), \textit{Jetty at Lake} ($p < .001$), and \textit{Tunnel} ($p < .001$), and also exceeded \textit{Shouting Man with Gun} ($p = .012$). By contrast, the populated \textit{Surrounded by Elephants} scene ($M = 94.68 \pm 8.82$~s) did not significantly differ from \textit{Puppies} in the revised post-hoc results. 

Direct group comparisons on the response scale further showed significantly longer engagement in the Validation condition than in the Designer pre-design baseline across every environment (all $p < .001$). The largest relative increase occurred in \textit{Puppies}, where Validation participants remained in the scene about 2.31 times as long as the Designer baseline ($104.08$ vs.~$45.15$~s), whereas the smallest increase was observed in \textit{Surrounded by Elephants} ($94.68$ vs.~$53.25$~s). 

Finally, among the covariates, both age ($\chi^2(1) = 8.34$, $p = .004$) and VR experience ($\chi^2(3) = 8.52$, $p = .036$) significantly influenced the overall time spent in the scenes. Gender, however, did not show a significant effect on engagement time ($\chi^2(1) = 2.91$, $p = .088$). The omnibus fixed effects and the scene-wise response-scale estimates for the \textit{Validation group}, alongside validation-to-design time ratios, are summarized in Tables~\ref{tab:scene_time_glmm_results} and~\ref{tab:scene_time_pairwise_results}.  

\begin{table}[t]
\centering
\begin{minipage}[t]{0.35\linewidth}

\centering
\footnotesize
\renewcommand{\arraystretch}{1.1}
\setlength{\tabcolsep}{4pt}
\begin{threeparttable}
\caption{Type III GLMM results for scene engagement time.}
\label{tab:scene_time_glmm_results}
\begin{tabular}{lccc}
\toprule
\multicolumn{1}{c}{\multirow{2}{*}{Estimator}} & \multirow{2}{*}{$\chi^2$} & \multirow{2}{*}{df} & \multirow{2}{*}{$p$-value} \\
 & & & \\ 
\midrule
Group & 82.19 & 1 & $< .001$*** \\
Scene & 46.64 & 5 & $< .001$*** \\
Age & 8.34 & 1 & .004** \\
VR Experience & 8.52 & 3 & .036* \\
Gender & 2.91 & 1 & .088 \\
Group $\times$ Scene & 11.83 & 5 & .037* \\
\bottomrule
\end{tabular}
\begin{tablenotes}[flushleft]
\footnotesize
\item * $p < .05$, ** $p < .01$, *** $p < .001$.
\end{tablenotes}
\end{threeparttable}
\end{minipage}
\begin{minipage}[t]{0.60\linewidth}

\centering
\footnotesize
\renewcommand{\arraystretch}{1.1}
\setlength{\tabcolsep}{2pt}
\begin{threeparttable}
\caption{Scene-wise response-scale engagement estimates for the Validation group, with between-group contrasts.}
\label{tab:scene_time_pairwise_results}
\begin{tabular}{lcccc}
\toprule
Scene & Mean (sec) & SE & 95\% CI & With-/without-assets \\
 &  &  & (sec) & time ratio \\
\midrule
Puppies & \textbf{104.08***} & 9.70 & [85.07, 123.09] & 2.31$\times$ \\
Solitary Confinement & \textbf{68.95***} & 6.40 & [56.41, 81.49] & 1.90$\times$ \\
Jetty at Lake & \textbf{79.52***} & 7.41 & [65.00, 94.04] & 1.90$\times$ \\
Shouting Man with Gun & \textbf{88.32***} & 8.23 & [72.19, 104.45] & 1.87$\times$ \\
Tunnel & \textbf{81.22***} & 7.60 & [66.32, 96.12] & 1.83$\times$ \\
Surrounded by Elephants & \textbf{94.68***} & 8.82 & [77.39, 111.97] & 1.78$\times$ \\
\bottomrule
\end{tabular}
\begin{tablenotes}[flushleft]
\footnotesize
\item Values are converted to the linear scale. * $p < .05$, ** $p < .01$, *** $p < .001$.
\end{tablenotes}
\end{threeparttable}
\end{minipage}
\end{table}

\begin{figure}[t]
  \centering
  \includegraphics[height=0.16\textheight]{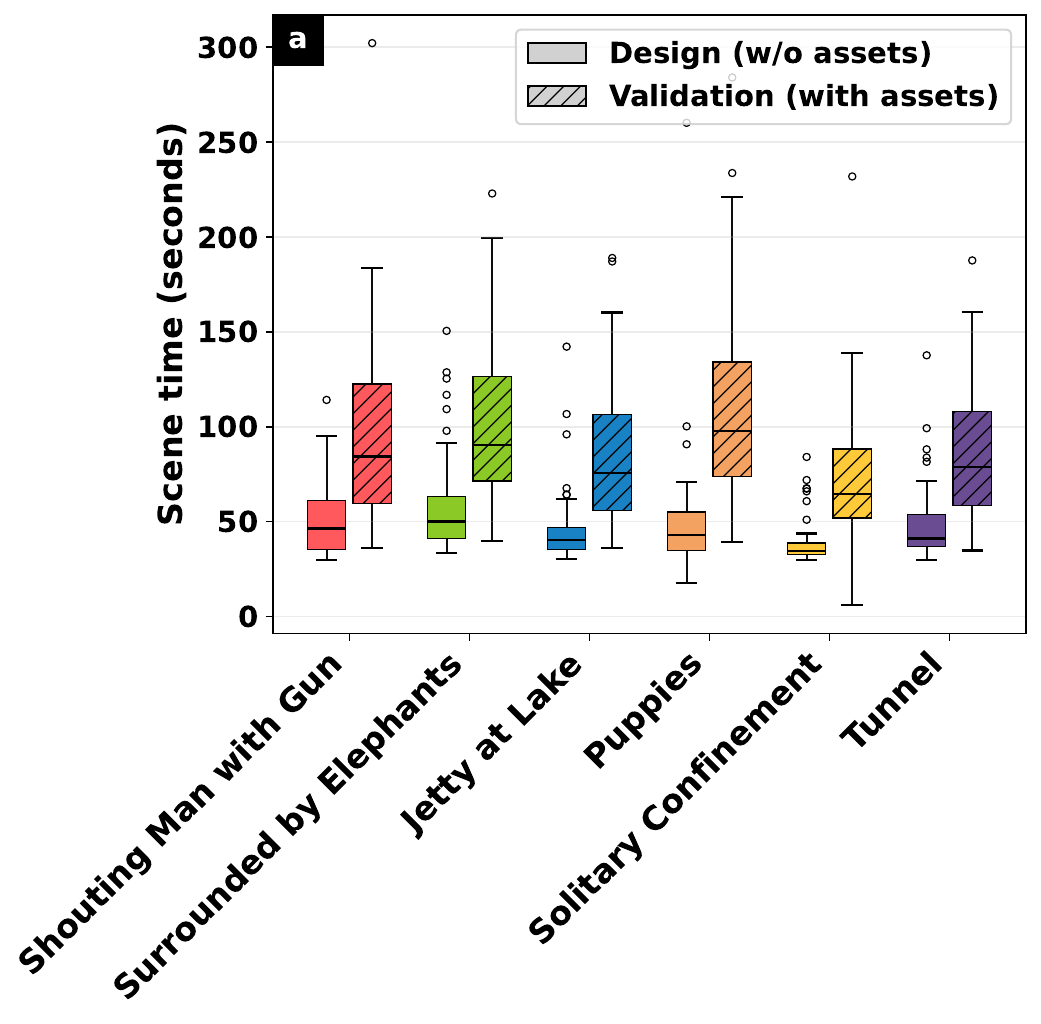}
  \hspace{-1em}
  \includegraphics[height=0.16\textheight]{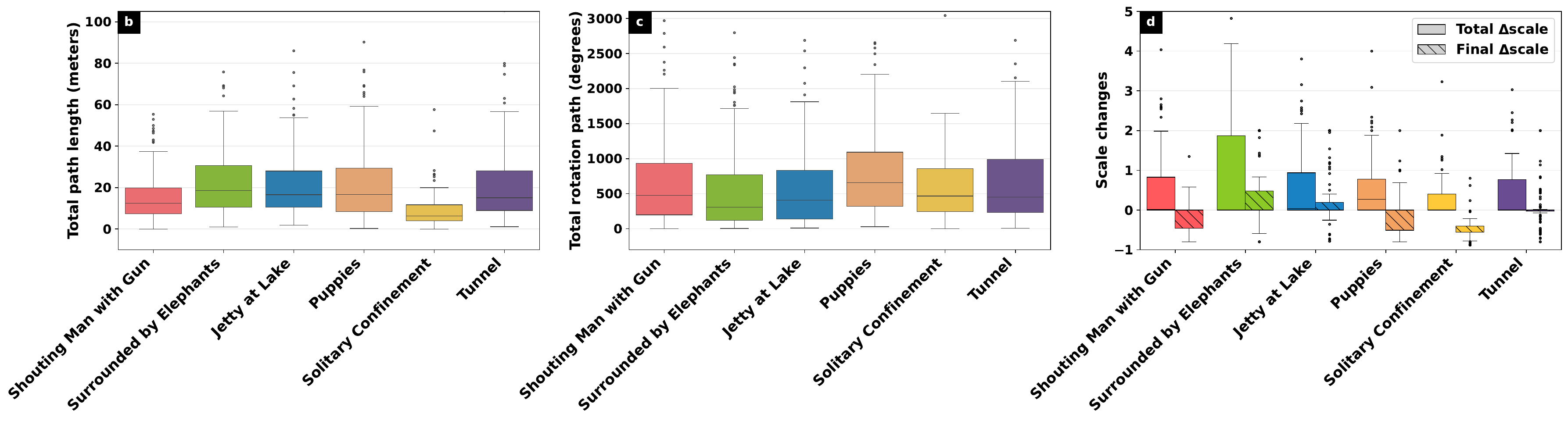}
  \caption{Scene engagement and validator interaction metrics. a) scene engagement time across environments. b)-d) Total path length of asset translation, rotation, and scaling across environments. d) also includes the final scale delta (final\_asset\_scale - 1.0) across environments.}
  \Description{A composite of four panels arranged in two columns. Panel a on the left is a grouped box-and-whisker plot of scene engagement time in seconds, with the six scenes along the x-axis and two boxes per scene (Design pre-design in a lighter hue and Validation in a darker hue). Validation boxes sit clearly above the Design boxes in every scene; Puppies shows the largest Validation median (around 104 seconds) while Solitary Confinement shows the smallest Validation median (around 69 seconds). Panels b, c, and d on the right form a vertical strip of three box plots for the \textit{Validation group} only, showing (b) cumulative translation path length in meters, (c) cumulative rotation path in degrees, and (d) cumulative scale effort together with the final scale delta relative to the initial spawn scale, each with the six scenes on the x-axis. Surrounded by Elephants and Puppies show the largest translation paths, Puppies and Shouting Man with Gun show the largest rotation paths, and Solitary Confinement shows the smallest values in all three interaction metrics. In panel d, the final scale-delta points hover near zero for most scenes, indicating modest net resizing.}
  \label{fig:engagement_interaction}
\end{figure}

These findings demonstrate that populating VR environments with user-guided, generated 3D assets substantially increases user engagement and retention, while also reshaping scene preferences within the populated condition itself: \textit{Puppies} emerged as the most attention-holding environment, whereas \textit{Solitary Confinement} remained the least visited even after assets were introduced.

\vspace{1em}
\subsubsection{Asset Interaction Dynamics} 
~\\ 
Beyond engagement duration, we analyzed how \textit{Validation group} participants manipulated the generated 3D assets within each virtual environment. Continuous spatial telemetry capturing translation, rotation, and scaling was recorded throughout each VR session, yielding a total of 476,842 discrete spatial updates across all participants and scenes. To ensure data integrity, instantaneous velocities were computed between consecutive frames, and positional updates exceeding a 10.0~m/s threshold were filtered to mitigate hardware tracking anomalies. From these cleaned trajectories, three cumulative interaction effort metrics were derived for each asset: \emph{total translation path length} (accumulated spatial displacement in meters), \emph{total rotation path} (accumulated angular deviation via sequential quaternion dot products, in degrees), and \emph{total scale effort} (accumulated magnitude of scale change relative to initial spawn scale). We conceptually decoupled the spatial manipulation process (interaction effort) from the final design outcome to provide a comprehensive account of user interaction behavior.

\textbf{Overall Interaction Effort.}
Each effort metric was modeled using a Gamma GLMM with a log link, with Scene as a fixed effect and Participant as a random intercept, adjusting for age, gender, and VR experience. The full fixed-effects estimates are reported in \autoref{tab:asset_interaction_fixed_effects}, and the estimated marginal means on the response scale are shown in \autoref{fig:engagement_interaction} (right).


\begin{table*}[t]
    \centering
    \caption{Fixed-effects estimates from Gamma GLMMs (log link) for spatial interaction effort metrics among Validation participants. Each metric was modeled as \textit{Outcome} $\sim$ Scene + Age + Gender + VR~Experience + (1\,|\,Participant). Estimates are on the log-link scale; exponentiate to obtain multiplicative effects on the response scale. Baseline: Scene = Shouting Man with Gun.}
    \label{tab:asset_interaction_fixed_effects}
\resizebox{\linewidth}{!}{%
\begin{tabular}{l ccc ccc ccc}
\toprule
\multirow{2}{*}{\textbf{Parameter}} & 
\multicolumn{3}{c}{\textbf{Translation Path (meter$\Delta$)}} & 
\multicolumn{3}{c}{\textbf{Rotation Path (deg$\Delta$)}} & 
\multicolumn{3}{c}{\textbf{Scale Change (linear ratio$\Delta$)}} \\
\cmidrule(lr){2-4}\cmidrule(lr){5-7}\cmidrule(l){8-10}
 & Est. & SE & 95\% CI & Est. & SE & 95\% CI & Est. & SE & 95\% CI \\
\midrule
(Intercept)              & \textbf{2.742\textsuperscript{***}}   & 0.633 & [1.50, 3.98] & \textbf{6.639\textsuperscript{***}}   & 0.840 & [4.99, 8.29] & 5.039                                  & 4.276 & [$-$3.34, 13.42] \\
Age                      & +0.003                                & 0.018 & [$-$0.03, 0.04] & +0.006                                & 0.024 & [$-$0.04, 0.05] & \textbf{$-$0.273\textsuperscript{*}}   & 0.124 & [$-$0.52, $-$0.03] \\
Gender (Male)            & +0.203                                & 0.128 & [$-$0.05, 0.45] & +0.306                                & 0.171 & [$-$0.03, 0.64] & +0.003                                 & 0.869 & [$-$1.70, 1.71] \\
VR Exp.: Never           & $-$0.299                              & 0.478 & [$-$1.24, 0.64] & $-$0.608                              & 0.637 & [$-$1.86, 0.64] & $-$2.089                               & 3.238 & [$-$8.44, 4.26] \\
VR Exp.: Few times       & $-$0.033                              & 0.478 & [$-$0.97, 0.90] & $-$0.345                              & 0.637 & [$-$1.60, 0.90] & +0.545                                 & 3.237 & [$-$5.80, 6.89] \\
Scene: Surrounded by Elephants & \textbf{+0.308\textsuperscript{**}}   & 0.099 & [0.11, 0.50] & \textbf{$-$0.331\textsuperscript{**}} & 0.120 & [$-$0.57, $-$0.09] & +0.053                                 & 0.362 & [$-$0.66, 0.76] \\
Scene: Jetty at Lake     & \textbf{+0.213\textsuperscript{*}}    & 0.098 & [0.02, 0.41] & \textbf{$-$0.258\textsuperscript{*}}  & 0.120 & [$-$0.49, $-$0.02] & $-$0.081                               & 0.372 & [$-$0.81, 0.65] \\
Scene: Puppies           & \textbf{+0.294\textsuperscript{**}}   & 0.099 & [0.10, 0.49] & +0.197                                & 0.121 & [$-$0.04, 0.43] & $-$0.284                               & 0.360 & [$-$0.99, 0.42] \\
Scene: Solitary Confinement & \textbf{$-$0.688\textsuperscript{***}}& 0.099 & [$-$0.88, $-$0.49] & $-$0.116                              & 0.120 & [$-$0.35, 0.12] & \textbf{$-$1.203\textsuperscript{**}}  & 0.367 & [$-$1.92, $-$0.48] \\
Scene: Tunnel            & \textbf{+0.201\textsuperscript{*}}    & 0.099 & [0.01, 0.40] & $-$0.044                              & 0.121 & [$-$0.28, 0.19] & $-$0.578                               & 0.359 & [$-$1.28, 0.13] \\
\bottomrule
\end{tabular}%
}
\vspace{0.5ex}
\raggedright\footnotesize

\textsuperscript{*}\,$p<.05$, 
\textsuperscript{**}\,$p<.01$, 
\textsuperscript{***}\,$p<.001$. 
Significant estimates are bolded.
\end{table*}

\textbf{Moving Assets.} 
Scene had a highly significant effect on the total distance participants moved the assets ($\chi^2(5) = 146.25$, $p < .001$). The \textit{Solitary Confinement} scene elicited the shortest translation paths ($M = 8.31 \pm 1.43$~m), significantly lower than all other environments (all FDR-adjusted $p < .001$). This is expected as the \textit{Solitary Confinement} scene has limited spatial dimensions. In contrast, \textit{Surrounded by Elephants} ($M = 22.51 \pm 3.92$~m) and \textit{Puppies} ($M = 22.19 \pm 3.88$~m) encouraged the most expansive spatial manipulation, while \textit{Jetty at Lake} ($M = 20.47 \pm 3.52$~m) and \textit{Tunnel} ($M = 20.22 \pm 3.54$~m) fell between these extremes. 

\textbf{Rotating Assets.} 
Scene also significantly influenced rotational inspection ($\chi^2(5) = 24.35$, $p < .001$). Participants rotated assets the most in the \textit{Puppies} scene ($M = 905.67 \pm 207.87^{\circ}$) and the \textit{Shouting Man with Gun} scene ($M = 743.56 \pm 168.11^{\circ}$), while the least rotational effort was observed in \textit{Surrounded by Elephants} ($M = 533.89 \pm 122.48^{\circ}$). Post-hoc contrasts confirmed that rotational effort in \textit{Puppies} significantly exceeded \textit{Surrounded by Elephants} ($z = 4.37$, FDR $p < .001$) and \textit{Jetty at Lake} ($z = 3.76$, FDR $p = .001$). 

\textbf{Scaling Assets.} 
Cumulative resizing effort also varied significantly by Scene ($\chi^2(5) = 18.42$, $p = .002$). \textit{Solitary Confinement} again elicited the least effort ($M = 0.06 \pm 0.07$), significantly lower than \textit{Surrounded by Elephants} ($M = 0.22 \pm 0.25$; FDR $p = .004$) and \textit{Shouting Man with Gun} ($M = 0.21 \pm 0.23$; FDR $p = .007$). Among the covariates, age had a significant negative effect on scale effort (Estimate = $-0.273 \pm 0.124$, $p = .028$), indicating that older participants invested less cumulative effort in resizing.

\textbf{Final Scale Decisions.}
To complement the analysis of manipulation effort, we examined users' final sizing decisions as an indicator of design intent. We employed a two-part hurdle GLMM to model the final scale delta (final asset scale minus initial spawn scale), separating the binary decision to scale from the continuous magnitude of adjustment. The binomial hurdle component (logit link) revealed that Scene did not significantly predict whether a user chose to scale the asset at all ($\chi^2(5) = 4.19$, $p = .523$): the likelihood of leaving the default scale was uniform across environments. However, the Gaussian component (identity link) for non-zero deltas showed that, once a sizing decision was made, Scene heavily dictated the final direction and magnitude ($\chi^2(5) = 232.30$, $p < .001$). Users systematically scaled objects \emph{up} in vast outdoor environments -- \textit{Surrounded by Elephants} ($M = +0.74 \pm 0.15$) and \textit{Jetty at Lake} ($M = +0.51 \pm 0.14$) -- and scaled objects \emph{down} in indoor or constrained scenes -- \textit{Solitary Confinement} ($M = -0.52 \pm 0.13$) and \textit{Shouting Man with Gun} ($M = -0.45 \pm 0.15$). Post-hoc contrasts confirmed the largest divergences: \textit{Surrounded by Elephants} vs.\ \textit{Solitary Confinement} ($t = 12.35$, FDR $p < .001$), \textit{Jetty at Lake} vs.\ \textit{Solitary Confinement} ($t = 10.26$, FDR $p < .001$), and \textit{Surrounded by Elephants} vs.\ \textit{Puppies} ($t = 8.97$, FDR $p < .001$). Among the covariates, male participants demonstrated a significant tendency to scale assets larger than female participants (Estimate = $+0.246 \pm 0.067$, $p < .001$).

These interaction patterns reveal a coherent relationship between scene properties and spatial manipulation behavior. Wide-open, positively valenced natural scenes (\textit{Surrounded by Elephants}, \textit{Jetty at Lake}, \textit{Puppies}) encouraged extensive translation as participants moved assets across expansive spaces, while confined, negatively valenced environments (\textit{Solitary Confinement}) suppressed gross motor movement, requiring highly localized adjustments. The dissociation between rotation and translation is also notable: \textit{Puppies} elicited the highest rotational inspection despite moderate translation, suggesting that the nature of the assets (animate, detail-rich) invited close scrutiny. The systematic scaling pattern -- enlarging objects in outdoor spaces and shrinking them indoors -- demonstrates that users reliably detect when an asset's default scale violates the spatial logic of a scene and will actively correct it, suggesting that auto-scaling heuristics based on scene boundary information could improve initial spawn sizes for VR asset generation pipelines.

\subsection{Human-AI Co-Design Conversation Analysis}
\label{sec:conversation_analysis}
We then analyzed the temporal dynamics of the co-design conversations, focusing on how scene context influenced the patterns of human and AI responses, including the number of conversational messages, the number of design iterations, the response time and the generation time for the \textit{Design group}.

\begin{table}[t]
  \centering
  \begin{minipage}[t]{0.38\linewidth}
  \centering
\footnotesize
\renewcommand{\arraystretch}{1.1}
\setlength{\tabcolsep}{8pt}
\begin{threeparttable}
\caption{Type III GLMM results for chat response time.}
\label{tab:response_time_glmm_type3_results}
\begin{tabular}{lccc}
\toprule
\multicolumn{1}{c}{\multirow{2}{*}{Estimator}} & \multirow{2}{*}{$\chi^2$} & \multirow{2}{*}{df} & \multirow{2}{*}{$p$-value} \\
 & & & \\
\midrule
Role                & 1197.82 & 1 & $< .001$*** \\
Scene               & 17.57 & 5 & .004** \\
Age                 & 0.34 & 1 & .562 \\
Gender              & 0.35 & 1 & .552 \\
VR Experience       & 0.09 & 2 & .957 \\
Role $\times$ Scene & 24.95 & 5 & $< .001$*** \\
\bottomrule
\end{tabular}
\begin{tablenotes}[flushleft]
\footnotesize
\item * $p < .05$, ** $p < .01$, *** $p < .001$.
\end{tablenotes}
\end{threeparttable}
  \end{minipage}\hfill
  \begin{minipage}[t]{0.60\linewidth}
  \centering
\footnotesize
\renewcommand{\arraystretch}{1.1}
\setlength{\tabcolsep}{2.5pt}
\begin{threeparttable}
\caption{Scene-wise response-scale chat response-time estimates for the Human role, with within-scene Human-versus-AI contrasts.}
\label{tab:response_time_pairwise_results}
\begin{tabular}{lccccc}
\toprule
Scene & Mean (sec) & SE & 95\% CI & Human/AI & FDR-adjusted \\
 &  &  & (sec) & time ratio & $p$-value \\
\midrule
Shouting Man with Gun & 30.92 & 2.46 & [26.46, 36.14] & 4.95$\times$ & $< .001$*** \\
Solitary Confinement & 27.93 & 2.21 & [23.91, 32.62] & 5.10$\times$ & $< .001$*** \\
Tunnel & 25.27 & 2.00 & [21.65, 29.51] & 4.21$\times$ & $< .001$*** \\
Surrounded by Elephants & 23.33 & 1.84 & [19.98, 27.24] & 4.05$\times$ & $< .001$*** \\
Jetty at Lake & 22.72 & 1.81 & [19.44, 26.55] & 4.13$\times$ & $< .001$*** \\
Puppies & 22.18 & 1.76 & [18.97, 25.92] & 4.12$\times$ & $< .001$*** \\
\bottomrule
\end{tabular}
\begin{tablenotes}[flushleft]
\footnotesize
\item Values are converted to the linear scale. * $p < .05$, ** $p < .01$, *** $p < .001$.
\end{tablenotes}
\end{threeparttable}
  \end{minipage}
  \end{table}
  
  \begin{figure}[t]
    \centering
    \begin{minipage}[t]{0.34\linewidth}
      \centering
      \includegraphics[width=\linewidth]{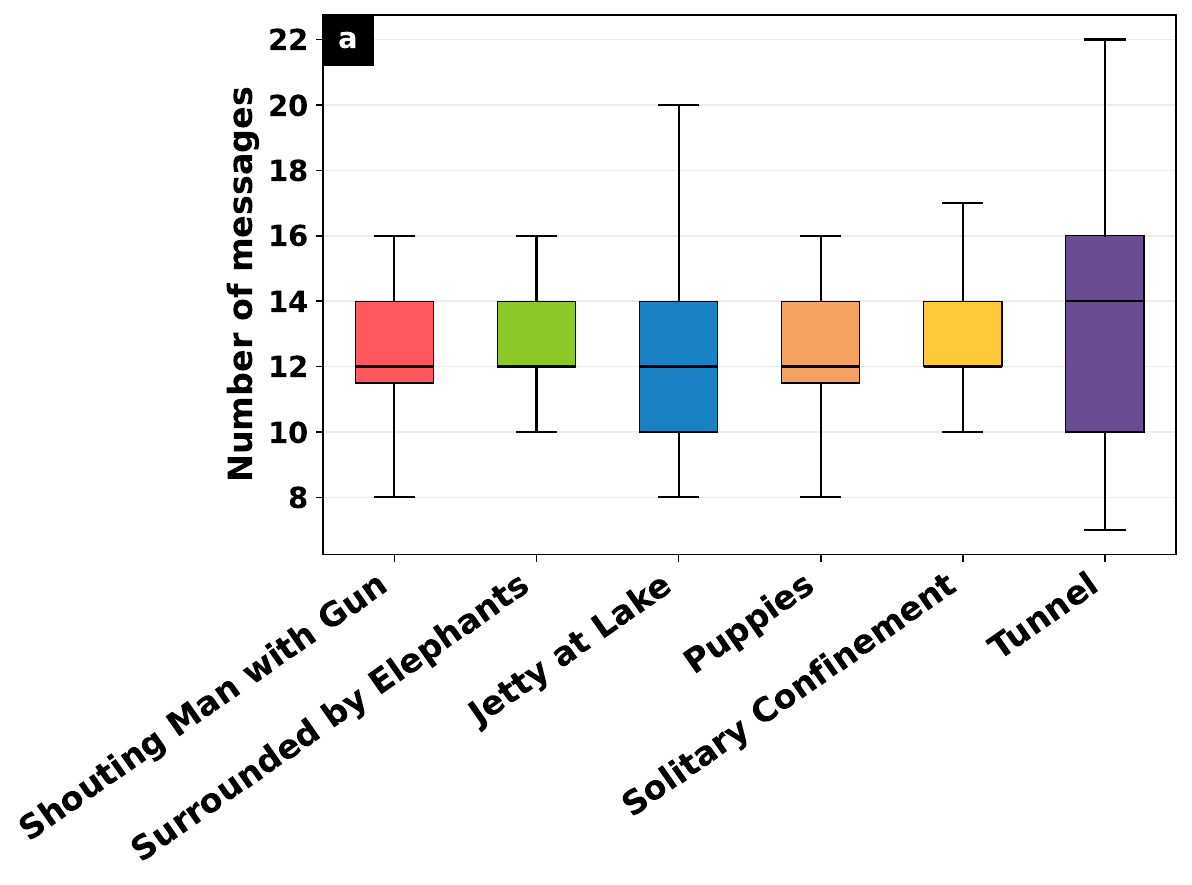}
    \end{minipage}\hspace{-1.5em}
    \begin{minipage}[t]{0.34\linewidth}
      \centering
      \includegraphics[width=\linewidth]{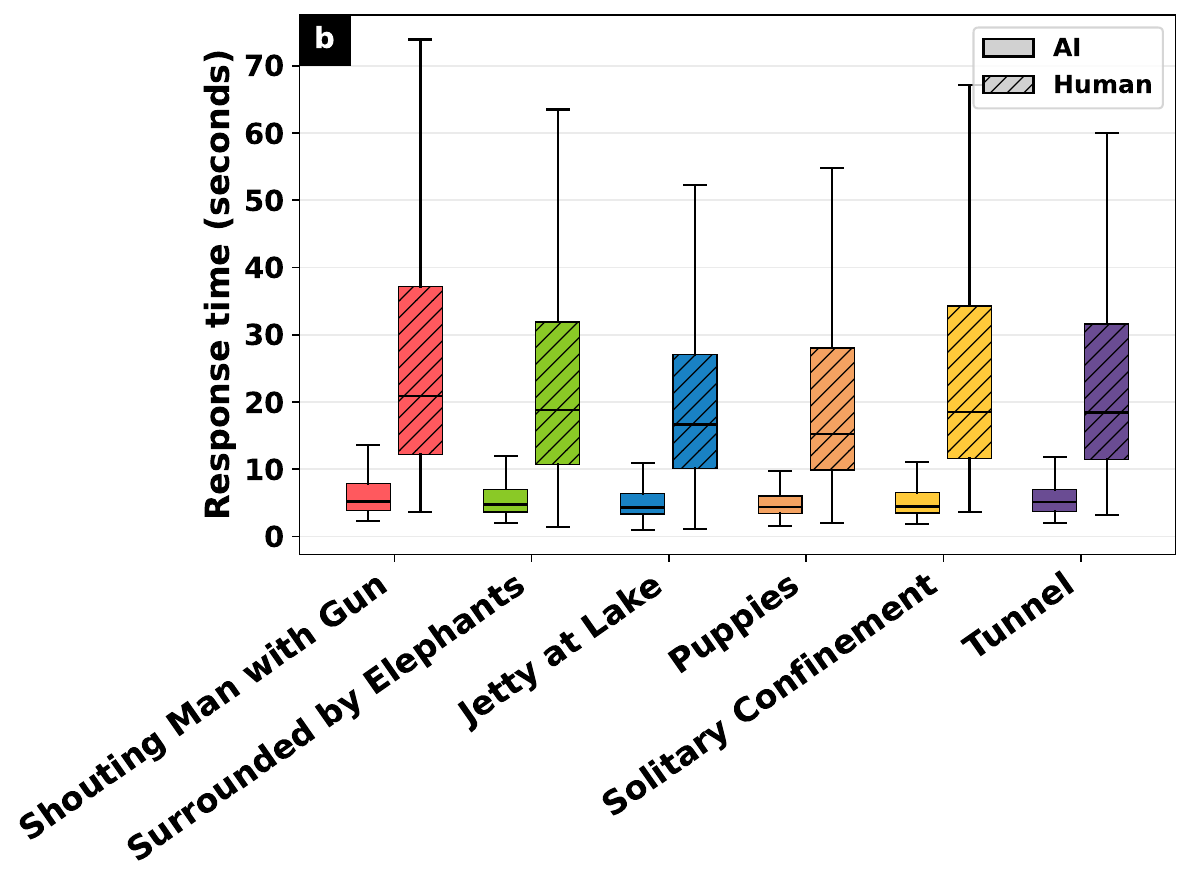}
    \end{minipage}\hspace{-1.5em}
    \begin{minipage}[t]{0.34\linewidth}
      \centering
      \includegraphics[width=\linewidth]{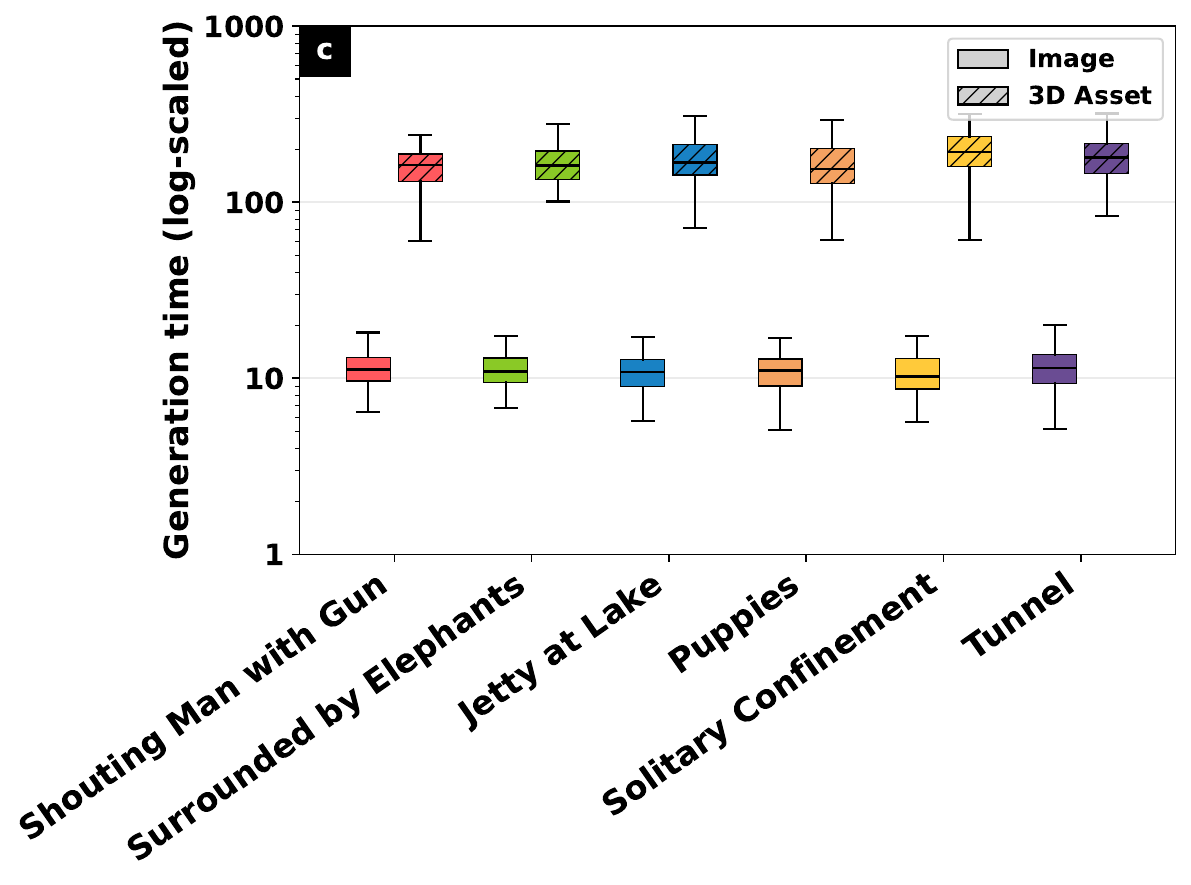}
    \end{minipage}
    \caption{Temporal characteristics of the co-design pipeline across scenes. a) Number of messages in the co-design session. b) AI and human chat response time by scene. c) image and 3D asset generation time by scene (log-scaled seconds).}
    \Description{A three-panel figure showing temporal characteristics of the co-design pipeline across six scenes. Panel a is a box plot of the number of conversational messages per scene. Panel b is a grouped box plot of chat response times in seconds, with two boxes per scene (AI agent and human designer): AI responses are tightly clustered at lower values, while human response times are longer and more variable, especially in Shouting Man with Gun, Solitary Confinement, and Tunnel. Panel c is a grouped box plot of generation times on a log-scaled vertical axis (seconds), comparing 2D image generation via FLUX.1 and 3D asset generation via Hunyuan3D-2. Image generation clusters around tens of seconds, whereas 3D generation is roughly an order of magnitude slower with heavier right tails.}
    \label{fig:response_generation_time}
  \end{figure}

\textbf{Conversation volume and design iterations.}
Before response-time modeling, we examined whether scene context altered the total number of conversational messages and the number of 2D image concept iterations before 3D generation. Shapiro-Wilk tests indicated non-normality across all scenes for both outcomes (all $p \le .05$). Hence, we used Friedman tests for repeated within-participant scene comparisons. Neither metric showed a significant scene effect: \texttt{num of messages}, $\chi^2(5)=5.18$, $p=.395$; \texttt{num of iterations}, $\chi^2(5)=3.90$, $p=.563$. Overall, conversation volume was stable across scenes (overall $M=13.21$, $SD=3.27$, median$=12$; scene means $13.00$--$13.63$), and iteration counts were similarly consistent (overall $M=1.23$, $SD=0.53$, median$=1$; scene means $1.18$--$1.30$). This indicates that most participants were satisfied with the initial concept images generated in the co-design session. 

\textbf{Response time.}
To formally evaluate the temporal demands of the co-design process, we analyzed per-turn chat response times using a Gamma generalized linear mixed-effects model (GLMM) with a log link. We evaluated the fixed effects of Role (AI vs.~Human), Scene, and their interaction, while including participant as a random intercept to account for repeated measures. Covariates including age, gender, and prior VR experience were non-significant (all $p > .50$). 

As shown in Table~\ref{tab:response_time_glmm_type3_results} and illustrated in \autoref{fig:response_generation_time}, the results confirmed a significant main effect of Role ($\chi^2(1) = 1197.82$, $p < .001$), establishing that AI responses were systematically faster than human deliberation times throughout the co-design process. A significant main effect of Scene ($\chi^2(5) = 17.57$, $p = .004$) and a highly significant Role $\times$ Scene interaction ($\chi^2(5) = 24.95$, $p < .001$) indicated that the scene's affective character directly shaped conversational pacing -- but differently for humans than for AI.

Focusing on the human designers, the estimated marginal means on the linear time scale reveal a clear affective gradient. The \textit{Shouting Man with Gun} scene imposed the highest deliberation cost ($M = 30.92 \pm 2.46$~s), and post-hoc FDR-adjusted pairwise contrasts confirmed it was significantly slower than all other scenes (all $p \leq .050$). \textit{Solitary Confinement} was the second most demanding ($M = 27.93 \pm 2.21$~s), significantly slower than the three positive scenes -- \textit{Puppies} ($p < .001$), \textit{Jetty at Lake} ($p < .001$), and \textit{Surrounded by Elephants} ($p < .001$) -- as well as \textit{Tunnel} ($p = .050$). In contrast, human designers established the fastest, most fluid cadence in \textit{Puppies} ($M = 22.18$~s), \textit{Jetty at Lake} ($M = 22.72$~s), and \textit{Surrounded by Elephants} ($M = 23.33$~s), with no significant differences among these three positive environments. AI response times remained comparatively stable across all scenes, ranging from $5.39$~s (\textit{Puppies}) to $6.25$~s (\textit{Shouting Man with Gun}).

\textbf{Generation time.}
Finally, we analyzed the system's generation performance as shown in \autoref{fig:response_generation_time} (right). Generating the initial 2D concept images was highly efficient and consistent, with an overall mean generation time of 11.46 seconds (SD = 4.40 seconds). Conversely, transforming these 2D concepts into deployable 3D assets introduced a significant computational bottleneck. Overall, 3D asset generation was substantially longer and highly variable, demanding a mean time of 194.08 seconds (SD = 123.75 seconds) across all environments. This validates the justification of our concept-image-to-3D pipeline. While the text-to-image stage is sufficiently fast to support real-time co-design interactions, the image-to-3D conversion currently requires a more extended processing period that may disrupt conversational flow and user engagement. The justification remains even with more efficient 3D generation models, as its complexity and resource demands are expected to remain significantly higher than those of text-to-image generation for the foreseeable future.

Overall, these results demonstrate that the emotional valence of a target VR environment directly influences the cognitive effort -- and thus the conversational pacing -- that participants in the \textit{Design group} invest in co-designing contextually appropriate 3D assets. The omnibus fixed effects and scene-wise human response-time estimates with within-scene Human-versus-AI contrasts are summarized in Tables~\ref{tab:response_time_glmm_type3_results} and~\ref{tab:response_time_pairwise_results}.

\subsection{Generation Prompt Analysis}
\label{sec:generation_prompt_analysis}

\begin{figure}[t]
  \centering
  \includegraphics[width=\linewidth]{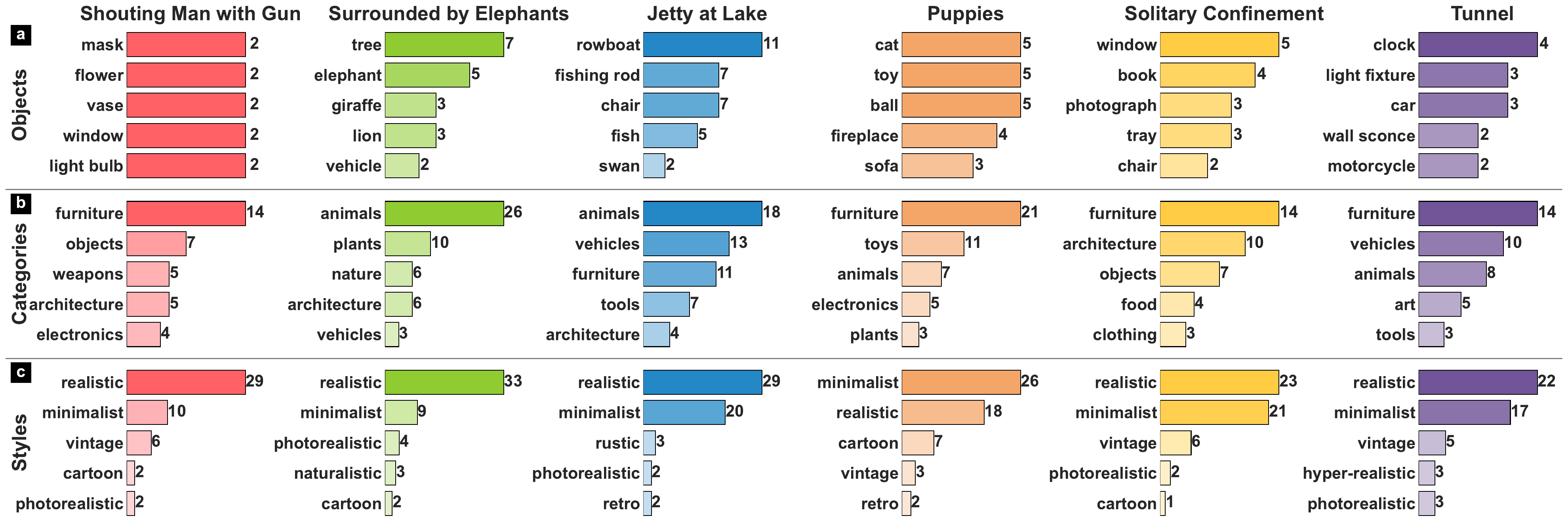}
  \caption{Text-to-image prompt analysis. a) the Top-5 objects designed by participants for each scene; b) the Top-5 categories of the designed objects; c) the Top-5 styles of the designed objects.}
  \Description{A three-panel bar-chart figure summarizing LLM-parsed structured attributes extracted from the 360 confirmed text-to-image prompts. Panel a is a set of six small horizontal bar charts, one per VR scene, each listing the top five most frequently designed primary objects in that scene (for example, motorcycles and lanterns for Tunnel; balls and chew toys for Puppies; boats and stones for Jetty at Lake; books and cups for Solitary Confinement; shields and handguns for Shouting Man with Gun; bananas and grass bundles for Surrounded by Elephants), with bar length encoding count. Panel b is a horizontal bar chart showing the top five semantic categories aggregated across all scenes (for instance, furniture and props, vehicles, tools, natural objects, and animals), sorted in decreasing order. Panel c is a horizontal bar chart of the top five aesthetic styles across all 360 prompts: realistic appears by far the longest bar (154 occurrences), followed by minimalist (103), then vintage (20), photorealistic (13), and cartoon (12).}
  \label{fig:generation_prompt_analysis}
\end{figure}

To better understand the semantic content and creative directions participants pursued during the generative 3D co-design process, we conducted an exploratory analysis of the textual prompts submitted by particpants in the \textit{Design group}. We utilized a large language model (DeepSeek V3.2 API) to parse the raw text prompts ($n=360$) and extract structured, descriptive properties for the generated assets. Specifically, we categorized the extractions into three primary metrics: aesthetic style, semantic category, and primary object. The outputs were manually verified by the researchers. These structured extraction results are illustrated in \autoref{fig:generation_prompt_analysis} with Top-5 most frequent terms for each metric broken down by scene.

Across all six VR scenes, participants exhibited clear trend on preferences for specific aesthetic styles. The most frequently requested style was realistic ($n=154$), followed by minimalist ($n=103$). Other styles such as vintage ($n=20$), photorealistic ($n=13$), and cartoon ($n=12$) appeared significantly less often. This strong preference for realism and minimalism suggests that participants aimed to maintain presence and immersion in the VR environments through grounded, believable assets, rather than introducing highly stylized, distracting, or abstract designs.

For the semantic categories and primary objects, the generated content appropriately reflected the contextual needs and targeted emotions of the specific scenes. The most common object categories across the entire dataset were furniture ($n=74$), animals ($n=59$), and vehicles ($n=26$). A clearer pattern emerges when grouping the scenes by affective tone. In the more positive scenes, participants tended to request playful, natural, and scene-congruent content. In the \textit{Puppies} scene, they combined domestic and interactive elements -- furniture was the dominant category ($n=21$), followed by toys ($n=11$) and animals ($n=7$), while the most frequent primary objects were toys, balls, and cats ($n=5$ each), followed by fireplaces ($n=4$) and sofas ($n=3$). This playful emphasis is consistent with participants in the \textit{Design group}'s reflections that \pquote{a very cute little white cat gave the scene a very happy feeling} (D10) and that the added object \pquote{could be used to interact with the dogs, which was very fun} (D49). A similar scene-congruent pattern appeared in the outdoor positive environments. For \textit{Jetty at Lake}, animals ($n=18$), vehicles ($n=13$), furniture ($n=11$), and tools ($n=7$) were the most common categories, with rowboats ($n=11$), fishing rods ($n=7$), and chairs ($n=7$) leading the primary objects. A participant remarked that \pquote{the small boat model made the whole scene feel calmer and more relaxing} (D23). In \textit{Surrounded by Elephants}, animals ($n=26$) and plants ($n=10$) dominated, and trees ($n=7$) and elephants ($n=5$) were the most frequent objects, matching participants' accounts that adding a tree made \pquote{the whole scene feel more realistic and increased immersion} (D23) and that \pquote{giving the elephant a hat made it feel happier} (D01).

By contrast, the lower-valence or more tense scenes elicited more atmospheric, functional, and transitional props. In the \textit{Tunnel} scene, furniture ($n=14$), vehicles ($n=10$), and animals ($n=8$) were the leading categories, while clocks ($n=4$), light fixtures ($n=3$), cars ($n=3$), and motorcycles ($n=2$) were among the most frequent primary objects, suggesting an effort to intensify the scene's uncanny, in-between atmosphere through both ambient cues and mobile artifacts. This interpretation is also reflected in one participant's remark that \pquote{this element fit the scene better, the motorcycle looked good and appropriate} (D24). This same justification extended to the other negative scenes, but with different emphases. In \textit{Shouting Man with Gun}, furniture ($n=14$), objects ($n=7$), weapons ($n=5$), and architecture ($n=5$) dominated, while light bulbs, windows, vases, flowers, and masks (all $n=2$) appeared as the most frequent primary objects, indicating that participants shaped the threatening environment through staging details rather than weapons alone. A participant, for instance, noted that \pquote{with the incandescent light bulb, when I was picking up the object and a stranger suddenly came in, it felt even scarier} (D23). In \textit{Solitary Confinement}, furniture ($n=14$) and architecture ($n=10$) were again dominant, with windows ($n=5$), books ($n=4$), photographs ($n=3$), and trays ($n=3$) emerging most often, pointing to a design strategy centered on deprivation, enclosure, and sparse narrative cues. This reading is supported by participants' comments that \pquote{the black iron window fit the scene quite well, though it would be even better if there were scenery outside the window} (D22) and that \pquote{this round wall clock matched the other elements in the scene very well, giving a lonelier and more frightening feeling} (D60).

These findings highlight how participants in the \textit{Design group} systematically adapted their text prompts to populate the environments with contextually coherent 3D assets that reinforced the target affective state of each scene.

\subsection{Design Experience Feedback}
\label{sec:results_design_experience}

\begin{figure}[t]
  \centering
  \includegraphics[width=0.525\linewidth]{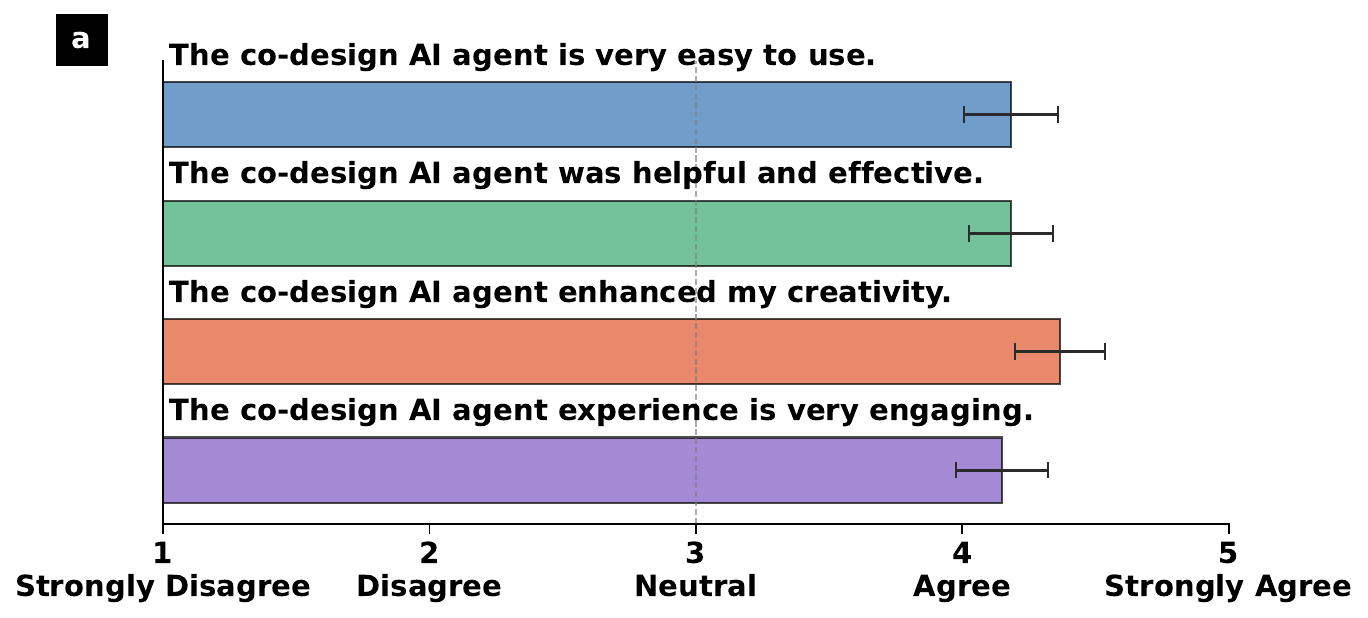}
  \includegraphics[width=0.465\linewidth]{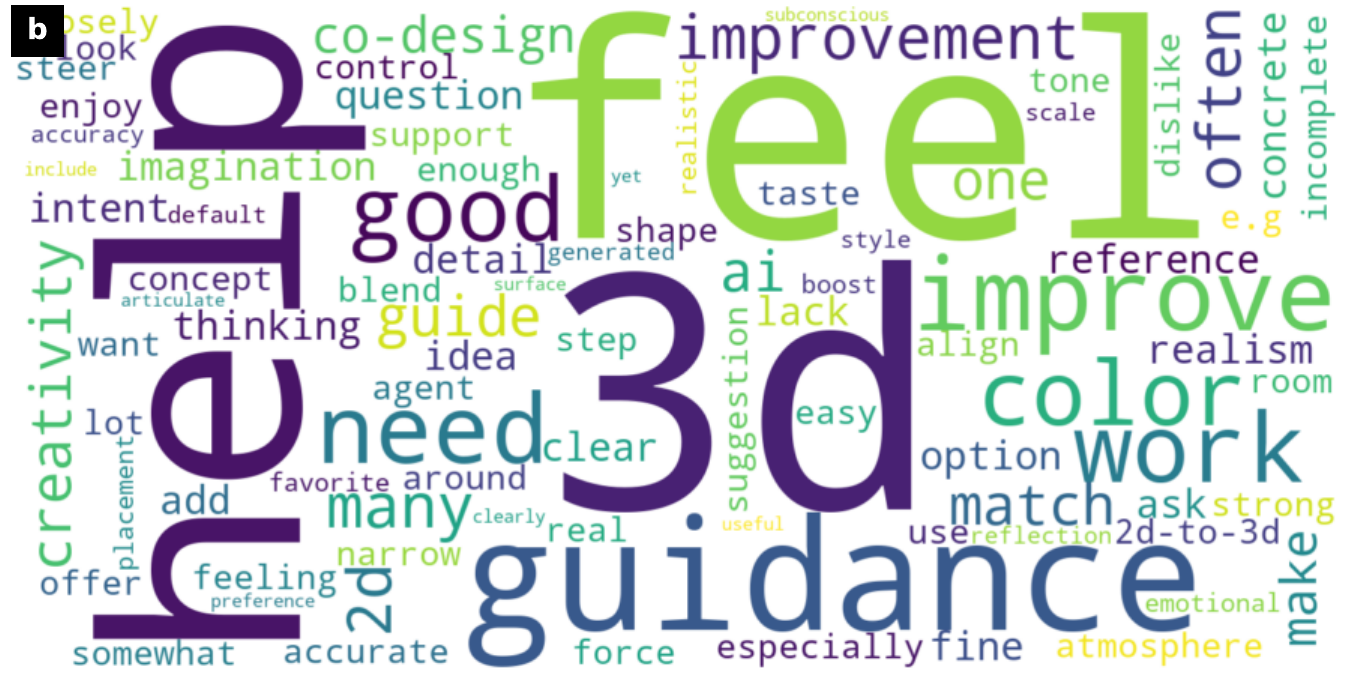}
  \caption{Design experience feedback: a) Designer-reported ease, helpfulness, creativity support, and engagement with the co-design agent (post-study, $n{=}60$). b) Word cloud of the most frequent lemmas in \textit{Design group} participants' open-ended feedback about their co-design experience.}
  \Description{Bar chart with four categories on the x-axis and mean rating 1--5 on the y-axis, values near 4.2.}
  \label{fig:designer_feedback}
\end{figure}

Beyond per-scene satisfaction with individual outputs, participants from the \textit{Design group} also answered four post-study 5-point Likert items about the overall co-design agent: ease of use, helpfulness, perceived creativity support, and engagement, with an open-ended question regarding their general experience. The results are illustrated in \autoref{fig:designer_feedback}.

Overall, the Likert ratings for the co-design agent's ease of use, helpfulness, creativity support, and engagement all showed strong endorsement, with mean values clustered between 4.15 and 4.37 on the 1--5 scale (all $SD < 0.71$), indicating broadly positive impressions. In particular, 56 of 60 \textit{Design group} pariticipants (93.3\%) rated creativity support as 4 or 5, with 27 top-box ratings of 5, while helpfulness/effectiveness received 53 ratings (88.3\%) at 4 or 5, and ease of use had 52 ratings (86.7\%) at 4 or 5, with engagement similarly high at 50 ratings (83.3\%) at 4 or 5. 

The open-ended feedback converged on the same broad pattern. Designers frequently described the agent as helpful for clarifying vague intentions and stimulating imagination, which is also visible in the word-cloud counts: the most frequent lemmas were ``3d'' ($n=14$), ``feel'' ($n=11$), ``help'' ($n=9$), ``guidance'' ($n=8$), ``improve'' ($n=8$), ``work'' ($n=7$), ``need'' ($n=7$), and ``color'' ($n=6$). These terms suggest that participants primarily framed the experience around two linked concerns: the usefulness of the guided conversation and the remaining limitations of 3D realization. Several quotes directly support this interpretation. One participant wrote, \pquote{I really liked this co-design assistant, especially because it helped surface preferences that even I was not fully aware of yet} and added that the iterative process was useful not only for design but also for \pquote{looking inward and discovering habits and preferences I had not noticed before} (D02). Others emphasized that the agentic questioning structure was especially valuable when they lacked initial ideas: \pquote{The guided questioning worked well and was very helpful when I had no inspiration} (D06), while another participant reported that \pquote{the co-design assistant helped me better imagine the details of the object, including material, shape, and color, which improved my creativity} (D38), demonstrating the value of a proactive co-creator over a passive text box. Participants also highlighted novelty and atmosphere as core strengths, describing the experience as \pquote{very creative} and \pquote{quite novel} (D22), and noting that \pquote{I really liked the atmosphere... being able to create by using my imagination was very satisfying} (D10).

At the same time, the qualitative feedback clarifies why engagement scores, although still high, were slightly lower than the other Likert dimensions. The main critiques concerned constrained interaction structure, limited scene-aware control, and the persistent fidelity gap in the final 3D outputs. Some participants preferred more explicit interface scaffolding over pure conversation, suggesting that \pquote{it might be better if the AI assistant could provide selectable options instead of only dialogue} (D22), or remarking that the assistant's guidance was \pquote{very clear but also quite limited} and should better reflect scene-specific emotional goals, for example by guiding frightening scenes toward disliked rather than liked colors (D30). Others asked for more practical control over scale, placement, and physical realism: \pquote{It would be even better if the created objects matched real-world proportions more closely} (D01), and \pquote{add useful customization functions, such as letting the AI preset object size based on the scene} (D49). Finally, multiple comments returned to the 2D-to-3D bottleneck already seen in our quantitative satisfaction analyses. Participants noted that \pquote{the 2D images were still quite stunning, but the quality of the 3D models was somewhat low} (D57), that \pquote{accuracy still needs improvement; gloss handling is not very good. Cartoon-style designs work well, but realistic ones do not} (D53), and that some indoor assets looked overly artificial, with \pquote{furniture and panels [having] low realism and too much of a model-like appearance} (D56). Another participant explicitly connected this limitation to affective authoring scope, observing that \pquote{only being able to generate one object is limiting, because a single object is often not enough to create a clear emotional shift visually} (D37). Together, these comments position the agent as a useful creativity scaffold whose main remaining weaknesses lie less in preference elicitation than in scene-aware control and downstream 3D execution fidelity.

\section{Discussion}
\label{sec:discussion}

\subsection{Findings}
\label{sec:disc:findings}

Our empirical evaluation of \sysname{} demonstrates the efficacy of a scene-anchored, human-AI co-design pipeline for immersive authoring. Specifically, our research questions can be answered as below:

\textbf{Answering RQ1 (Pipeline Effectiveness):} Our findings demonstrate that \sysname{} successfully enables non-expert users to translate conversational intent into deployable VR assets. As evidenced by the prompt analysis (\autoref{sec:generation_prompt_analysis}) and subjective feedback (\autoref{sec:results_design_experience}), the LLM agent's structured elicitation process effectively scaffolded users. Participants reported that the conversation helped clarify vague intentions and stimulated creativity, allowing them to generate contextually appropriate, scene-anchored designs without requiring professional 3D modeling expertise. 

\textbf{Answering RQ2 (Output Quality \& Authorship):} While the text-to-image stage produced highly rated concept images, we identified a persistent quality bottleneck during the image-to-3D conversion phase (\autoref{sec:results_satisfaction}). In particular, our satisfaction models revealed that this perceived quality gap was \textit{not} moderated by authorship context. Participants who co-designed the assets were just as critical of the degraded 3D meshes as independent validators, indicating an absence of the traditional ``IKEA effect'' within this specific generative workflow.

\textbf{Answering RQ3 (Affective \& Behavioral Impact):} Integrating generated assets into immersive scenes significantly altered user behavior and emotional perception. Behaviorally, scene engagement (dwell time) effectively doubled in populated-scene conditions relative to the asset-free baselines (\autoref{sec:results_engagement}). Affectively, the presence of these assets shifted emotional responses—most notably by increasing valence, arousal, and dominance in neutral or negatively valenced environments (e.g., \textit{Tunnel} and \textit{Solitary Confinement}) (\autoref{sec:results_sam}). Furthermore, users exhibited highly scene-dependent interaction patterns, dynamically adjusting their spatial manipulation (scaling and translating) based on the environmental context.

\subsection{Scene-Driven Design Reasoning and Applications}
\label{sec:disc:reasoning}

Expanding on our findings for \textbf{RQ1}, we observed that the success of the co-design pipeline relies heavily on how users contextualize their intent. While our study leveraged scenes that render particular affective outcomes to establish a controlled baseline, the resulting generative behaviors reveal a broader phenomenon: users naturally perform \textit{scene-driven design reasoning}. Participants did not generate arbitrary objects, instead, they adapted their intents to fit the spatial, narrative, and atmospheric constraints of their specific environment. This was evident in their strong preference for ``realism'' over stylization to maintain immersion, and in their selection of minimalist, functional furniture for confined spaces versus organic, playful objects for expansive outdoor scenes.

This scene-driven reasoning underscores the broad generalizability of the \sysname{} pipeline. Because the LLM agent successfully anchors the user's conversational intent to a specific environmental brief, this architecture can scale well beyond affective modulation into diverse XR domains. For instance, in educational VR, the agent could guide students in co-designing period-accurate historical artifacts. In industrial training, workers could generate context-specific tools on the fly to simulate edge-case scenarios. In social VR, users could collaboratively design theme-congruent personalized artifacts, shifting the paradigm from passive consumption to situated, continuous creation.

\subsection{The Illusion of the ``IKEA Effect'' in AI Co-Creation}
\label{sec:disc:ikea}
In exploring how authorship moderates quality perception (\textbf{RQ2}), , traditional HCI and psychology literature suggests that users attribute higher value to artifacts they help create compared to identical artifacts created entirely by others -- a phenomenon known as the ``IKEA effect''~\cite{norton2012ikea}. However, our satisfaction models reveal a disruption of this psychological ownership in generative AI co-design. Across most scenes, both the designers (who co-authored the prompts) and the validators (who merely experienced the assets) rated degraded 3D models with similarly critical scrutiny, with no significant Group $\times$ Modality interaction indicating generalized author leniency, as shown in \autoref{sec:results_satisfaction}.

This absence of the IKEA effect can be attributed to the ``black box'' nature of the image-to-3D execution. Because the AI handles the final, computationally opaque stages of meshing and texturing, the user's sense of ownership is fundamentally mediated. They operate as \textit{art directors} rather than \textit{crafters}. Consequently, participants maintained a critical distance from the final output, judging the 3D asset purely on its objective visual fidelity rather than rationalizing its topological flaws through a lens of personal effort. For future co-design tools to reinstate psychological ownership, they may need to integrate post-generation manual refinement tools, allowing users to physically ``touch up'' the AI's output.

\subsection{The Need for Generative World Models in Spatial Co-Design}
\label{sec:disc:worldmodels}
Based on the behavioral interaction patterns uncovered in \textbf{RQ3}, we observed that participants systematically scaled objects up in vast outdoor environments (\textit{Surrounded by Elephants}, \textit{Jetty at Lake}) and scaled them down in confined indoor environments (\textit{Solitary Confinement}). This highly consistent manipulation demonstrates that users possess strict mental models of spatial logic and proportions; models that current text-to-3D pipelines lack. Previous work such as GestuProp~\cite{yao2026gestuprop} has shown that users can adapt their gestures to spatially control generative outputs, but the underlying system still treats each generation as an isolated event without spatial awareness.

Presently, generative models synthesize isolated objects in a vacuum, forcing users to act as manual spatial correctors upon deployment. To alleviate this friction, future generative pipelines should evolve from isolated asset generation toward ``Generative World Models.'' By ingesting spatial contextual data, such as scene bounding boxes, collision meshes, and semantic surfaces (e.g., recognizing a tabletop versus a floor), the system could automatically infer spatially logical default transforms and spawn points. This would allow the AI to seamlessly ground assets in the environment based on the user's spatial intent, without requiring tedious manual recalibration.

\subsection{Generating Kinematics and Interactivity Beyond Static Meshes}
\label{sec:disc:kinematics}

Furthermore, the extensive physical manipulation (translation and rotation) observed in \textbf{RQ3} highlights a secondary behavioral implication. Specifically, we note that while achieving high-fidelity geometry and textures is a significant milestone, it is insufficient for total VR immersion if the asset behaves like a static statue~\cite{ashtari2020creating}. In our study, a visually flawless bird suspended rigidly in mid-air or a machine devoid of moving parts inevitably risks breaking the illusion of presence, plunging the experience into an interactive uncanny valley.

As visual generation improves, the research community could pivot toward authoring interactive affordances. Co-generating a mesh should be coupled with co-generating its kinematics. Future workflows will need to synthesize underlying skeletons, joint constraints, and physical properties (such as mass, center of gravity, and precise colliders) alongside the visual geometry. This will ensure that generated objects possess inherent interactive affordances and behave logically within the VR physics engine, transitioning them from mere decorative set dressing to fully functional interactive props.

\subsection{Limitations and Future Work}
\label{sec:disc:limitations}
Our study has several limitations that present opportunities for future research. First, our participant pool was recruited via convenience sampling from universities, resulting in a culturally and demographically homogeneous samples. Future work can replicate these findings across broader demographics and diverse headset ecosystems. Second, the cross-sectional nature of the study makes it difficult to untangle whether the doubling of scene engagement time was driven by the inherent value of the co-designed assets or by a ``novelty effect'' of experiencing AI-generated content in VR. Longitudinal studies are needed to see if these interaction patterns persist over time. Additionally, as generative model architectures evolve rapidly, the latency and fidelity metrics reported here represent a snapshot of current capabilities, though we anticipate the 2D-to-3D bottleneck will persist as a relative constraint in the near term.

In future work, we aim to expand \sysname{} from generating single props to facilitating multi-asset ecosystems. Populating an entire room will likely require batch conversational planning and staged execution to keep multi-object authoring manageable, as suggested by prior XR scene-authoring and world-building systems~\cite{lee2025imaginatear,zhang2024vrcopilot,vachha2025dreamcrafter}. Furthermore, multi-asset scenes require \textit{cross-object interaction} -- for example, generating a bird that possesses the semantic and physical logic to land on a dynamically generated chair. This will require tighter integration between conversational intent, spatial reasoning, and physics-aware generation. Finally, we plan to explore the potential of generative world models that can ingest spatial context to automatically infer logical placement and scaling, reducing the burden of manual corrections for users and further streamlining the co-design process.

\section{Conclusion}
\label{sec:conclusion}
In this paper, we presented \sysname{}, an agentic human-AI co-design pipeline that enables non-expert users to generate and deploy 3D assets for VR environments through a proactive, structured conversational interface. Our empirical evaluation with 120 participants across six affectively diverse VR scenes revealed higher engagement and shifted affective responses in populated-scene conditions, particularly in neutral and negatively valenced contexts. However, we also identified a persistent quality bottleneck in the 2D-to-3D generation pipeline, which impacted user satisfaction with the final assets. Our analysis of user prompts and feedback highlighted a strong preference for realism and scene-congruent designs, as well as a desire for more scene-aware control and manual refinement tools to mitigate the fidelity gap. These findings suggest that while the \sysname{} pipeline effectively facilitates human-AI co-design in VR, there remain challenges in generation quality and spatial reasoning to fully realize this approach's potential for immersive content creation. Future work will explore multi-asset co-design, generative world models, and tighter integration between conversational intent and spatial context to further enhance the co-design experience and outcomes.



\bibliographystyle{ACM-Reference-Format}
\bibliography{main}

@inproceedings{zhang2024vrcopilot,
  title        = {VRCopilot: Authoring 3D Layouts with Generative AI Models in VR},
  author       = {Zhang, Lei and Pan, Jin and Gettig, Jacob and Oney, Steve and Guo, Anhong},
  year         = 2024,
  booktitle    = {Proceedings of the 37th Annual ACM Symposium on User Interface Software and Technology},
  location     = {Pittsburgh, PA, USA},
  publisher    = {Association for Computing Machinery},
  address      = {New York, NY, USA},
  series       = {UIST '24},
  doi          = {10.1145/3654777.3676451},
  isbn         = 9798400706288,
  url          = {https://doi.org/10.1145/3654777.3676451},
  articleno    = 96,
  numpages     = 13
}

@article{rahimi2025generative,
  title        = {Generative AI Meets Virtual Reality: A Comprehensive Survey on Applications, Challenges, and Future Direction},
  author       = {Rahimi, Fatema and Sadeghi-Niaraki, Abolghasem and Choi, Soo-Mi},
  year         = 2025,
  journal      = {IEEE Access},
  volume       = 13,
  pages        = {94893--94909},
  doi          = {10.1109/ACCESS.2025.3574779}
}

@article{caetano2025agentic,
  title        = {Agentic workflows for conversational human-ai interaction design},
  author       = {Caetano, Arthur and Verma, Kavya and Taheri, Atieh and Kumaran, Radha and Chen, Zichen and Chen, Jiaao and H{\"o}llerer, Tobias and Sra, Misha},
  year         = 2025,
  journal      = {arXiv preprint arXiv:2501.18002}
}

@article{li2026safetybuilder,
  title        = {SafetyBuilder: An AR-based Framework for In-situ AI-assisted Creation of Child Safety Protection},
  author       = {Li, Jiawei and Li, Zisu and Chen, Siyu and Wang, Ziyan and Zhang, Yukai and Fan, Mingming and He, Liang},
  year         = 2026,
  month        = mar,
  journal      = {Proc. ACM Interact. Mob. Wearable Ubiquitous Technol.},
  publisher    = {Association for Computing Machinery},
  address      = {New York, NY, USA},
  volume       = 10,
  number       = 1,
  doi          = {10.1145/3789690},
  url          = {https://doi.org/10.1145/3789690},
  articleno    = 8,
  issue_date   = {March 2026},
  numpages     = 29
}

@inproceedings{lee2025imaginatear,
  title        = {ImaginateAR: AI-Assisted In-Situ Authoring in Augmented Reality},
  author       = {Lee, Jaewook and Aleotti, Filippo and Mazala, Diego and Garcia-Hernando, Guillermo and Vicente, Sara and Johnston, Oliver James and Kraus-Liang, Isabel and Powierza, Jakub and Shin, Donghoon and Froehlich, Jon E. and Brostow, Gabriel and Van Brummelen, Jessica},
  year         = 2025,
  booktitle    = {Proceedings of the 38th Annual ACM Symposium on User Interface Software and Technology},
  publisher    = {Association for Computing Machinery},
  address      = {New York, NY, USA},
  series       = {UIST '25},
  doi          = {10.1145/3746059.3747635},
  isbn         = 9798400720376,
  url          = {https://doi.org/10.1145/3746059.3747635},
  articleno    = 52,
  numpages     = 21
}

@inproceedings{yao2026gestuprop,
  title        = {GestuProp: 3D Virtual Reality Prop Generation with Co-Speech Gestures},
  author       = {Yao, Zhihao and Yao, Xiwen and Xiong, Haowei and Feng, Yuan-Ling and Sun, Qirui and Guo, Yijie and Mi, Haipeng},
  year         = 2026,
  booktitle    = {Proceedings of the 2026 CHI Conference on Human Factors in Computing Systems},
  publisher    = {Association for Computing Machinery},
  address      = {New York, NY, USA},
  series       = {CHI '26},
  doi          = {10.1145/3772318.3790491},
  isbn         = 9798400722783,
  url          = {https://doi.org/10.1145/3772318.3790491},
  articleno    = 1107,
  numpages     = 17
}

@inproceedings{kim2025realitycrafter,
  title        = {RealityCrafter: User-guided Editable 3D Scene Generation from a Single Image in Mixed Reality},
  author       = {Kim, Seokyoung and Kim, Dooyoung and Son, Taejun and Woo, Woontack},
  year         = 2025,
  booktitle    = {Adjunct Proceedings of the 38th Annual ACM Symposium on User Interface Software and Technology},
  publisher    = {Association for Computing Machinery},
  address      = {New York, NY, USA},
  series       = {UIST Adjunct '25},
  doi          = {10.1145/3746058.3758405},
  isbn         = 9798400720369,
  url          = {https://doi.org/10.1145/3746058.3758405},
  articleno    = 109,
  numpages     = 3
}

@article{safaribazargani20262d,
  title        = {From 2D to XR: enhancing object placement in design simulations through spatial relations},
  author       = {Safari Bazargani, Jalal and Sadeghi-Niaraki, Abolghasem and Choi, Soo-Mi},
  year         = 2026,
  month        = {Feb},
  day          = 24,
  journal      = {Virtual Reality},
  volume       = 30,
  number       = 2,
  pages        = 76,
  doi          = {10.1007/s10055-026-01327-0},
  issn         = {1434-9957},
  url          = {https://doi.org/10.1007/s10055-026-01327-0}
}

@inproceedings{chen2026ai4xr,
  title        = {AI4XR: AI in Extended Reality for 3D Scene Editing and Accessibility Design},
  author       = {Chen, Junlong},
  year         = 2026,
  booktitle    = {Proceedings of the Extended Abstracts of the 2026 CHI Conference on Human Factors in Computing Systems},
  publisher    = {Association for Computing Machinery},
  address      = {New York, NY, USA},
  series       = {CHI EA '26},
  doi          = {10.1145/3772363.3799187},
  isbn         = 9798400722813,
  url          = {https://doi.org/10.1145/3772363.3799187},
  articleno    = 879,
  numpages     = 6
}

@inproceedings{wen2026prompt,
  title        = {From Prompt to Presence: Co-Creating Personalised Emotional Sanctuaries in VR with Generative AI},
  author       = {Wen, Ruoyu and Gupta, Kunal and Hoermann, Simon and Billinghurst, Mark and Nassani, Alaeddin and Piumsomboon, Thammathip},
  year         = 2026,
  booktitle    = {Proceedings of the 31st International Conference on Intelligent User Interfaces},
  publisher    = {Association for Computing Machinery},
  address      = {New York, NY, USA},
  series       = {IUI '26},
  pages        = {955–971},
  doi          = {10.1145/3742413.3789067},
  isbn         = 9798400719844,
  url          = {https://doi.org/10.1145/3742413.3789067},
  numpages     = 17
}

@inproceedings{vachha2025dreamcrafter,
  title        = {Dreamcrafter: Immersive Editing of 3D Radiance Fields Through Flexible, Generative Inputs and Outputs},
  author       = {Vachha, Cyrus and Kang, Yixiao and Dive, Zach and Chidambaram, Ashwat and Gupta, Anik and Jun, Eunice and Hartmann, Bj\"{o}rn},
  year         = 2025,
  booktitle    = {Proceedings of the 2025 CHI Conference on Human Factors in Computing Systems},
  publisher    = {Association for Computing Machinery},
  address      = {New York, NY, USA},
  series       = {CHI '25},
  doi          = {10.1145/3706598.3714312},
  isbn         = 9798400713941,
  url          = {https://doi.org/10.1145/3706598.3714312},
  articleno    = 3,
  numpages     = 13
}

@inproceedings{rombach2022high-resolution,
  title        = {High-Resolution Image Synthesis with Latent Diffusion Models},
  author       = {Rombach, Robin and Blattmann, Andreas and Lorenz, Dominik and Esser, Patrick and Ommer, Bj{\"o}rn},
  year         = 2022,
  booktitle    = {Proceedings of the IEEE/CVF Conference on Computer Vision and Pattern Recognition (CVPR)},
  pages        = {10684--10695}
}

@inproceedings{esser2024scaling,
  title        = {Scaling Rectified Flow Transformers for High-Resolution Image Synthesis},
  author       = {Esser, Patrick and Kulal, Sumith and Blattmann, Andreas and Entezari, Rahim and M{\"u}ller, Jonas and Saini, Harry and Levi, Yam and Lorenz, Dominik and Sauer, Axel and Boesel, Frederic and others},
  year         = 2024,
  booktitle    = {Proceedings of the 41st International Conference on Machine Learning (ICML)}
}

@inproceedings{liu2023zero-1-to-3,
  title        = {Zero-1-to-3: Zero-shot One Image to {3D} Object},
  author       = {Liu, Ruoshi and Ramesh, Aditya and Liu, Ce and Zhu, Chenfanfu},
  year         = 2023,
  booktitle    = {Proceedings of the IEEE/CVF International Conference on Computer Vision (ICCV)}
}

@article{zhao2025hunyuan3d,
  title        = {Hunyuan3D 2.0: Scaling Diffusion Models for High Resolution Textured {3D} Assets Generation},
  author       = {Zhao, Zibo and Chen, Zeqiang and Yang, Qingxu and others},
  year         = 2025,
  journal      = {arXiv preprint arXiv:2501.12202}
}

@misc{labs2025flux1kontextflowmatching,
      title={FLUX.1 Kontext: Flow Matching for In-Context Image Generation and Editing in Latent Space},
      author={Black Forest Labs and Stephen Batifol and Andreas Blattmann and Frederic Boesel and Saksham Consul and Cyril Diagne and Tim Dockhorn and Jack English and Zion English and Patrick Esser and Sumith Kulal and Kyle Lacey and Yam Levi and Cheng Li and Dominik Lorenz and Jonas Müller and Dustin Podell and Robin Rombach and Harry Saini and Axel Sauer and Luke Smith},
      year={2025},
      eprint={2506.15742},
      archivePrefix={arXiv},
      primaryClass={cs.GR},
      url={https://arxiv.org/abs/2506.15742},
}

@article{shneiderman2007creativity,
  title        = {Creativity Support Tools: Accelerating Discovery and Innovation},
  author       = {Shneiderman, Ben},
  year         = 2007,
  journal      = {Communications of the ACM},
  volume       = 50,
  number       = 12,
  pages        = {20--32}
}

@inproceedings{davis2013human-computer,
  title        = {Human-Computer Co-Creativity: Blending Human and Computational Creativity},
  author       = {Davis, Nicholas},
  year         = 2013,
  booktitle    = {Proceedings of the AAAI Workshop on Intelligent Narrative Technologies}
}

@article{bradley1994measuring,
  title        = {Measuring Emotion: The Self-Assessment Manikin and the Semantic Differential},
  author       = {Bradley, Margaret M. and Lang, Peter J.},
  year         = 1994,
  journal      = {Journal of Behavior Therapy and Experimental Psychiatry},
  volume       = 25,
  number       = 1,
  pages        = {49--59}
}

@article{russell1977evidence,
  title        = {Evidence for a Three-Factor Theory of Emotions},
  author       = {Russell, James A. and Mehrabian, Albert},
  year         = 1977,
  journal      = {Journal of Research in Personality},
  volume       = 11,
  number       = 3,
  pages        = {273--294}
}

@article{chirico2018designing,
  title        = {Designing Awe in Virtual Reality: An Experimental Study},
  author       = {Chirico, Alice and Ferrise, Francesco and Cordella, Lorenzo and Gaggioli, Andrea},
  year         = 2018,
  journal      = {Frontiers in Psychology},
  volume       = 8,
  pages        = 2351
}

@misc{deepseek-ai2025deepseek-v3,
  title        = {DeepSeek-V3 Technical Report},
  author       = {DeepSeek-AI and Aixin Liu and Bei Feng and Bing Xue and Bingxuan Wang and Bochao Wu and Chengda Lu and Chenggang Zhao and Chengqi Deng and Chenyu Zhang and Chong Ruan and Damai Dai and Daya Guo and Dejian Yang and Deli Chen and Dongjie Ji and Erhang Li and Fangyun Lin and Fucong Dai and Fuli Luo and Guangbo Hao and Guanting Chen and Guowei Li and H. Zhang and Han Bao and Hanwei Xu and Haocheng Wang and Haowei Zhang and Honghui Ding and Huajian Xin and Huazuo Gao and Hui Li and Hui Qu and J. L. Cai and Jian Liang and Jianzhong Guo and Jiaqi Ni and Jiashi Li and Jiawei Wang and Jin Chen and Jingchang Chen and Jingyang Yuan and Junjie Qiu and Junlong Li and Junxiao Song and Kai Dong and Kai Hu and Kaige Gao and Kang Guan and Kexin Huang and Kuai Yu and Lean Wang and Lecong Zhang and Lei Xu and Leyi Xia and Liang Zhao and Litong Wang and Liyue Zhang and Meng Li and Miaojun Wang and Mingchuan Zhang and Minghua Zhang and Minghui Tang and Mingming Li and Ning Tian and Panpan Huang and Peiyi Wang and Peng Zhang and Qiancheng Wang and Qihao Zhu and Qinyu Chen and Qiushi Du and R. J. Chen and R. L. Jin and Ruiqi Ge and Ruisong Zhang and Ruizhe Pan and Runji Wang and Runxin Xu and Ruoyu Zhang and Ruyi Chen and S. S. Li and Shanghao Lu and Shangyan Zhou and Shanhuang Chen and Shaoqing Wu and Shengfeng Ye and Shengfeng Ye and Shirong Ma and Shiyu Wang and Shuang Zhou and Shuiping Yu and Shunfeng Zhou and Shuting Pan and T. Wang and Tao Yun and Tian Pei and Tianyu Sun and W. L. Xiao and Wangding Zeng and Wanjia Zhao and Wei An and Wen Liu and Wenfeng Liang and Wenjun Gao and Wenqin Yu and Wentao Zhang and X. Q. Li and Xiangyue Jin and Xianzu Wang and Xiao Bi and Xiaodong Liu and Xiaohan Wang and Xiaojin Shen and Xiaokang Chen and Xiaokang Zhang and Xiaosha Chen and Xiaotao Nie and Xiaowen Sun and Xiaoxiang Wang and Xin Cheng and Xin Liu and Xin Xie and Xingchao Liu and Xingkai Yu and Xinnan Song and Xinxia Shan and Xinyi Zhou and Xinyu Yang and Xinyuan Li and Xuecheng Su and Xuheng Lin and Y. K. Li and Y. Q. Wang and Y. X. Wei and Y. X. Zhu and Yang Zhang and Yanhong Xu and Yanhong Xu and Yanping Huang and Yao Li and Yao Zhao and Yaofeng Sun and Yaohui Li and Yaohui Wang and Yi Yu and Yi Zheng and Yichao Zhang and Yifan Shi and Yiliang Xiong and Ying He and Ying Tang and Yishi Piao and Yisong Wang and Yixuan Tan and Yiyang Ma and Yiyuan Liu and Yongqiang Guo and Yu Wu and Yuan Ou and Yuchen Zhu and Yuduan Wang and Yue Gong and Yuheng Zou and Yujia He and Yukun Zha and Yunfan Xiong and Yunxian Ma and Yuting Yan and Yuxiang Luo and Yuxiang You and Yuxuan Liu and Yuyang Zhou and Z. F. Wu and Z. Z. Ren and Zehui Ren and Zhangli Sha and Zhe Fu and Zhean Xu and Zhen Huang and Zhen Zhang and Zhenda Xie and Zhengyan Zhang and Zhewen Hao and Zhibin Gou and Zhicheng Ma and Zhigang Yan and Zhihong Shao and Zhipeng Xu and Zhiyu Wu and Zhongyu Zhang and Zhuoshu Li and Zihui Gu and Zijia Zhu and Zijun Liu and Zilin Li and Ziwei Xie and Ziyang Song and Ziyi Gao and Zizheng Pan},
  year         = 2025,
  url          = {https://arxiv.org/abs/2412.19437},
  archiveprefix = {arXiv},
  eprint       = {2412.19437},
  primaryclass = {cs.CL}
}

@article{russell1980circumplex,
  title        = {A Circumplex Model of Affect},
  author       = {Russell, James A.},
  year         = 1980,
  journal      = {Journal of Personality and Social Psychology},
  publisher    = {American Psychological Association},
  volume       = 39,
  number       = 6,
  pages        = {1161--1178}
}

@article{jiang2024immersive,
  title        = {An Immersive and Interactive {VR} Dataset to Elicit Emotions},
  author       = {Jiang, Weiwei and Windl, Maximiliane and Tag, Benjamin and Sarsenbayeva, Zhanna and Mayer, Sven},
  year         = 2024,
  journal      = {IEEE Transactions on Visualization and Computer Graphics}
}

@article{li2017public,
  title        = {A Public Database of Immersive {VR} Videos with Corresponding Ratings of Arousal, Valence, and Correlations between Head Movements and Self-Report Measures},
  author       = {Li, Benjamin J. and Bailenson, Jeremy N. and Pines, Adam and Greenleaf, Walter J. and Williams, Leanne M.},
  year         = 2017,
  journal      = {Frontiers in Psychology},
  volume       = 8,
  pages        = 2116
}

@article{zhao2025clear,
  title        = {CleAR: Robust Context-Guided Generative Lighting Estimation for Mobile Augmented Reality},
  author       = {Zhao, Yiqin and Dasari, Mallesham and Guo, Tian},
  year         = 2025,
  month        = sep,
  journal      = {Proc. ACM Interact. Mob. Wearable Ubiquitous Technol.},
  publisher    = {Association for Computing Machinery},
  address      = {New York, NY, USA},
  volume       = 9,
  number       = 3,
  doi          = {10.1145/3749535},
  url          = {https://doi.org/10.1145/3749535},
  articleno    = 154,
  issue_date   = {September 2025},
  numpages     = 26
}

@article{li2025remverse,
  title        = {RemVerse: Supporting Reminiscence Activities for Older Adults through AI-Assisted Virtual Reality},
  author       = {Li, Ruohao and Li, Jiawei and Sun, Jia and Wu, Zhiqing and Li, Zisu and Wang, Ziyan and Lin, Ge and Fan, Mingming},
  year         = 2025,
  month        = sep,
  journal      = {Proc. ACM Interact. Mob. Wearable Ubiquitous Technol.},
  publisher    = {Association for Computing Machinery},
  address      = {New York, NY, USA},
  volume       = 9,
  number       = 3,
  doi          = {10.1145/3749505},
  url          = {https://doi.org/10.1145/3749505},
  articleno    = 103,
  issue_date   = {September 2025},
  numpages     = 25
}

@article{huang2024surfshare,
  title        = {SurfShare: Lightweight Spatially Consistent Physical Surface and Virtual Replica Sharing with Head-mounted Mixed-Reality},
  author       = {Huang, Xincheng and Xiao, Robert},
  year         = 2024,
  month        = jan,
  journal      = {Proc. ACM Interact. Mob. Wearable Ubiquitous Technol.},
  publisher    = {Association for Computing Machinery},
  address      = {New York, NY, USA},
  volume       = 7,
  number       = 4,
  doi          = {10.1145/3631418},
  url          = {https://doi.org/10.1145/3631418},
  articleno    = 162,
  issue_date   = {December 2023},
  numpages     = 24
}

@article{cai2025hpipainting,
  title        = {HPIPainting: A Hand-Pen Interaction for VR Painting},
  author       = {Cai, Ang and Yao, Chao and Liu, Hongjun and Li, Changsheng and Zhang, Yongyue and Guo, Yu and Wang, Xiaokun and Ban, Xiaojuan},
  year         = 2025,
  month        = sep,
  journal      = {Proc. ACM Interact. Mob. Wearable Ubiquitous Technol.},
  publisher    = {Association for Computing Machinery},
  address      = {New York, NY, USA},
  volume       = 9,
  number       = 3,
  doi          = {10.1145/3749538},
  url          = {https://doi.org/10.1145/3749538},
  articleno    = 71,
  issue_date   = {September 2025},
  numpages     = 26
}

@article{wanniarachchi2025mimic,
  title        = {MIMIC: AI and AR-enhanced Multi-Modal, Immersive, Relative Instruction Comprehension},
  author       = {Wanniarachchi, Dhanuja and Misra, Archan},
  year         = 2025,
  month        = mar,
  journal      = {Proc. ACM Interact. Mob. Wearable Ubiquitous Technol.},
  publisher    = {Association for Computing Machinery},
  address      = {New York, NY, USA},
  volume       = 9,
  number       = 1,
  doi          = {10.1145/3712268},
  url          = {https://doi.org/10.1145/3712268},
  articleno    = 21,
  issue_date   = {March 2025},
  numpages     = 34
}

@inproceedings{tang2023delicate,
  title        = {Delicate textured mesh recovery from nerf via adaptive surface refinement},
  author       = {Tang, Jiaxiang and Zhou, Hang and Chen, Xiaokang and Hu, Tianshu and Ding, Errui and Wang, Jingdong and Zeng, Gang},
  year         = 2023,
  booktitle    = {Proceedings of the IEEE/CVF International Conference on Computer Vision},
  pages        = {17739--17749},
  doi          = {10.1109/iccv51070.2023.01626}
}

@inproceedings{yang2022vox-fusion,
  title        = {Vox-fusion: Dense tracking and mapping with voxel-based neural implicit representation},
  author       = {Yang, Xingrui and Li, Hai and Zhai, Hongjia and Ming, Yuhang and Liu, Yuqian and Zhang, Guofeng},
  year         = 2022,
  booktitle    = {2022 IEEE International Symposium on Mixed and Augmented Reality (ISMAR)},
  pages        = {499--507},
  doi          = {10.1109/ismar55827.2022.00066},
  organization = {IEEE}
}

@inproceedings{kania2022conerf,
  title        = {Conerf: Controllable neural radiance fields},
  author       = {Kania, Kacper and Yi, Kwang Moo and Kowalski, Marek and Trzci{\'n}ski, Tomasz and Tagliasacchi, Andrea},
  year         = 2022,
  booktitle    = {Proceedings of the IEEE/CVF Conference on Computer Vision and Pattern Recognition},
  pages        = {18623--18632},
  doi          = {10.1109/cvpr52688.2022.01807}
}

@article{bao20253d,
  title        = {3d gaussian splatting: Survey, technologies, challenges, and opportunities},
  author       = {Bao, Yanqi and Ding, Tianyu and Huo, Jing and Liu, Yaoli and Li, Yuxin and Li, Wenbin and Gao, Yang and Luo, Jiebo},
  year         = 2025,
  journal      = {IEEE Transactions on Circuits and Systems for Video Technology},
  publisher    = {IEEE},
  volume       = 35,
  number       = 7,
  pages        = {6832--6852},
  doi          = {10.1109/tcsvt.2025.3538684}
}

@inproceedings{carlon2025bayesian,
  title        = {A Bayesian Exploration on the Motivational and Behavioral Impacts of Chatbots in Language Learning},
  author       = {Carlon, May Kristine Jonson and Matthews, Julian Rodney and Kuniyoshi, Yasuo},
  year         = 2025,
  booktitle    = {Proceedings of the Extended Abstracts of the CHI Conference on Human Factors in Computing Systems},
  pages        = {1--7},
  doi          = {10.1145/3706599.3720088}
}

@article{li2026separategen,
  title        = {SeparateGen: Semantic Component-based 3D Character Generation from Single Images},
  author       = {Li, Dong-Yang and Liu, Yi-Long and Liu, Zi-Xian and Cao, Yan-Pei and Guo, Meng-Hao and Hu, Shi-Min},
  year         = 2026,
  journal      = {IEEE Transactions on Visualization and Computer Graphics},
  publisher    = {IEEE},
  doi          = {10.1109/tvcg.2026.3652452}
}

@article{wen2022pixel2mesh++,
  title        = {Pixel2Mesh++: 3D mesh generation and refinement from multi-view images},
  author       = {Wen, Chao and Zhang, Yinda and Cao, Chenjie and Li, Zhuwen and Xue, Xiangyang and Fu, Yanwei},
  year         = 2022,
  journal      = {IEEE Transactions on Pattern Analysis and Machine Intelligence},
  publisher    = {IEEE},
  volume       = 45,
  number       = 2,
  pages        = {2166--2180},
  doi          = {10.1109/tpami.2022.3169735}
}

@inproceedings{tsalicoglou2024textmesh,
  title        = {Textmesh: Generation of realistic 3d meshes from text prompts},
  author       = {Tsalicoglou, Christina and Manhardt, Fabian and Tonioni, Alessio and Niemeyer, Michael and Tombari, Federico},
  year         = 2024,
  booktitle    = {2024 International Conference on 3D Vision (3DV)},
  pages        = {1554--1563},
  doi          = {10.1109/3dv62453.2024.00154},
  organization = {IEEE}
}

@inproceedings{henderson2020leveraging,
  title        = {Leveraging 2d data to learn textured 3d mesh generation},
  author       = {Henderson, Paul and Tsiminaki, Vagia and Lampert, Christoph H},
  year         = 2020,
  booktitle    = {Proceedings of the IEEE/CVF conference on computer vision and pattern recognition},
  pages        = {7498--7507},
  doi          = {10.1109/cvpr42600.2020.00752}
}

@inproceedings{raj2023dreambooth3d,
  title        = {Dreambooth3d: Subject-driven text-to-3d generation},
  author       = {Raj, Amit and Kaza, Srinivas and Poole, Ben and Niemeyer, Michael and Ruiz, Nataniel and Mildenhall, Ben and Zada, Shiran and Aberman, Kfir and Rubinstein, Michael and Barron, Jonathan and others},
  year         = 2023,
  booktitle    = {Proceedings of the IEEE/CVF international conference on computer vision},
  pages        = {2349--2359},
  doi          = {10.1109/iccv51070.2023.00223}
}

@inproceedings{chen2023single-stage,
  title        = {Single-stage diffusion nerf: A unified approach to 3d generation and reconstruction},
  author       = {Chen, Hansheng and Gu, Jiatao and Chen, Anpei and Tian, Wei and Tu, Zhuowen and Liu, Lingjie and Su, Hao},
  year         = 2023,
  booktitle    = {Proceedings of the IEEE/CVF international conference on computer vision},
  pages        = {2416--2425},
  doi          = {10.1109/iccv51070.2023.00229}
}

@inproceedings{lin2023magic3d,
  title        = {Magic3d: High-resolution text-to-3d content creation},
  author       = {Lin, Chen-Hsuan and Gao, Jun and Tang, Luming and Takikawa, Towaki and Zeng, Xiaohui and Huang, Xun and Kreis, Karsten and Fidler, Sanja and Liu, Ming-Yu and Lin, Tsung-Yi},
  year         = 2023,
  booktitle    = {Proceedings of the IEEE/CVF conference on computer vision and pattern recognition},
  pages        = {300--309},
  doi          = {10.1109/cvpr52729.2023.00037}
}

@inproceedings{xiang2025structured,
  title        = {Structured 3d latents for scalable and versatile 3d generation},
  author       = {Xiang, Jianfeng and Lv, Zelong and Xu, Sicheng and Deng, Yu and Wang, Ruicheng and Zhang, Bowen and Chen, Dong and Tong, Xin and Yang, Jiaolong},
  year         = 2025,
  booktitle    = {Proceedings of the IEEE/CVF conference on computer vision and pattern recognition},
  pages        = {21469--21480},
  doi          = {10.1109/cvpr52734.2025.02000}
}

@inproceedings{hllein2023text2room,
  title        = {Text2room: Extracting textured 3d meshes from 2d text-to-image models},
  author       = {H{\"o}llein, Lukas and Cao, Ang and Owens, Andrew and Johnson, Justin and Nie{\ss}ner, Matthias},
  year         = 2023,
  booktitle    = {Proceedings of the IEEE/CVF International Conference on Computer Vision},
  pages        = {7909--7920},
  doi          = {10.1109/iccv51070.2023.00727}
}

@article{liu2023one-2-3-45,
  title        = {One-2-3-45: Any single image to 3d mesh in 45 seconds without per-shape optimization},
  author       = {Liu, Minghua and Xu, Chao and Jin, Haian and Chen, Linghao and Varma T, Mukund and Xu, Zexiang and Su, Hao},
  year         = 2023,
  journal      = {Advances in Neural Information Processing Systems},
  volume       = 36,
  pages        = {22226--22246},
  doi          = {10.52202/075280-0976}
}

@inproceedings{matsuki20254dtam,
  title        = {4dtam: Non-rigid tracking and mapping via dynamic surface gaussians},
  author       = {Matsuki, Hidenobu and Bae, Gwangbin and Davison, Andrew J},
  year         = 2025,
  booktitle    = {Proceedings of the Computer Vision and Pattern Recognition Conference},
  pages        = {26921--26932},
  doi          = {10.1109/cvpr52734.2025.02507}
}

@inproceedings{xiong2025texgaussian,
  title        = {Texgaussian: Generating high-quality pbr material via octree-based 3d gaussian splatting},
  author       = {Xiong, Bojun and Liu, Jialun and Hu, Jiakui and Wu, Chenming and Wu, Jinbo and Liu, Xing and Zhao, Chen and Ding, Errui and Lian, Zhouhui},
  year         = 2025,
  booktitle    = {Proceedings of the Computer Vision and Pattern Recognition Conference},
  pages        = {551--561},
  doi          = {10.1109/cvpr52734.2025.00060}
}

@inproceedings{zhou2024understanding,
  title        = {Understanding nonlinear collaboration between human and AI agents: A co-design framework for creative design},
  author       = {Zhou, Jiayi and Li, Renzhong and Tang, Junxiu and Tang, Tan and Li, Haotian and Cui, Weiwei and Wu, Yingcai},
  year         = 2024,
  booktitle    = {Proceedings of the 2024 CHI conference on human factors in computing systems},
  pages        = {1--16},
  doi          = {10.1145/3613904.3642812}
}

@inproceedings{gmeiner2023exploring,
  title        = {Exploring challenges and opportunities to support designers in learning to co-create with AI-based manufacturing design tools},
  author       = {Gmeiner, Frederic and Yang, Humphrey and Yao, Lining and Holstein, Kenneth and Martelaro, Nikolas},
  year         = 2023,
  booktitle    = {Proceedings of the 2023 CHI conference on human factors in computing systems},
  pages        = {1--20},
  doi          = {10.1145/3544548.3580999}
}

@inproceedings{lin2021engaging,
  title        = {Engaging teachers to co-design integrated AI curriculum for K-12 classrooms},
  author       = {Lin, Phoebe and Van Brummelen, Jessica},
  year         = 2021,
  booktitle    = {Proceedings of the 2021 CHI conference on human factors in computing systems},
  pages        = {1--12},
  doi          = {10.1145/3411764.3445377}
}

@inproceedings{yang2020re-examining,
  title        = {Re-examining whether, why, and how human-AI interaction is uniquely difficult to design},
  author       = {Yang, Qian and Steinfeld, Aaron and Ros{\'e}, Carolyn and Zimmerman, John},
  year         = 2020,
  booktitle    = {Proceedings of the 2020 chi conference on human factors in computing systems},
  pages        = {1--13},
  doi          = {10.1145/3313831.3376301}
}

@inproceedings{krol2025exploring,
  title        = {Exploring the needs of practising musicians in co-creative ai through co-design},
  author       = {Krol, Stephen James and Llano Rodriguez, Maria Teresa and Loor Paredes, Miguel J},
  year         = 2025,
  booktitle    = {Proceedings of the 2025 CHI Conference on Human Factors in Computing Systems},
  pages        = {1--13},
  doi          = {10.1145/3706598.3713894}
}

@article{fouad2026human-centered,
  title        = {Human-Centered User Interface Design for Explainable AI in Chest Radiology: A Multi-Phase Co-Design Approach},
  author       = {Fouad, Shereen and Hakobyan, Lilit and Ihongbe, Izegbua E and Kavakli-Thorne, Manolya and Atkins, Sarah and Bhatia, Bahadar},
  year         = 2026,
  journal      = {IEEE Access},
  publisher    = {IEEE},
  doi          = {10.1109/access.2026.3653233}
}

@article{christensen2026discretionary,
  title        = {Discretionary Freedom in Social Work? Co-Design of AI-Enabled Case Management System in Trouble},
  author       = {Christensen, Lars Rune and Petersen, Anette CM},
  year         = 2026,
  journal      = {Computer Supported Cooperative Work (CSCW)},
  publisher    = {Springer},
  volume       = 35,
  number       = 2,
  pages        = 3,
  doi          = {10.1007/s10606-026-09539-3}
}

@inproceedings{madaio2020co-designing,
  title        = {Co-designing checklists to understand organizational challenges and opportunities around fairness in AI},
  author       = {Madaio, Michael A and Stark, Luke and Wortman Vaughan, Jennifer and Wallach, Hanna},
  year         = 2020,
  booktitle    = {Proceedings of the 2020 CHI conference on human factors in computing systems},
  pages        = {1--14},
  doi          = {10.1145/3313831.3376445}
}

@inproceedings{jing2026req2cad,
author = {Jing, Qianzhi and Lu, Hankai and Huang, Shuojin and Childs, Peter and Chen, Liuqing},
title = {Req2CAD: bridging functional requirements and parametric CAD models to support conceptual 3D design},
year = {2026},
isbn = {9798400722783},
publisher = {Association for Computing Machinery},
address = {New York, NY, USA},
url = {https://doi.org/10.1145/3772318.3791949},
doi = {10.1145/3772318.3791949},
booktitle = {Proceedings of the 2026 CHI Conference on Human Factors in Computing Systems},
articleno = {152},
numpages = {25},
location = {
},
series = {CHI '26}
}

@inproceedings{kuang2026understanding,
author = {Kuang, Zheyuan and Li, Tinghui and Jiang, Weiwei and Mayer, Sven and D. Salim, Flora and Tag, Benjamin and Withana, Anusha and Sarsenbayeva, Zhanna},
title = {Understanding the Effects of Interaction on Emotional Experiences in VR},
year = {2026},
isbn = {9798400722783},
publisher = {Association for Computing Machinery},
address = {New York, NY, USA},
url = {https://doi.org/10.1145/3772318.3790313},
doi = {10.1145/3772318.3790313},
booktitle = {Proceedings of the 2026 CHI Conference on Human Factors in Computing Systems},
articleno = {631},
numpages = {20},
location = {
},
series = {CHI '26}
}

@article{schone2023library,
  title={Library for universal virtual reality experiments (luVRe): A standardized immersive 3D/360 picture and video database for VR based research},
  author={Sch{\"o}ne, Benjamin and Kisker, Joanna and Sylvester, Rebecca Sophia and Radtke, Elise Leila and Gruber, Thomas},
  journal={Current Psychology},
  volume={42},
  number={7},
  pages={5366--5384},
  year={2023}
}

@article{somarathna2022virtual,
	title        = {Virtual Reality for Emotion Elicitation: A Review},
	author       = {Somarathna, Rukshani and Bednarz, Tomasz and Mohammadi, Gelareh},
	year         = 2022,
	journal      = {IEEE Transactions on Affective Computing},
	volume       = {},
	number       = {},
	pages        = {1--21},
	doi          = {10.1109/TAFFC.2022.3181053}
}

@inproceedings{xie2020applying,
  title={Applying self-assessment manikin (sam) to evaluate the affective arousal effects of vr games},
  author={Xie, Tianhua and Cao, Mingliang and Pan, Zhigeng},
  booktitle={Proceedings of the 2020 3rd International Conference on Image and Graphics Processing},
  pages={134--138},
  year={2020}
}

@inproceedings{bartram2017affective,
	title        = {Affective color in visualization},
	author       = {Bartram, Lyn and Patra, Abhisekh and Stone, Maureen},
	year         = 2017,
	booktitle    = {Proceedings of the 2017 CHI conference on human factors in computing systems},
	pages        = {1364--1374}
}

@inproceedings{jun2018full,
	title        = {Full-body ownership illusion can change our emotion},
	author       = {Jun, Joohee and Jung, Myeongul and Kim, So-Yeon and Kim, Kwanguk},
	year         = 2018,
	booktitle    = {Proceedings of the 2018 CHI conference on human factors in computing systems},
	pages        = {1--11}
}

@inproceedings{kern2020influence,
	title        = {The influence of mood induction by music or a soundscape on presence and emotions in a virtual reality park scenario},
	author       = {Kern, Angelika C and Ellermeier, Wolfgang and Jost, Lina},
	year         = 2020,
	booktitle    = {Proceedings of the 15th International Conference on Audio Mostly},
	pages        = {233--236}
}

@inproceedings{wagener2024moodshaper,
  title={MoodShaper: A Virtual Reality Experience to Support Managing Negative Emotions},
  author={Wagener, Nadine and Kiesewetter, Arne and Reicherts, Leon and Wo{\'z}niak, Pawe{\l} W and Sch{\"o}ning, Johannes and Rogers, Yvonne and Niess, Jasmin},
  booktitle={Proceedings of the 2024 ACM Designing Interactive Systems Conference},
  pages={2286--2304},
  year={2024}
}

@inproceedings{elor2021understanding,
  title={Understanding emotional expression with haptic feedback vest patterns and immersive virtual reality},
  author={Elor, Aviv and Song, Asiiah and Kurniawan, Sri},
  booktitle={2021 IEEE Conference on Virtual Reality and 3D User Interfaces Abstracts and Workshops (VRW)},
  pages={183--188},
  year={2021}
}

@article{lottridge2011affective,
  title={Affective interaction: understanding, evaluating, and designing for human emotion},
  author={Lottridge, Danielle and Chignell, Mark and Jovicic, Aleksandra},
  journal={Reviews of Human Factors and Ergonomics},
  volume={7},
  number={1},
  pages={197--217},
  year={2011}
}

@article{riva2007affective,
  title={Affective interactions using virtual reality: the link between presence and emotions},
  author={Riva, Giuseppe and Mantovani, Fabrizia and Capideville, Claret Samantha and Preziosa, Alessandra and Morganti, Francesca and Villani, Daniela and Gaggioli, Andrea and Botella, Cristina and Alca{\~n}iz, Mariano},
  journal={Cyberpsychology \& behavior},
  volume={10},
  number={1},
  pages={45--56},
  year={2007}
}

@inproceedings{estupinan2014can,
  title={Can virtual reality increase emotional responses (arousal and valence)? A pilot study},
  author={Estupi{\~n}{\'a}n, Sergio and Rebelo, Francisco and Noriega, Paulo and Ferreira, Carlos and Duarte, Em{\'\i}lia},
  booktitle={International conference of design, user experience, and usability},
  pages={541--549},
  year={2014}
}

@inproceedings{jicol2021effects,
  title={Effects of emotion and agency on presence in virtual reality},
  author={Jicol, Crescent and Wan, Chun Hin and Doling, Benjamin and Illingworth, Caitlin H and Yoon, Jinha and Headey, Charlotte and Lutteroth, Christof and Proulx, Michael J and Petrini, Karin and O'Neill, Eamonn},
  booktitle={Proceedings of the 2021 CHI conference on human factors in computing systems},
  pages={1--13},
  year={2021}
}

@article{csikszentmihalyi1987validity,
  title={Validity and reliability of the experience-sampling method},
  author={Csikszentmihalyi, Mihaly and Larson, Reed},
  journal={The Journal of nervous and mental disease},
  volume={175},
  number={9},
  pages={526--536},
  year={1987}
}

@inproceedings{ashtari2020creating,
author = {Ashtari, Narges and Bunt, Andrea and McGrenere, Joanna and Nebeling, Michael and Chilana, Parmit K.},
title = {Creating Augmented and Virtual Reality Applications: Current Practices, Challenges, and Opportunities},
year = {2020},
isbn = {9781450367080},
publisher = {Association for Computing Machinery},
address = {New York, NY, USA},
url = {https://doi.org/10.1145/3313831.3376722},
doi = {10.1145/3313831.3376722},
booktitle = {Proceedings of the 2020 CHI Conference on Human Factors in Computing Systems},
pages = {1–13},
numpages = {13},
location = {Honolulu, HI, USA},
series = {CHI '20}
}

@article{dresp2016affine,
  title={Affine geometry, visual sensation, and preference for symmetry of things in a thing},
  author={Dresp-Langley, Birgitta},
  journal={Symmetry},
  volume={8},
  number={11},
  pages={127},
  year={2016},
  publisher={MDPI}
}

@article{schloss2018modeling,
  title={Modeling color preference using color space metrics},
  author={Schloss, Karen B and Lessard, Laurent and Racey, Chris and Hurlbert, Anya C},
  journal={Vision Research},
  volume={151},
  pages={99--116},
  year={2018},
  publisher={Elsevier}
}

@article{norton2012ikea,
title = {The IKEA effect: When labor leads to love},
journal = {Journal of Consumer Psychology},
volume = {22},
number = {3},
pages = {453-460},
year = {2012},
issn = {1057-7408},
doi = {https://doi.org/10.1016/j.jcps.2011.08.002},
url = {https://www.sciencedirect.com/science/article/pii/S1057740811000829},
author = {Michael I. Norton and Daniel Mochon and Dan Ariely}
}

%
\end{document}